\documentclass[leqno,draft,11pt]{article}
\usepackage{amssymb,amsmath,latexsym,theorem,a4wide}
\usepackage[final]{graphicx}

\def\sameenum{}

\ifx\mydefs\undefined\else \fi
\let\mydefs\relax

\ifx\slide\undefined
  \allowdisplaybreaks[2]
\fi

\ifx\loadcyr\undefined\else
\input cyracc.def
\makeatletter
 at 1\@ptsize pt
 at 1\@ptsize pt
 at 1\@ptsize pt
\makeatother

\fi

\let\M\mathit
\def\gobble#1{}
\def\fixsup#1#2{{#1\let\dp\gobble\mathstrut}^#2_}

\def\bme{\hskip.75em\relax}

\let\eq\leftrightarrow
\let\EQ\Leftrightarrow

\def\iff{\quad\text{iff}\quad}
\let\LOR\bigvee
\let\ET\bigwedge
\let\TO\Rightarrow
\def\?{\mathbin?}

\let\model\vDash
\let\nmodel\nvDash
\def\Par{\mathrm{Par}}

\newbox\circlebox
\setbox\circlebox\hbox{$\bigcirc$}
\def\circled#1{%
  \setbox0\hbox to\wd\circlebox{\hss$#1$\hss}\wd0=0pt
  \box0\copy\circlebox}

\let\fii\varphi
\let\tet\vartheta
\let\ep\varepsilon

\let\roo\varrho
\ifx\slide\empty
  \def\greek#1{$\expandafter\greeknum\csname c@#1\endcsname$}
\else
  \def\greek#1{$\mathop{\boldsymbol{\expandafter\greeknum\csname c@#1\endcsname}}$}
\fi
\makeatletter
\def\greeknum#1{\ifcase#1\or\alpha\or\beta\or\gamma\or\delta\or\ep
      \or\digamma\or\zeta\or\eta\or\tet\or\iota\else\@ctrerr\fi}
\makeatother

\def\p#1{\langle#1\rangle}

\def\lh#1{\lvert#1\rvert}

\let\cat\smallfrown

\let\bez\smallsetminus

\let\sset\subseteq
\let\nsset\nsubseteq
\let\ssset\subsetneq
\let\Sset\supseteq
\let\nSset\nsupseteq
\let\sSset\supsetneq
\let\onto\twoheadrightarrow

\def\cupd{\mathbin{\dot\cup}}
\def\bigcupd{\mathop{\dot\bigcup}}
\def\pw#1{\mathcal P(#1)}
\let\nul\varnothing
\def\res{\mathbin\restriction}

\def\two{\mathbf2}
\def\lex{\mathrm{Lex}}
\let\nulstr\epsilon

\newcommand\rpair[3][3em]{\mathrel{%
   \begin{matrix}%
     \strut\smash{\xrightonto{\hbox to#1{\hss$#2$\hss}}}\\[-1.7ex]%
     \strut\smash{\xleftembed[\hbox to#1{\hss$#3$\hss}]{}}%
   \end{matrix}}}
\makeatletter
\newcommand\xrightonto[2][]{\ext@arrow 0359\rightontofill{#1}{#2}}
\newcommand\xleftembed[2][]{\ext@arrow 3095\leftembedfill{#1}{#2}}
\def\leftembedfill{\arrowfill@\leftarrow\relbar\hookleftnoarrow}
\def\rightontofill{\arrowfill@\relbar\relbar\onto}
\def\hookleftnoarrow{\DOTSB\relbar\joinrel\rhook}
\makeatother

\def\fracd#1#2{\frac{\displaystyle#1}{\displaystyle#2}}

\mathchardef\#="2023 

\def\tr#1{\lVert#1\rVert}

\ifx\busspf\undefined\else
\usepackage{bussproofs}
\EnableBpAbbreviations

\fi

\let\dia\Diamond
\def\diadot{\centerdot\dia}
\ifx\symlasy\undefined
  
  \def\boxdot{{\origboxdot}}
  \def\centerdot#1{{%
    \setbox0\hbox{$\mathop{#1}$}\dimen0 \ht0
    \setbox0\hbox{$#1$}\advance\dimen0 -\ht0
    \setbox2\hbox to\wd0{\hss$\mathop{\cdot}$\hss}\wd2=0pt
    \lower\dimen0\box2\box0 }}
\else
  \def\boxdot{\centerdot\Box}
  \def\centerdot#1{{%
     \setbox0\hbox{$#1$}%
     \raise0.206\ht0\hbox to\wd0{\hss$\cdot$\hss}%
     \kern-\wd0 \box0 }}
\fi

\def\ru{\mathrel/}
\def\Ru{\Bigm/}
\let\sls|
\def\up{\mathord\uparrow}

\def\Up{{\setbox0\hbox{$\uparrow$}%
         \lower\dp0\hbox to\wd0{\hss\vrule width4pt height.4pt\hss}%
         \kern-\wd0\box0}}
\def\UP{{\setbox0\hbox{$\uparrow$}%
         \lower\dp0\hbox to\wd0{\hss\vrule width4pt height.4pt\hss}%
         \kern-\wd0\copy0\kern-\wd0\raise.35ex\box0}}
\def\Down{{\setbox0\hbox{$\downarrow$}%
         \raise\ht0\hbox to\wd0{\hss\vrule width4pt depth.4pt\hss}%
         \kern-\wd0\box0}}

\def\frx{\mathrm{Fr}}
\def\clx{\lgc{Clx}}
\def\CLX{\mathrm{CLX}}
\def\pext{\mathrm{PExt}}
\def\ECI{\EC^\infty}
\def\EC{\M{EC}}

\newif\ifnadm
\DeclareRobustCommand*\adm{\nadmfalse\doadm}
\DeclareRobustCommand*\nadm{\nadmtrue\doadm}
\def\doadm{\mathrel{%
   \setbox0 \hbox{$\mathop\vdash$}\dimen0 \ht0
   \setbox0 \hbox{$\vdash$}\advance\dimen0 -\ht0
   \vrule width.8\fontdimen8 \textfont3 height\ht0 depth\dp0
   \mkern-1mu
   \lower\dimen0 \hbox{$\vcenter{%
      \ifnadm
        \setbox0 \hbox{$\scriptstyle\sim\mathstrut$}%
        \hbox{\hbox to\wd0{\hss$\scriptstyle/$\hss}\kern-\wd0 \box0 }%
      \else
        \hbox{$\scriptstyle\sim\mathstrut$}%
      \fi}$}}}

\def\nr#1{\mathchoice
  {\nrstyle\displaystyle\scriptstyle{#1}}%
  {\nrstyle\textstyle\scriptstyle{#1}}%
  {\nrstyle\scriptstyle\scriptscriptstyle{#1}}%
  {\nrstyle\scriptscriptstyle\scriptscriptstyle{#1}}}
\def\nrstyle#1#2#3{%
  \setbox0\hbox{$#1\bigcirc$}%
  \vcenter{\hbox to\wd0{\hss$#2#3$\hss}}%
  \kern-\wd0\box0 }

\DeclareMathOperator\dom{dom}

\DeclareMathOperator\id{id}

\DeclareMathOperator\cls{cl}

\DeclareMathOperator\Mod{Mod}

\DeclareMathOperator\Var{Var}
\ifx\slide\undefined
  \DeclareMathOperator\Form{Form}
\fi

\DeclareMathOperator\Ext{Ext}
\DeclareMathOperator\NExt{NExt}

\DeclareMathOperator\rcl{rcl}

\DeclareMathOperator\tp{tp}
\DeclareMathOperator\base{bas}
\DeclareMathOperator\xcl{xcl}
\DeclareMathOperator\xcb{xcb}
\DeclareMathOperator\bdt{bd}
\DeclareMathOperator\crit{crit}
\DeclareMathOperator\Sub{Sub}
\DeclareMathOperator\sat{Sat}

\let\cxt\mathrm

\def\st{\expandafter\hat}

\let\lgc\mathbf

\def\CPC{\lgc{CPC}}

\def\I{{\bullet}}
\def\R{{\circ}}

\def\FL#1{\lgc{FL_{#1}}}
\def\DP{\mathrm{DP}}

\mathcode`\*="0203
\let\ob\overline

\mathchardef\mhyphen="2D

\ifx\slide\undefined

\def\noproof{\leavevmode\unskip\bme\vadjust{}\nobreak\hfill$\qed$\par}
\let\qed\Box
\newenvironment{Pf}[1][]
  {\par\noindent\textit{Proof\optpar{#1}:}\bme\ignorespaces}
  {\noproof\pagebreak[2]\vskip\medskipamount\ignorespacesafterend}
\def\optpar#1{\ifx\relax#1\relax\else\ #1\fi}
\def\qedhere{\relax\ifmmode\eqno\qed\expandafter\aftergroup
                   \else\noproof\fi\noqed}
\def\noqed{\let\noproof\relax}

\theoremstyle{plain}
\ifx\shortthm\undefined
\newtheorem{Thm}{Theorem}[section]
\else
\newtheorem{Thm}{Theorem}
\fi
\newtheorem{Prop}[Thm]{Proposition}
\newtheorem{Cor}[Thm]{Corollary}
\newtheorem{Lem}[Thm]{Lemma}

\newtheorem{Obs}[Thm]{Observation}

\newtheorem{Prob}[Thm]{Problem}

\newtheorem{Cl}{Claim}[Thm]
\ifx\shortthm\undefined
\def\theCl{\arabic{Cl}}
\fi

\theorembodyfont\upshape
\newtheorem{Def}[Thm]{Definition}
\newtheorem{Rem}[Thm]{Remark}
\newtheorem{Exm}[Thm]{Example}

\newenvironment{Pf*}{\let\qed\qedCl\Pf}\endPf

\fi

\ifx\slide\undefined
  \usepackage[reftex]{theoremref}
\fi

\newif\iflinenumbers
\linenumberstrue

{\catcode`\^^I=13 \catcode`\^^M=13
\gdef\doalgo#1#2\end#{\hbox to\hsize{\hss \let^^I\qquad%
  \def\\^^M{\nobreak\hfil\break\vadjust{}\qquad}%
  \fboxsep1em \linenum0 %
  \fbox{\hsize#1\vbox{%
  \everypar{\advance\linenum1 %
      \hbox to1.2em{%
           \hss\iflinenumbers$\scriptstyle\the\linenum$\hskip.6em\fi}}%
  #2}}\hss}\end}}
\newcount\linenum

\def\key{\relax\ifmmode\expandafter\mathbf\else\expandafter\textbf\fi}

\def\allowhyphens{\nobreak\hskip0pt\relax}
\def\hyph{\allowhyphens-\hskip0pt\relax}
\DeclareRobustCommand*\magiclparen{\ifmmode(\else\textup(\allowhyphens\fi}
\DeclareRobustCommand*\magicrparen{\ifmmode)\else\textup)\fi}
\let\lparen=(  \let\rparen=)
\def\magicparon{\catcode`\(\active\catcode`\)\active}
\def\magicparoff{\catcode`\(12 \catcode`\)12 }
\AtBeginDocument{\ifx\ifPreview\iftrue\else\magicparon\fi}
\magicparon
\let (=\magiclparen  \let )=\magicrparen

\ifx\sameenum\undefined
  \def\theenumi{\roman{enumi}}
  \ifx\enumup\undefined
    
  \else
    
  \fi

\else
  \def\theenumi{(\roman{enumi})}

\fi

\magicparoff

\mathchardef\comma=\mathcode`\,
{\catcode`\,=\active \gdef,{\comma\penalty\relpenalty}}

\ifx\slide\undefined

\providecommand\dedic{\thanks{Supported by
grant IAA100190902 of GA AV \v CR, Center of Excellence CE-ITI under the grant
P202/12/G061 of GA \v CR, and RVO: 67985840.}}

\author{Emil Je\v r\'abek\dedic\\[\medskipamount]
Institute of Mathematics of the Academy of Sciences\\
\small \v Zitn\'a 25,
115\:67 Praha 1,
Czech Republic,
email: \texttt{jerabek@math.cas.cz}
}

\else\ifx

\author{Emil Je\v r\'abek}
\email{jerabek@math.cas.cz\\[-1em]http://math.cas.cz/\string~jerabek/}
\institution{Institute of Mathematics of the Academy of Sciences, Prague}

\else 

\author[Emil Je\v r\'abek]{Emil Je\v r\'abek\\[\medskipamount]
   \scriptsize\texttt{jerabek@math.cas.cz}\\\texttt{http://math.cas.cz/\string~jerabek/}}
\institute{Institute of Mathematics of the Academy of Sciences, Prague}

\fi\fi

\def\cput(#1)#2{\put(#1){\hbox to0pt{\hss#2\hss}}}

\title{Rules with parameters in modal logic~I}

\begin{document}
\maketitle

\begin{abstract}
We study admissibility of inference rules and unification with
parameters in transitive modal logics (extensions of~$\lgc{K4}$), in
particular we generalize various results on parameter-free
admissibility and unification to the setting with parameters.

Specifically, we give a characterization of projective formulas
generalizing Ghilardi's characterization in the parameter-free case,
leading to new proofs of Rybakov's results that
admissibility with parameters is decidable and unification is finitary
for logics satisfying suitable frame extension properties (called
cluster-extensible logics in this paper). We construct explicit bases
of admissible rules with parameters for cluster-extensible logics, and
give their semantic description. We show that in the case of finitely
many parameters, these logics have independent bases of admissible
rules, and determine which logics have finite bases.

As a sideline, we show that cluster-extensible logics have various
nice properties: in particular, they are finitely axiomatizable, and
have an exponential-size model property. We also give a rather general
characterization of logics with directed (filtering) unification.

In the sequel, we will use the same machinery to investigate the
computational complexity of admissibility and unification with
parameters in cluster-extensible logics, and we will adapt the results
to logics with unique 
top cluster (e.g., $\lgc{S4.2}$) and superintuitionistic logics.
\end{abstract}

\section{Introduction}\label{sec:introduction}

Admissibility of inference rules is among the fundamental properties of
nonclassical propositional logic: a rule is admissible if the set of
tautologies of the logic is closed under the rule, or equivalently, if
adjunction of the rule to the logic does not lead to derivation of new
tautologies. Admissible rules of basic transitive modal logics
($\lgc{K4}$, $\lgc{S4}$, $\lgc{GL}$, $\lgc{Grz}$, $\lgc{S4.3}$, \dots)
are fairly well understood. Rybakov proved that admissibility in a
large class of modal logics is decidable and provided semantic
description of their admissible rules, see~\cite{ryb:bk} for a
detailed treatment. 
Ghilardi~\cite{ghil} gave a characterization of projective
formulas in terms of extension properties of their models, and
proved the existence of finite projective approximations. This led
to an alternative proof of some of Rybakov's results, and it was
utilized by Je\v r\'abek~\cite{ejadm,ej:indep} to construct explicit
bases of admissible rules, and to determine the computational
complexity of admissibility~\cite{ej:admcomp}. A sequent calculus for
admissible rules was developed by Iemhoff and Metcalfe~\cite{iem-met}.
Methods used for transitive modal logics were paralleled by a similar
treatment of intuitionistic and intermediate logics, see
e.g.~\cite{ryb:bk,ghilil,iem:aripc,iem:imed2}.

Admissibility is closely related to
unification~\cite{baa-sny,baa-ghi}: for equational
theories corresponding to algebraizable propositional logics,
$E$-unification can be stated purely in terms of logic, namely a unifier
of a formula is a substitution which makes it a tautology. Thus, a
rule is admissible iff every unifier of the premises of the rule
also unifies its conclusion, and conversely the unifiability of a
formula can be expressed as nonadmissibility of a rule with
inconsistent conclusion. In fact, the primary purpose of
Ghilardi~\cite{ghil} was to prove that unification in the modal logics
in question is finitary.

In unification theory, it is customary to work in a more general
setting that allows for extension of the basic equational theory by
free constants. In logical terms, formulas may include atoms
(variously called parameters, constants, coefficients, or
metavariables) that behave like ordinary propositional variables for
most purposes, but are required to be left fixed by substitutions.
Some of the above-mentioned results on admissibility
in transitive modal logics also apply to admissibility and unification
with parameters, in particular Rybakov~\cite{ryb:s4con,ryb:provlog,ryb:grz,ryb:bk} proved the
decidability of admissibility with parameters in basic transitive
logics, and he has recently extended his method to show that unification with
parameters is finitary in these logics~\cite{ryb:modunifcoef,ryb:intunifcoef2}. Nevertheless, a significant part of the
theory only deals with parameter-free rules and unifiers.

The purpose of this paper is to (at least partially) remedy this
situation by extending some of the results on admissibility in
transitive modal logics to the setup with parameters. Our basic
methodology is similar to the parameter-free case, however the
presence of parameters brings in new phenomena leading to nontrivial
technical difficulties that we have to deal with.

For a more detailed overview of the content of the paper, after
reviewing basic concepts and notation (Section~\ref{sec:prel-notat})
and establishing some elementary background on multiple-conclusion
consequence relations with parameters
(Section~\ref{sec:param-cons-relat}), we start in
Section~\ref{sec:projective-formulas} with a parametric
version of Ghilardi's characterization of projective formulas in
transitive modal logics with the finite model property in terms of
a suitable model extension property on finite models. In
Section~\ref{sec:clx-logics}, we introduce the class of
cluster-extensible (clx) logics (and more generally, $\Par$-extensible
logics for the case when the set~$\Par$ of allowed parameters is finite),
and we use the results from Section~\ref{sec:projective-formulas} to
show that in clx (or $\Par$-extensible) logics, all formulas have
projective approximations. As a corollary, this reproves results of Rybakov~\cite{ryb:bk,ryb:modunifcoef} that such logics~$L$ have finitary
unification type, and if $L$ is decidable, then admissibility in~$L$
is also decidable, and one can compute a finite complete set of
unifiers of a given formula. In order to determine which of these
logics have unitary unification, we include in
Section~\ref{sec:directed-unification} a simple syntactic
criterion for directed (filtering) unification, vastly generalizing
the result of Ghilardi and Sacchetti~\cite{ghi-sacc}. In
Section~\ref{sec:struct-clx}, we look more closely at semantic and
structural properties of clx logics: we show that
every clx logic is finitely axiomatizable, decidable,
$\forall\exists$-definable on finite frames, and has an exponential-size
model property. Moreover, the class of clx logics is closed under
joins in the lattice
of normal extensions of~$\lgc{K4}$. (These results mostly do not have
good analogues in the parameter-free case; they exploit the fact that
the extension conditions designed to make the other results on
admissibility and unification work need to be more restrictive when
parameters are considered.)

In Section~\ref{sec:extension-rules}, we introduce
(multiple-conclusion) rules corresponding to the existence of a
parametric version of tight predecessors, generalizing the
parameter-free rules considered in~\cite{ejadm,ej:indep}. We
investigate their semantic properties, and as the main result of this
section, we show that these extension rules form bases of admissible
rules for clx or $\Par$-extensible logics. We present
single-conclusion variants of these bases in
Section~\ref{sec:single-concl-bases}. Finally, in
Section~\ref{sec:finite-indep-bases}, we modify the extension rules
further to provide independent bases of admissible rules with finitely
many parameters for $\Par$-extensible logics, and we show that finite
bases exist if and only if the logic has bounded branching.

As the name suggests, this paper is to be continued by a sequel, where
we will address the computational complexity of admissibility and
unification with parameters in clx logics, and modifications of our
results to related classes of logics: modal logics whose finite rooted
frames have a single top cluster (such as $\lgc{K4.2}$
and~$\lgc{S4.2}$), and intuitionistic and intermediate logics.

\section{Preliminaries and notation}\label{sec:prel-notat}

The purpose of this section is to review basic definitions and
standard facts we are going to use in order to fix our terminology and
notation. For more detailed information, we refer the reader to
\cite{cha-zax} (modal logic), \cite{ryb:bk} (admissible rules),
\cite{baa-sny,baa-ghi} (unification), \cite{jans-sep} (propositional
consequence relations), \cite{sh-sm} (multiple-conclusion consequence
relations).

We will work with \emph{propositional languages}~$\Form$ consisting of
formulas
freely built from a (usually countably infinite) set of \emph{atoms}
using a fixed set of finitary \emph{connectives}. (We distinguish
two types of atoms: variables and parameters. We will elaborate on
this in Section~\ref{sec:param-cons-relat}.) We will usually denote
formulas by lowercase Greek letters $\fii,\psi,\chi,\dots$. We write
$\psi\sset\fii$ if $\psi$ is a subformula of~$\fii$. $\Sub(\fii)$
denotes the set of all subformulas of a formula~$\fii$, and $\lh\fii$
the length (i.e., the number of symbols) of~$\fii$. Finite sets of
formulas will be usually denoted by uppercase Greek letters
$\Gamma,\Delta,\dots$.

Let us fix a propositional language~$\Form$.
An \emph{atomic substitution} is a mapping $\sigma\colon\Form\to\Form$
that commutes with connectives, hence it is uniquely determined by its
values on atoms. (We reserve the term ``substitution'' for
parameter-preserving substitutions, see below.) A
\emph{(propositional) logic}~$L$ is an atomic
single-conclusion consequence relation: a binary relation between
finite sets of formulas and formulas (written in infix notation as~$\Gamma\vdash_L\fii$), satisfying
\begin{enumerate}
\item\label{item:53}
(identity) $\fii\vdash_L\fii$,
\item\label{item:54}
(weakening) $\Gamma\vdash_L\fii$ implies $\Gamma,\Delta\vdash_L\fii$,
\item\label{item:55}
(cut) $\Gamma\vdash_L\fii$ and $\Gamma,\fii\vdash_L\psi$ implies
$\Gamma\vdash_L\psi$,
\item\label{item:56}
(substitution) $\Gamma\vdash_L\fii$ implies
$\sigma(\Gamma)\vdash_L\sigma(\fii)$ for every atomic
substitution~$\sigma$.
\end{enumerate}
Here we employ common conventions for sets of formulas:
$\Gamma,\Delta$ denotes $\Gamma\cup\Delta$, $\fii$ can stand
for~$\{\fii\}$, and $\sigma(\Gamma)=\{\sigma(\fii):\fii\in\Gamma\}$.
We also write ${}\vdash_L\fii$ instead of~$\nul\vdash_L\fii$; such
formulas~$\fii$ are called \emph{$L$-tautologies}. A logic is
\emph{inconsistent} if all formulas are tautologies, otherwise it is
\emph{consistent}. Note that our
consequence relations are by definition finitary, or more precisely,
they are finitary fragments of consequence relations under a more
conventional definition; we consider our choice more convenient for
the purpose of investigation of admissible rules and unification, as
these only concern the finitary fragment of a given logic.

Being a binary relation, a logic is a
set of pairs $\p{\Gamma,\fii}$, where $\Gamma$ is a finite set of
formulas, and $\fii$ a formula. Such pairs are called
\emph{single-conclusion rules}, and we write them as~$\Gamma\ru\fii$.
In this context, a formula~$\fii$ can be identified with the
rule~$\nul\ru\fii$ (an \emph{axiom}). If $L$ is a logic and $X$ a set
of single-conclusion rules (or formulas), $L\oplus X$ denotes the smallest
logic including $L$ and~$X$. (We reserve $+$ for parametric
consequence relations, see below.) 

In this paper, we will mostly work with normal modal logics. The
basic modal language is generated by the connectives $\to,\bot,\Box$;
other common
connectives ($\dia,\land,\lor,\neg,\eq,\top$) are defined as
abbreviations in the usual way. We also put
$\boxdot\fii=\fii\land\Box\fii$, $\diadot\fii=\fii\lor\dia\fii$.
$\lgc{K4}$ is the smallest logic in the basic modal language that
includes classical
propositional tautologies, and the axioms and rules
\begin{gather*}
\Box(\fii\to\psi)\to(\Box\fii\to\Box\psi),\\
\Box\fii\to\Box\Box\fii,\\
\fii,\fii\to\psi\ru\psi,\\
\fii\ru\Box\fii.
\end{gather*}
A \emph{transitive modal logic} is an axiomatic extension
of~$\lgc{K4}$, i.e., a logic of the form $\lgc{K4}\oplus X$, where $X$
is a set of formulas. (Under our definition,
normal modal logics are identified with their
global consequence relations, whereas in most modal literature they
are identified with their sets of tautologies. Nevertheless, we will
abuse the notation and write $L\Sset\lgc{K4}$ as a short-hand for
``$L$ is a transitive modal logic''. Local consequence relations or
non-normal modal logics do not appear in this paper, hence our usage
of~$\oplus$ agrees with its standard meaning.)

The set of all transitive modal logics ordered by
inclusion is a complete lattice, denoted $\NExt\lgc{K4}$. The meet of a family of logics is
just their intersection, and we will write it as such. We will write
joins with~$\LOR$, though in the case of binary joins we also have
$L_0\lor L_1=L_0\oplus L_1$.

A \emph{(transitive) Kripke frame} is a pair $\p{W,<}$, where $<$ is a
transitive binary relation on a (possibly empty) set~$W$. We will
generally use the same symbol to denote both the frame and its
underlying set. We write $u\le v$ for $u<v\lor
u=v$, $u\sim v$ for $u\le v\le u$, and $u\lnsim v$ for $u<v\land v\nless
u$. We read $u<v$ as ``$v$ is accessible from~$u$'', or for short, ``$u$ sees~$v$''.
A point~$u\in W$ is \emph{reflexive} if $u<u$, and
\emph{irreflexive} otherwise. If $X\sset W$, we define $X\up=\{v\in W:\exists u\in X\,u<v\}$,
$X\Up=\{v\in W:\exists u\in X\,u\le v\}$. The \emph{cluster} of~$u\in W$ is $\cls(u):=\{v\in
W:u\sim v\}$. If $u\in W$, then $W_u$ is
the frame~$\p{u\Up,<}$. If $W=W_u$ for some~$u\in W$, $W$ is called a
\emph{rooted frame}, any such~$u$ is its \emph{root}, and
$\rcl(W):=\cls(u)$ its \emph{root cluster}. If $X\sset W$, an $x\in X$
is called a \emph{maximal} (or \emph{$<$-maximal}) point of~$X$, if
$v\notin X$ for every~$v\gnsim u$. A cluster is \emph{final} if its
points are maximal in~$W$, and \emph{inner} otherwise.
An $X\sset W$ is an
\emph{antichain} if $u\nless v$ for any $u,v\in X$ such that $u\ne v$.

A \emph{(Kripke) model} is a triple $\p{W,<,\model}$, where $\p{W,<}$ is
a Kripke frame, and the \emph{valuation}~$\model$ is a relation between
elements of~$W$ and 
formulas, satisfying the usual conditions for compound formulas. Again,
we often use the same symbol for a model and its underlying set,
and we write $W,u\model\fii$ instead of~$u\model\fii$ when we need to
stress which model the~$\model$ belongs to. We write $W\model\fii$ if
$W,u\model\fii$ for every~$u\in W$.

If $L\Sset\lgc{K4}$, a \emph{Kripke $L$-frame} is a Kripke frame
$\p{W,<}$ such that $W\model\fii$ for every model $\p{W,<,\model}$ and
every $L$-tautology~$\fii$. An \emph{$L$-model} is a model
$\p{W,<,\model}$ such that $\p{W,<}$ is an $L$-frame. $\Mod_L$ denotes the set of all finite
rooted $L$-models. For a formula~$\fii$, we put
$\Mod_L(\fii):=\{W\in\Mod_L:W\model\fii\}$. Notice that
$\Mod_L(\fii)=\Mod_L(\boxdot\fii)$. $L$ has the \emph{finite model
property (fmp)} if $\Mod_L(\fii)=\Mod_L$ implies $\vdash_L\fii$ for
every formula~$\fii$.

If $W$ is a model and $\sigma$ a substitution, we define $\sigma(W)$
to be the model based on the same frame such that
$\sigma(W),u\model\fii$ iff $W,u\model\sigma(\fii)$ for every
formula~$\fii$ and~$u\in W$. Notice that
$(\sigma\circ\tau)(W)=\tau(\sigma(W))$.

Let $W$ be a finite model. The \emph{depth} of~$W$ is the maximal
length of a chain $x_1\lnsim x_2\lnsim\dots\lnsim x_n$ in~$W$. The
\emph{branching} of~$W$ is the maximal number of immediate successor
clusters of any node~$u\in W$. The \emph{width} of~$W$ is the maximal
size of an antichain in any rooted subframe of~$W$.

For any formula~$\fii$, we put $\fii^1=\fii$, $\fii^0=\neg\fii$. If
$\Gamma$ is a set of formulas, $\two^\Gamma$ denotes the set of
all assignments $e\colon\Gamma\to\two$, where $\two:=\{0,1\}$. If
$\Gamma$ is finite and $e\in\two^\Gamma$, we put
$\Gamma^e:=\ET_{\fii\in\Gamma}\fii^{e(\fii)}$. (Here and elsewhere, the
empty conjunction is defined as~$\top$, and the empty disjunction as~$\bot$.)
Conversely, if
$W$ is a Kripke model and $u\in W$, $\sat_\Gamma(W,u)$
(shortened to $\sat_\Gamma(u)$ if $W$ is understood from the context)
denotes the assignment $e\in\two^\Gamma$ such that $e(\fii)=1$ iff
$W,u\model\fii$. If $W$ is a model and $\fii$ a formula,
$W\res\fii$ denotes $\{u\in W:W,u\model\fii\}$.

A \emph{general frame} is $\p{W,<,A}$, where $\p{W,<}$ is a Kripke
frame, and $A\sset\pw W$ is a Boolean algebra of sets closed under the
operation $\Box X:=\{u\in W:\forall v\,(u<v\TO v\in X)\}$, or equivalently,
under $\dia X:=\{u\in W:\exists v\in X\,u<v\}$. Sets from~$A$ are
called \emph{admissible} (or \emph{definable}), and their arbitrary
intersections are \emph{closed sets}.
A Kripke frame
$\p{W,<}$ can be identified with the general frame $\p{W,<,\pw W}$. We
will sometimes write just \emph{frame} instead of general frame,
however finite frames are always assumed to be Kripke frames.
An \emph{admissible valuation} in~$\p{W,<,A}$ is a valuation~$\model$
in~$\p{W,<}$ satisfying $W\res\fii\in A$ for every~$\fii$.

If $\kappa$ is a cardinal number, a general frame $\p{W,<,A}$ is
\emph{$\kappa$-generated} if $A$ is generated as a modal algebra by a
subset of size at most~$\kappa$. (Note that this notion is unrelated
to the similarly named generated subframes.)

A general frame $\p{W,<,A}$ is \emph{refined} if for every $u,v\in W$,
\begin{gather*}
\forall X\in A\,(u\in\Box X\TO v\in X)\TO u<v,\\
\forall X\in A\,(u\in X\TO v\in X)\TO u=v.
\end{gather*}
(In other words, all sets of the form $\{u\}$ or~$u\up$ are closed.)
A family of sets has the \emph{finite intersection property (fip)} if
any its finite subfamily has a nonempty intersection. A frame is
\emph{compact} if every family of admissible (or closed) sets with fip
has a nonempty intersection.
Compact refined frames are called \emph{descriptive}.

If $L$ is a transitive modal logic, a \emph{(general) $L$-frame} is a
frame $\p{W,<,A}$ such that $W\model\fii$ for every admissible
valuation~$\model$ and $L$-tautology~$\fii$. Every~$L$ is complete
with respect to descriptive $L$-frames.

We will use the following well-known property:
\begin{Lem}\th\label{lem:descrmax}
If $\p{W,<,A}$ is a  descriptive frame, $X\sset W$ is closed, and $u\in
X$, then there exists a $<$-maximal $v\in X$ such that $u\le v$.
\end{Lem}
\begin{Pf}
If $C\sset X$ is a nonempty chain, the set $S=\{X\}\cup\{v\Up:v\in
C\}$ has fip. Since each $v\Up$ is closed, $S$ has a nonempty
intersection, and any $w\in\bigcap S$ is an element of~$X$
majorizing~$C$. Thus, $\p{X,\le}$ satisfies the assumptions of Zorn's
lemma, and the result follows.
\end{Pf}

Let $\p{W,<,A}$ and $\p{V,\prec,B}$ be general frames. $V$ is a
\emph{generated subframe} of~$W$ if $V\sset W$,
${\prec}={<}\cap(V\times W)$ (which implies $V\up\sset V$), and
$B=\{X\cap V:X\in A\}$. A \emph{p-morphism} from~$W$ to~$V$ is a
mapping $f\colon W\to V$ such that
\begin{itemize}
\item $f^{-1}(X)\in A$,
\item $f(u)\prec v$ iff there is $u'>u$ such that $f(u')=v$,
\end{itemize}
for every $u\in W$, $v\in V$, and~$X\in B$. The \emph{disjoint union}
of frames $\p{W_i,<_i,A_i}$, $i\in I$, is the frame $\p{W,<,A}$ whose
underlying set $W$ is the disjoint union $\bigcupd_{i\in I}W_i$,
${<}=\bigcup_{i\in I}<_i$, and $A=\{X\sset W:\forall i\in I\,(X\cap
W_i\in A_i)\}$. Generated submodels, and p-morphisms and disjoint
unions of models, are defined similarly.

We will usually index sequences of formulas, frames, points, and other
objects by nonnegative integers, whose set is denoted~$\omega$. In
particular, if $n\in\omega$, then $i<n$ (without further qualification
such as $1\le i<n$) means $i=0,\dots,n-1$.

\subsection{Parametric consequence relations}\label{sec:param-cons-relat}

As already mentioned, we consider atoms of two kinds: variables and
parameters (in unification literature, the latter are usually called
constants). The set of all variables is denoted~$\Var$, and we assume
it is countably infinite. We can enumerate $\Var=\{x_n:n\in\omega\}$,
but for ease of reading we will also use letters $x,y,z,\dots$ for
variables. The set of all parameters is denoted~$\Par$, and we will
use letters such as $p,q,r,\dots$ for parameters. We assume that
$\Par$ is at most countable, but we allow it to be infinite or finite
(or even empty, so that our results subsume the parameter-free case).
If $P\sset\Par$ and $V\sset\Var$, $\Form(P,V)$ denotes the set of
modal formulas in parameters~$P$ and variables~$V$.

A \emph{substitution} is an atomic substitution $\sigma$ such that
$\sigma(p)=p$ for every parameter~$p$. A \emph{single-conclusion
consequence relation} is a relation between finite sets of formulas
and formulas (or equivalently, a set of single-conclusion rules) which
satisfies conditions \ref{item:53}--\ref{item:55} from the definition
of a logic, as well as
\begin{enumerate}
\def\theenumi{(iv')}
\item\label{item:57}
$\Gamma\vdash\fii$ implies $\sigma(\Gamma)\vdash\sigma(\fii)$ for every
substitution~$\sigma$.
\end{enumerate}
More generally, a \emph{multiple-conclusion consequence relation} (or
just \emph{consequence relation} for short) is a binary relation
between finite sets of formulas, satisfying
\begin{enumerate}
\item\label{item:58}
(identity) $\fii\vdash\fii$,
\item\label{item:59}
(weakening) $\Gamma\vdash\Delta$ implies $\Gamma,\Gamma'\vdash\Delta,\Delta'$,
\item\label{item:60}
(cut) $\Gamma\vdash\fii,\Delta$ and $\Gamma,\fii\vdash\Delta$ implies
$\Gamma\vdash\Delta$,
\item\label{item:61}
(substitution) $\Gamma\vdash\Delta$ implies
$\sigma(\Gamma)\vdash\sigma(\Delta)$ for every substitution~$\sigma$.
\end{enumerate}
This definition implies that the following more general form
of the cut property holds for all finite sets of formulas~$\Theta$:
\begin{enumerate}
\def\theenumi{(iii')}
\item\label{item:65}
(general cut) If $\Gamma,\Pi\vdash\Lambda,\Delta$ for every partition
$\Theta=\Pi\cupd\Lambda$, then $\Gamma\vdash\Delta$.
\end{enumerate}
(We consider here only finitary consequence
relations, however we remark that if we allowed $\Gamma\vdash\Delta$ with $\Gamma,\Delta$ infinite, the proper
definition of consequence relations would need to include \ref{item:65} for arbitrary sets~$\Theta$ in place of the
weaker condition~\ref{item:60}; see~\cite{sh-sm} for details.)

A consequence relation is thus a set of pairs of finite sets of
formulas. We will call such pairs \emph{multiple-conclusion rules}, or
just \emph{rules}, and we will write them as $\Gamma\ru\Delta$.

For every consequence relation~$\vdash$, the set of single-conclusion
rules $\Gamma\ru\fii$ such that $\Gamma\vdash\fii$ is a
single-conclusion consequence relation, the \emph{single-conclusion
fragment of\/~$\vdash$}. Conversely, for every single-conclusion
consequence relation~$\vdash_1$, there is a smallest consequence
relation~$\vdash_m$ whose single-conclusion fragment is $\vdash_1$,
namely $\Gamma\vdash_m\Delta$ iff $\Gamma\vdash_1\fii$ for
some~$\fii\in\Delta$. In particular, if $L$ is a logic, a rule is
\emph{$L$-derivable} if it belongs to the smallest consequence
relation extending~$L$ (which we identify with~$L$ itself).

If $L$ is a consequence relation and $X$ a set of rules, $L+X$ denotes
the smallest consequence relation containing $L$ and~$X$.

Let $L$ be a logic. An \emph{$L$-unifier} of a set of
formulas~$\Gamma$ is a substitution~$\sigma$ such that
$\vdash_L\sigma(\fii)$ for every~$\fii\in\Gamma$. A rule
$\Gamma\ru\Delta$ is \emph{$L$-admissible} if every unifier
of~$\Gamma$ also unifies some~$\fii\in\Delta$. The set of all
$L$-admissible rules forms a consequence relation which we
denote~$\adm_L$. A \emph{basis} of $L$-admissible rules is a set of
rules~$B$ such that ${\adm_L}=L+B$.

A logic $L$ is \emph{(finitely) equivalential} if there is a finite set of
formulas~$E(x,y)$ such that
\begin{align*}
&\vdash_L\ep(x,x)\quad\text{for all $\ep\in E$},\\
E(x,y),\fii(x)&\vdash_L\fii(y)
\end{align*}
for every formula~$\fii$, possibly involving other variables not
shown. Modal logics are equivalential with $E(x,y)=\{x\eq y\}$.
Substitutions $\sigma,\tau$ are \emph{equivalent}, written $\sigma=_L\tau$, if
$\vdash_L\ep(\sigma(x),\tau(x))$ for every variable~$x$ and $\ep\in E$. A substitution $\tau$ is
\emph{more general} than~$\sigma$, written $\sigma\le_L\tau$, if
$\sigma=_L\upsilon\circ\tau$ for some substitution~$\upsilon$. A
\emph{complete set of unifiers} of a set of formulas~$\Gamma$ is a
set~$C$ of unifiers of~$\Gamma$ such that every unifier of~$\Gamma$ is
less general than some~$\sigma\in C$. A complete set of unifiers is
\emph{minimal} if no $C'\ssset C$ is complete, or equivalently, if $C$
consists of pairwise incomparable $\le_L$-maximal unifiers. A
\emph{most general unifier (mgu)} of~$\Gamma$ is a unifier of~$\Gamma$
more general than any other unifier of~$\Gamma$.
If $\Gamma$ has a minimal complete set of unifiers~$C$, its
cardinality is an invariant of~$\Gamma$. We say that $\Gamma$ is of
\begin{itemize}
\item type $1$ (unary), if $\lh C=1$ (i.e., $\Gamma$ has a mgu),
\item type $\omega$ (finitary), if $1<\lh C<\aleph_0$,
\item type $\infty$ (infinitary), if $C$ is infinite,
\item type $0$ (nullary), if $\Gamma$ has no minimal complete set of
unifiers.
\end{itemize}
The \emph{unification type} of~$L$ is the maximum of types of unifiable
finite sets of formulas~$\Gamma$, where the types are ordered as
$1<\omega<\infty<0$. $L$ has \emph{at most finitary} unification, if
its unification type is unary or finitary.

A \emph{parametric Kripke frame} is $\p{W,<,\model_p}$, where
$\p{W,<}$ is a Kripke frame, and ${\model_p}\sset W\times\Par$.
Similarly, a \emph{parametric (general) frame} is
$\p{W,<,A,\model_p}$, where $\p{W,<,A}$ is a general frame, and
${\model_p}\sset W\times\Par$ satisfies $\{u\in W:u\model_pp\}\in A$
for every~$p\in P$. An \emph{admissible valuation}
in~$\p{W,<,A,\model_p}$ is an admissible valuation~$\model$
in~$\p{W,<,A}$ such that ${\model}\Sset{\model_p}$.

A rule $\roo=\Gamma\ru\Delta$ is \emph{satisfied} in a model $\p{W,<,\model}$
if $W\nmodel\fii$ for some~$\fii\in\Gamma$, or $W\model\fii$ for
some~$\fii\in\Delta$. A rule~$\roo$ is \emph{valid} in a parametric frame
$W$, written $W\model\roo$, if $\roo$ is satisfied in any model based
on an admissible valuation in~$W$.

\emph{Generated subframes} and \emph{disjoint unions} of parametric frames are
defined in the obvious way. \emph{P-morphisms} of parametric frames are
required to preserve the valuation of parameters in both directions.
Validity of rules in parametric frames is preserved by p-morphic
images. Single-conclusion rules are also preserved under disjoint
unions, and premise-free rules under generated subframes.

Note that parametric frames (and other parametrified notions in this subsection) are not technically any more demanding
than usual frames or models. The purpose they serve is to avoid endless repetition that various frames come with a
predefined valuation of parameters, which is supposed to be preserved by constructs such as p-morphisms. They are also
conceptually important in that they provide adequate semantics for modal logic in signature expanded with free constants, just
as usual frames give an adequate semantics for modal logic in its basic signature.

Let $L\Sset\lgc{K4}$, $P\sset\Par$, $V\sset\Var$.
The \emph{canonical frame $C_L(P,V)$} is the descriptive parametric
frame $\p{C,<,A,\model_p}$, where $C$ is the collection of maximal
$L$-consistent subsets of $\Form(P,V)$; for $X,Y\in C$, we put $X<Y$ iff
$\{\fii:\Box\fii\in X\}\sset Y$ iff $\{\dia\fii:\fii\in X\}\sset Y$;
$A$~consists of sets of the form $\{X\in C:\fii\in X\}$, where
$\fii\in\Form(P,V)$; and we put $X\model_pp$ iff $p\in X$ for $p\in\Par$ and $X\in C$.

We have $C_L(P,V)\model L$. On the other hand, if $\nvdash_L\fii$,
where $\fii\in\Form(P,V')$ and $\lh{V'}\le\lh V$, then
$C_L(P,V)\nmodel\fii$. The following standard lemma follows easily.
\begin{Lem}\th\label{lem:admcanon}
Let $\Gamma\ru\Delta$ be a rule whose parameters are included in~$P$.
\begin{enumerate}
\item\label{item:62}
If $\Gamma\adm_L\Delta$, then $C_L(P,V)\model\Gamma\ru\Delta$ for
every~$V\sset\Var$.
\item\label{item:63}
If $\Gamma\nadm_L\Delta$, there is $n\in\omega$ such that
$C_L(P,V)\nmodel\Gamma\ru\Delta$ whenever $\lh V\ge n$.
\end{enumerate}
\end{Lem}

If $\mathcal W$ is a class of parametric frames, the set of all rules valid
in~$\mathcal W$ is easily seen to be a consequence relation
extending~$\vdash_\lgc{K4}$. On the other hand, every consequence
relation ${\vdash}\Sset{\vdash_\lgc{K4}}$ is complete wrt a class of
(finitely generated) descriptive frames. In particular, if
$\Gamma\nvdash\Delta$, where $\Gamma\cup\Delta\sset\Form(P,V)$, then
the general cut property \ref{item:65} from p.~\pageref{item:65} and Zorn's lemma imply that there
exists a partition $\Gamma'\cupd\Delta'=\Form(P,V)$ such that
$\Gamma\sset\Gamma'$, $\Delta\sset\Delta'$, and
$\Gamma''\nvdash\Delta''$ for any finite $\Gamma''\sset\Gamma'$,
$\Delta''\sset\Delta'$. Then $W:=\{X\in C_\lgc{K4}(P,V):\Gamma'\sset
X\}$ is a closed (hence descriptive) generated subframe
of~$C_\lgc{K4}(P,V)$, and one checks easily that $W\models{\vdash}$
and $W\nmodel\Gamma\ru\Delta$ (see e.g.\ \cite[Thm.~2.2]{ej:canrules}
for the parameter-free case).

On a related note, descriptive parametric frames can be embedded in
canonical frames. The lemma below holds for arbitrary
cardinals~$\kappa$ if we allow uncountable sets of variables and
parameters, but we will only need it for finite (hence finitely
generated) frames.
\begin{Lem}\th\label{lem:descrcanon}
Let $L\Sset\lgc{K4}$, $P\sset\Par$, and $W$ be a parametric $\kappa$-generated
descriptive $L$-frame. If $V$ is a set of variables such that
$\lh V\ge\kappa$, there is a general frame isomorphism from~$W$
onto a closed generated subframe of~$C_L(P,V)$, preserving the
valuation of parameters from~$P$.
\end{Lem}
\begin{Pf}
Let $F$ be the free $L$-algebra generated by $P\cup V$, and $h$ a
homomorphism from~$F$ to the algebra~$A$ of admissible sets
of~$W$, mapping $V$ onto a set of generators of~$A$, and each~$p\in P$
to the element of~$A$ given by the valuation of~$p$ in~$W$. Since
$h$ is onto, the dual p-morphism from~$W$ to~$C_L(P,V)$ is injective,
and it has the required properties.
\end{Pf}

A \emph{nonstructural consequence relation} is a binary relation between
finite sets of formulas satisfying conditions \ref{item:58},
\ref{item:59}, \ref{item:60} from the definition of
multiple-conclusion consequence relations. We will not refer to
nonstructural consequence relations directly, but we will extend the
$\vdash$ notation to rules as follows. Let $\vdash$ be a (structural)
consequence relation, and $R\cup\{\roo\}$ a set of rules. We write
$R\vdash\roo$ if $\roo$ is in the least nonstructural consequence
relation containing $\vdash$ and~$R$.

Note that if the rules in $R$ and~$\roo$ are just axioms,
$R\vdash\roo$ iff the same holds for the corresponding formulas
under the original consequence relation~$\vdash$, thus this overloading of the
symbol~$\vdash$ does not lead to conflicts. Also, if
$\roo=\Gamma\ru\Delta$, then $\nul\vdash\roo$ iff $\Gamma\vdash\Delta$.

($\vdash$ defines a sort of a single-conclusion consequence relation
operating with multiple-conclusion rules instead of formulas, but we
will not use this terminology in order to avoid unnecessary
confusion.)

If $\vdash$ extends~$\vdash_\lgc{K4}$, and $W$ is a frame
validating~$\vdash$, then $R\vdash\roo$ implies that every admissible
valuation $\model$ in~$W$ that satisfies all rules from~$R$ also
satisfies~$\roo$. One can in fact show easily that $\vdash$ is
complete with respect to this semantics, but we will not need this.
Rather, we will use the following lemma which follows from
\cite[L.~2.3, 2.4]{ejadm}, but we include a direct proof for completeness.
\begin{Lem}\th\label{lem:fmprules}
Let $L\Sset\lgc{K4}$ have fmp, and $R\cup\{\roo\}$ be a finite
set of rules. If $R\nvdash_L\roo$, there is a finite $L$-model $W$
such that $W\model R$, and $W\nmodel\roo$.
\end{Lem}
\begin{Pf}
Write $\roo=\Gamma\ru\Delta$, and let $\Sigma$ be the set of all
formulas occurring in $R\cup\{\roo\}$. By the general cut property~\ref{item:65},
there is a partition $\Sigma=X\cupd Y$ such that $\Gamma\sset X$,
$\Delta\sset Y$, and $R\nvdash_LX\ru Y$. In particular, if
$\Gamma'\ru\Delta'\in R$, then $\Gamma'\cap Y\ne\nul$ or $\Delta'\cap
X\ne\nul$, and for every $\psi\in Y$, $\nvdash_L\boxdot\ET
X\to\psi$. The latter implies that there are models $W_\psi\in\Mod_L$ with
roots $u_\psi$ such that $W_\psi\model X$, and
$W_\psi,u_\psi\nmodel\psi$. Let $W$ be the disjoint union of all
$W_\psi$, $\psi\in Y$. Then $W\model\fii$ for every $\fii\in
X$, and $W\nmodel\psi$ for every $\psi\in Y$. In particular,
$W\model\roo'$ for every $\roo'\in R$, and $W\nmodel\roo$.
\end{Pf}

\section{Projective formulas}\label{sec:projective-formulas}

Let us fix a logic $L\Sset\lgc{K4}$ with the finite model property.
Recall that a formula $\fii$ is \emph{projective} if it has a \emph{projective
unifier}, which is an $L$-unifier $\sigma$ of~$\fii$ such that
\begin{equation}\label{eq:proj}
\fii\vdash_L\sigma(x)\eq x
\end{equation}
for every variable~$x$ (which implies
$\fii\vdash_L\sigma(\psi)\eq\psi$ for every formula~$\psi$). A
projective unifier of a formula is also its most general unifier.

Ghilardi~\cite{ghil} described projective formulas in
the parameter-free case: they are exactly the formulas whose finite
$L$-models have a certain extension property, and moreover, one can
explicitly define for any formula a substitution
satisfying~\eqref{eq:proj} which turns out to be a projective unifier
whenever the formula is projective. The goal of this section is to
generalize this result to projectivity with parameters. Let us start
by defining the relevant extension properties and substitutions.

\begin{Def}\th\label{def:proj}
Models $F,F'\in\Mod_L$ are \emph{variants} of each other if they are
based on the same frame, have the same valuation of parameters, and
their valuation of variables can only differ in $\rcl(F)$. A set of
models $M\sset\Mod_L$ has the \emph{model extension property,} if
every $F\in\Mod_L$ whose all rooted generated proper submodels belong
to~$M$ has a variant in~$M$. A formula~$\fii$ has the model extension
property if this holds for~$M=\Mod_L(\fii)$.

Let $\fii\in\Form(P,V)$, where $P$ and~$V$ are \emph{finite} sets of
parameters and variables, respectively. Let $D=\p{d_x:x\in V}$, where
each $d_x\colon\two^P\to\two$ is a Boolean function of the parameters. We
define the \emph{L\"owenheim substitution} $\theta_{\fii,D}$ by
\[\theta_{\fii,D}(x)=(\boxdot\fii\land x)\lor(\neg\boxdot\fii\land d_x)\]
for every~$x\in V$, where $d_x$ is identified with any Boolean formula
representing it. Notice that sequences $D$ as above
can be equivalently represented as assignments $D\colon\two^P\times
V\to\two$ or $D\colon\two^P\to\two^V$. Let $\theta_\fii$ be the
composition of all $2^{\lh V2^{\lh P}}$ substitutions of the
form~$\theta_{\fii,D}$, in an arbitrary order. We will also write
$\theta_D=\theta_{\fii,D}$ and $\theta=\theta_\fii$ when 
$\fii$ is clear from the context. 
\end{Def}

Notice that in the case $P=\nul$, $D$ can be identified with a
variable assignment $D\colon V\to\two$, and
$\theta_{\fii,D}$ is equivalent to the substitution
\[\theta_{\fii,D}(x)=\begin{cases}
\boxdot\fii\to x,&\text{if }D(x)=1,\\
\boxdot\fii\land x,&\text{if }D(x)=0
\end{cases}\]
considered by Ghilardi. 

It is easy to see that substitutions satisfying \eqref{eq:proj} are
closed under composition, hence $\theta_\fii^N$ satisfies
\eqref{eq:proj} for any $N\in\omega$.

The rest of this section is devoted to the proof of the following
characterization.
\begin{Thm}\th\label{thm:proj}
Let $L\Sset\lgc{K4}$ have the finite model property, and $\fii$ be a
formula in finitely many parameters~$P$ and variables~$V$. Then the
following are equivalent.
\begin{enumerate}
\item\label{item:proj} $\fii$ is projective.
\item\label{item:modext} $\fii$ has the model extension property.
\item\label{item:theta} $\theta_\fii^N$ is a unifier of~$\fii$, where
$N=(\lh B+1)\bigl(2^{\lh P}+1\bigr)$, $B=\{\psi:\Box\psi\sset\fii\}$.
\end{enumerate}
\end{Thm}

\begin{Rem}\th\label{rem:proj}
In the parameter-free case, we obtain $N=2(\lh B+1)\le2\lh\fii$. This
is a considerable improvement over Ghilardi's original proof, which
gives $N$ nonelementary (a tower of exponentials whose height is the
modal degree of~$\fii$).
\end{Rem}

For the next few lemmas, let us fix finite sets of parameters~$P$ and
variables~$V$, a formula $\fii\in\Form(P,V)$ with the model extension
property, and $B=\{\psi:\Box\psi\sset\fii\}$. We aim to show that
$\vdash_L\theta^N(\fii)$, which in view of the fmp of~$L$ amounts to
$\theta^N(\fii)$ being true in every finite rooted $L$-model~$F$.

The basic idea behind the $\theta_D$ substitutions is that their
successive application leaves unchanged the part of~$F$ where
$\fii$ already holds, while we are making progress on the rest of the
model: specifically, a maximal cluster where $\fii$ fails can be made
to satisfy~$\fii$ by applying~$\theta_D$ for a suitably chosen~$D$.

\pagebreak[2]
\begin{Lem}\th\label{lem:progress}
Let $F\in\Mod_L$ and $D\colon\two^P\to\two^V$.
\begin{enumerate}
\item\label{item:fiiholds}
If $F\model\fii$, then $\theta_D(F)=F$.
\item\label{item:fiifails}
If $F,u\nmodel\boxdot\fii$, then $\sat_V(\theta_D(F),u)=D(\sat_P(F,u))$.
\item\label{item:one cluster}
If $F'\model\fii$, where $F'$ is the variant of~$F$ such that
$\sat_V(F',u)=D(\sat_P(F,u))$ for every $u\in\rcl(F)$, then
$\theta_D(F)\model\fii$.
\item\label{item:64}
If ${F\bez\rcl(F)}\model\fii$, then $\theta(F)\model\fii$.
\end{enumerate}
\end{Lem}
\begin{Pf}
\ref{item:fiiholds} and~\ref{item:fiifails} are clear from the
definition of~$\theta_D$.
\ref{item:one cluster}: If $F\model\fii$, the result follows
from~\ref{item:fiiholds}. Otherwise ${F\bez\rcl(F)}\model\fii$, and
$\theta_D(F)=F'$ by \ref{item:fiiholds} and~\ref{item:fiifails}.

\ref{item:64}: By the model extension property, $F$ has a variant~$F'$ such that $F'\model\fii$. We may assume that
points in~$\rcl(F')$ with the same valuation of parameters have the same valuation of variables: we can first collapse
all $u\in\rcl(F)$ with the same $\sat_P(F,u)$ to a single point by a p-morphism, apply the extension property, and
lift back the valuation of variables to the original frame. Let $D$ be such that 
$D(\sat_P(F,u))=\sat_V(F',u)$ for each $u\in\rcl(F)$, and write $\theta=\sigma\circ\theta_D\circ\tau$. By repeated use
of~\ref{item:fiiholds}, we have $\sigma(F)\bez\rcl(\sigma(F))=F\bez\rcl(F)$, hence $\theta_D(\sigma(F))\model\fii$
by~\ref{item:one cluster}, and $\theta(F)=\tau(\theta_D(\sigma(F)))\model\fii$ using~\ref{item:fiiholds} again.
\end{Pf}

\th\ref{lem:progress} implies that $\theta^k(F)\model\fii$ for any
$F\in\Mod_L$ of depth at most~$k$. However, in order to show that some
power of~$\theta$ is a unifier of~$\fii$, we need a uniform bound
on~$k$ independent of~$F$. As in Ghilardi's proof, we will achieve this
by defining a rank function on models whose number of possible values
depends only on~$\fii$, and showing that sufficiently many
applications of~$\theta$ will strictly decrease the rank or make the
model satisfy~$\fii$.

Ghilardi's rank is based on Fine's $n$-equivalence~\cite{fine1}. It
seems that matters become more delicate when we need to deal with
valuation of parameters, hence we need a notion of a rank better
adapted to our particular situation in order to make the arguments go
through. We will use a rank function based on the satisfaction of some
formulas related to~$\fii$. As a side effect, this leads to much
smaller bounds, as already mentioned in \th\ref{rem:proj}. We will
also find it helpful to consider ranks to be the actual sets of formulas
rather than just their cardinality.

\begin{Def}\th\label{def:rank}
If $F\in\Mod_L$, we put
\begin{align*}
R_0(F)&=\{\psi\in B:F\model\Box(\boxdot\fii\to\psi)\},\\
R_1(F)&=\Bigl\{e\in\two^P:F\model P^e\to
  \LOR_{\psi\in B\bez R_0(F)}\Box(\boxdot\fii\to\psi)\Bigr\}.
\end{align*}
Notice that $R_1(F)$ is a proper subset of~$\two^P$, specifically
$\sat_P(F,u)\notin R_1(F)$ for any $u\in\rcl(F)$.
The \emph{crude rank} of $F$ is $R_0(F)$, and its
\emph{rank} is $R(F):=\p{R_0(F),R_1(F)}$. The rank of a point~$u\in F$
is defined as~$R(F_u)$. Ranks are ordered lexicographically:
if $X,X'\sset B$ and $Y,Y'\ssset\two^P$, we put
\begin{align*}
\p{X,Y}\sset_\lex\p{X',Y'}&\iff X\ssset X'\lor(X=X'\land Y\sset Y'),\\
\p{X,Y}\ssset_\lex\p{X',Y'}&\iff X\ssset X'\lor(X=X'\land Y\ssset Y').
\end{align*}
The \emph{numerical rank} of~$F$ is $\tr{R(F)}$, where $\tr{\p{X,Y}}:=2^{\lh
P}\lh{X}+\lh{Y}$. Notice that $\p{X,Y}\ssset_\lex\p{X',Y'}$ implies
$\tr{\p{X,Y}}<\tr{\p{X',Y'}}$.
\end{Def}

\pagebreak[2]
\begin{Lem}\th\label{lem:rank}\
\begin{enumerate}
\item\label{item:rankup}
If $u,v\in F$, $u<v$, then $R(F_u)\sset_\lex R(F_v)$.
\item\label{item:ranklater}
$R(\theta_D(F))\sset_\lex R(F)$.
\end{enumerate}
\end{Lem}
\begin{Pf}
\ref{item:rankup} is obvious from the definition.
\ref{item:ranklater} follows from \th\ref{lem:progress}: when
passing from $F$ to~$\theta_D(F)$, the set of points satisfying
$\boxdot\fii$ can only increase, and the valuation of all formulas
$\psi\in B$ in $F\res\boxdot\fii$ is preserved.
\end{Pf}

The argument for decreasing rank will be different depending on whether
maximal clusters where $\fii$ fails are reflexive or irreflexive. We
treat the reflexive case first.
\begin{Lem}\th\label{lem:rankrefl}
Let $F\in\Mod_L$ be such that all points of $F\res\neg\boxdot\fii$
have the same rank, $R(F)=R(\theta(F))$, and $F\res\neg\boxdot\fii$
has a reflexive $<$-maximal cluster. Then $\theta(F)\model\fii$.
\end{Lem}
\begin{Pf}
Put $R=\p{R_0,R_1}:=R(F)$, $G:=F\res\boxdot\fii$. If $u\in F\bez G$
and $\theta=\sigma\circ\tau$, where $\sigma$, $\tau$ are compositions
of $\theta_D$'s, we have
\begin{equation}\label{eq:constr}
R=R(\theta(F))\sset_\lex R(\theta(F)_u)=R(\tau(\sigma(F))_u)\sset_\lex R(\sigma(F)_u)
\sset_\lex R(F_u)=R
\end{equation}
by \th\ref{lem:rank}, hence $R(\sigma(F)_u)=R$.

Fix $w$ in a reflexive maximal cluster of~$F\bez G$.
We define $D\colon\two^P\to\two^V$ as follows. Let $e\in\two^P$. If $e\in R_1$, we pick
an arbitrary $D(e)\in\two^V$. Otherwise $e\notin R_1(\theta(F)_w)$, hence
there exists $w_e\ge w$ such that
\begin{equation}\label{eq:w_e-def}
\theta(F),w_e\model P^e\land\ET_{\psi\in B\bez R_0}
    \dia(\boxdot\fii\land\neg\psi).
\end{equation}
We define $D(e)=\sat_V(\theta(F),w_e)$. Notice that
\begin{equation}\label{eq:w_e-prop}
\theta(F),w_e\model\boxdot\fii\land P^e\land V^{D(e)}
  \land\ET_{\psi\in R_0}\boxdot\psi
  \land\ET_{\psi\in B\bez R_0}\neg\Box\psi:
\end{equation}
we have $\theta(F),w_e\model\boxdot\fii$ by \th\ref{lem:progress}~\ref{item:64}; if
$\psi\in R_0$, then
$\theta(F)_w\model\boxdot(\boxdot\fii\to\boxdot\psi)$ as
$R_0(\theta(F)_w)=R_0$ and $w$ is reflexive; and if $\psi\in B\bez
R_0$, then $\theta(F),w_e\nmodel\Box\psi$ by~\eqref{eq:w_e-def}.

We can write $\theta=\sigma\circ\theta_D\circ\tau$. We claim
$\theta_D(\sigma(F))\model\fii$, which implies
$\theta(F)=\tau(\theta_D(\sigma(F)))\model\fii$. Assume for
contradiction $\theta_D(\sigma(F)),u\nmodel\fii$; we may also assume
without loss of generality that $\theta_D(\sigma(F)),v\model\fii$ for
every $v\gnsim u$. Since $\sigma(F),u\nmodel\boxdot\fii$ and
$R_0(\theta_D(\sigma(F))_u)=R_0$, we have
\begin{equation}\label{eq:qwer}
\theta_D(\sigma(F)),u'\model P^{\sat_P(u')}\land V^{D(\sat_P(u'))}
  \land\ET_{\psi\in R_0}\Box(\boxdot\fii\to\psi)
  \land\ET_{\psi\in B\bez R_0}\neg\Box\psi
\end{equation}
for every $u'\sim u$ using \th\ref{lem:progress}. Notice that
$\sat_P(u')\notin R_1$. We will show
\begin{equation}\label{eq:tyui}
\theta_D(\sigma(F)),u'\model\chi\iff
  \theta(F),w_{\sat_P(u')}\model\chi
\end{equation}
for every $u'\sim u$ and $\chi\sset\fii$ by induction on the
complexity of $\chi$. If $\chi\in P\cup V$, or $\chi=\Box\psi$ with
$\psi\notin R_0$, then \eqref{eq:tyui} follows immediately from
\eqref{eq:w_e-prop} and~\eqref{eq:qwer}. The steps for Boolean
connectives are trivial. If $\chi=\Box\psi$, $\psi\in R_0$, we have
$\theta(F),w_{\sat_P(u')}\model\Box\psi$ by \eqref{eq:w_e-prop}. On the
other hand, $\theta_D(\sigma(F)),v\model\psi$ for every $v\gnsim u$
by \eqref{eq:qwer}, and for every $v\sim u$ by the induction
hypothesis, since $\theta(F),w_{\sat_P(v)}\model\psi$. Thus,
$\theta_D(\sigma(F)),u'\model\Box\psi$, irrespective of the
reflexivity or irreflexivity of~$\cls(u)$.

However, \eqref{eq:tyui} and \eqref{eq:w_e-prop} imply
$\theta_D(\sigma(F)),u\model\fii$, a contradiction.
\end{Pf}

\begin{Def}
If $F\in\Mod_L$, $F\nmodel\fii$, we define
\begin{align*}
r(F)&:=\max\bigl\{\tr{R(F_u)}:F,u\nmodel\fii\},\\
r_0(F)&:=\max\bigl\{\lh{R_0(F_u)}:F,u\nmodel\fii\},\\
r_1(F)&:=\max\bigl\{\lh{R_1(F_u)}:F,u\nmodel\fii\}.
\end{align*}
If $F\model\fii$, we put $r(F)=r_0(F)=r_1(F)=-\infty$.
\end{Def}

\begin{Cor}\th\label{cor:rankrefl}
If $F\in\Mod_L$ is such that $F\nmodel\fii$, and all $<$-maximal
clusters of $F\res\neg\boxdot\fii$ are reflexive, then
$r(\theta(F))<r(F)$.
\end{Cor}
\begin{Pf}
We have $r(\theta(F))\le r(F)$ by \th\ref{lem:rank}. If
$r(\theta(F))=r(F)$, choose $u\in F$ such that
$\theta(F),u\nmodel\fii$ and $\tr R=r(F)$, where $R=R(\theta(F)_u)$.
As in~\eqref{eq:constr}, we have $R(F_v)=R$ for every $v\in F_u\res\neg\boxdot\fii$. Thus $\theta(F)_u\model\fii$ by
\th\ref{lem:rankrefl}, a contradiction.
\end{Pf}

\begin{Lem}\th\label{lem:rankirr}
Let $F\in\Mod_L$ be such that all points of $F\res\neg\boxdot\fii$
have the same crude rank, $F\res\neg\boxdot\fii$ has an irreflexive
$<$-maximal point, and $R_0(F)=R_0\bigl(\theta^m(F)\bigr)$,
where $m=r_1(F)+2$. Then $\theta^m(F)\model\fii$.
\end{Lem}
\begin{Pf}
Put $R_0=R_0(F)$, and fix an irreflexive maximal point $w\in
F\res\neg\boxdot\fii$. For any $e\in\two^P$, we can change the
valuation of parameters in the root of $F_w$ to match $e$, and apply
the model extension property to obtain its variant $G^e$ such that
$G^e\model\fii$. Let $D(e)$ be the valuation of variables in the root
of~$G^e$. Since $R_0(F_w)=R_0$, $F,w\model\Box\fii$, and valuation
of boxed formulas in $w$ is unaffected by a change of variables or
parameters in~$w$, we have
\begin{equation}\label{eq:g^e-prop}
G^e,w\model\boxdot\fii\land P^e\land V^{D(e)}
   \land\ET_{\psi\in R_0}\Box\psi
   \land\ET_{\psi\in B\bez R_0}\neg\Box\psi.
\end{equation}
We can write $\theta=\sigma\circ\theta_D\circ\tau$, where $\sigma$
and~$\tau$ are compositions of some of the~$\theta_{D'}$. Put
$\eta=\tau\circ\sigma\circ\theta_D$ and
$\eta_k=\sigma\circ\theta_D\circ\eta^k$ for $k=0,\dots,m-1$, so that
$\theta^m=\eta_{m-1}\circ\tau$. Notice that
\begin{equation}\label{eq:constr0}
R_0(\eta_k(F)_u)=R_0
\end{equation}
for any $u\in F\res\neg\boxdot\fii$ and $k<m$, by the same argument
as in~\eqref{eq:constr}.
\begin{Cl}\th\label{cl:reflexive}
For any $k<m$, all $<$-maximal clusters of
$\eta_k(F)\res\neg\boxdot\fii$ are reflexive.
\end{Cl}
\begin{Pf*}
Assume for contradiction that $u$ is an irreflexive maximal point of
$\eta_k(F)\res\neg\boxdot\fii$. Let $e=\sat_P(\eta_k(F),u)$. Since
$\eta_k(F)$ is of the form $\theta_D(\cdots)$, we have
$\eta_k(F),u\model V^{D(e)}$ by \th\ref{lem:progress}. Also
$R_0(\eta_k(F)_u)=R_0$ by~\eqref{eq:constr0},
and $\eta_k(F),u\model\Box\fii$, hence
\[\eta_k(F),u\model P^e\land V^{D(e)}
   \land\ET_{\psi\in R_0}\Box\psi
   \land\ET_{\psi\in B\bez R_0}\neg\Box\psi.\]
By \eqref{eq:g^e-prop}, $\eta_k(F),u$ and $G^e,w$ satisfy the same
Boolean combinations of atoms and boxed subformulas of~$\fii$. In
particular, they agree on the satisfaction of~$\fii$ itself. However,
this contradicts $G^e,w\model\fii$ and $\eta_k(F),u\nmodel\fii$.
\end{Pf*}
Assume for contradiction $\theta^m(F)\nmodel\fii$. Since $\eta$ is a
composition of all $\theta_{D'}$ in some order, we can apply
\th\ref{cor:rankrefl} with $\eta$ in place of~$\theta$, obtaining
$r(\eta_{k+1}(F))<r(\eta_k(F))$ for every $k<m$. In view of
\eqref{eq:constr0}, this implies $r_1(\eta_{k+1}(F))<r_1(\eta_k(F))$.
However, this is impossible, as $r_1(\eta_0(F))\le r_1(F)=m-2$ and
$r_1(\eta_{m-1}(F))\ge0$.
\end{Pf}

\pagebreak[2]
\begin{Lem}\th\label{lem:rankdown}
If $F\in\Mod_L$ and $F\nmodel\fii$, then $r_0\bigl(\theta^{2^{\lh
P}+1}(F)\bigr)<r_0(F)$.
\end{Lem}
\begin{Pf}
Put $m=2^{\lh P}+1$. We always have $r_0(\theta^m(F))\le r_0(F)$.
Assume for contradiction $r_0(\theta^m(F))=r_0(F)$, and choose $u\in
F$ such that $\theta^m(F),u\nmodel\fii$ and $\lh{R_0}=r_0(F)$, where
$R_0=R_0(\theta^m(F)_u)$. As in~\eqref{eq:constr}, we have
$R_0(\theta^k(F)_v)=R_0$ for every $v\in F_u\res\neg\boxdot\fii$ and
$k\le m$.

If all maximal clusters of $\theta^k(F)_u\res\neg\boxdot\fii$ are
reflexive for every $k<m$, \th\ref{cor:rankrefl} implies that
$r(\theta^k(F)_u)$ is strictly decreasing, hence so is
$r_1(\theta^k(F)_u)$. However, this contradicts $r_1(F)<2^{\lh P}<m$ and
$r_1(\theta^m(F))\ge0$.

If some $\theta^k(F)_u\res\neg\boxdot\fii$ has an irreflexive maximal
cluster, let $k$ be the smallest such. As above, we have
$r_1(\theta^{i+1}(F)_u)<r_1(\theta^i(F)_u)$ for every $i<k$, hence
$r_1(\theta^k(F)_u)\le2^{\lh P}-1-k=m-k-2$. But then
\th\ref{lem:rankirr} gives
$\theta^m(F)_u=\theta^{m-k}(\theta^k(F))_u\model\fii$, a
contradiction.
\end{Pf}

\begin{Pf}[of \th\ref{thm:proj}]

\ref{item:proj}${}\to{}$\ref{item:modext}:
Let $\sigma$ be a projective unifier of~$\fii$, and $F\in\Mod_L$ be
such that $F\bez\rcl(F)\model\fii$. Since $\vdash_L\sigma(\fii)$, we
have $\sigma(F)\model\fii$, and \eqref{eq:proj} implies
$\sigma(F\bez\rcl(F))=F\bez\rcl(F)$, hence $\sigma(F)$ is a variant
of~$F$.

\ref{item:modext}${}\to{}$\ref{item:theta}: Assume
$\nvdash_L\theta^N(\fii)$, hence there exists $F\in\Mod _L$ such that
$\theta^N(F)\nmodel\fii$. Put $s(n)=r_0\bigl(\theta^{(2^{\lh
P}+1)n}(F)\bigr)$. We have $s(0)\le\lh B$ and $s(\lh
B+1)=r_0(\theta^N(F))\ge0$. However, $s$ is strictly decreasing by
\th\ref{lem:rankdown}, a contradiction.

\ref{item:theta}${}\to{}$\ref{item:proj} follows from the fact
that $\theta^N$ satisfies~\eqref{eq:proj}.
\end{Pf}

\section{Cluster-extensible logics}\label{sec:clx-logics}
Most of our results on admissibility and unification with parameters
will be stated for logics satisfying a suitable extensibility
condition on finite frames that we introduce in this section. We will
call logics satisfying the full condition \emph{cluster-extensible};
the condition can be somewhat relaxed if the set~$\Par$ of all
parameters is finite, leading to the definition of
\emph{$\Par$-extensible logics}. The primary reason our methods work
for these logics is that they have finite projective approximations, and
we will prove this in Section~\ref{sec:proj-apx}. We slightly digress in
Section~\ref{sec:directed-unification} to give a rather general
characterization of logics with directed unification, which enables us
to distinguish $\Par$-extensible logics with unary and finitary
unification type. In Section~\ref{sec:struct-clx}, we investigate
the behaviour of cluster-extensible logics in terms of various
properties commonly studied in modal logic; while these properties do
not directly involve admissibility or unification, we will
use them as tools later.

The $\Par$-extensibility condition is a restricted variant of the generalized property of branching below~$m$ from
Rybakov~\cite{ryb:bk}. It is similar to Assumption~1.2 of
Ghilardi~\cite{ghil}: the differences are that on the one hand, we need
to work with proper clusters as roots in order to accommodate
parameters, on the other hand we make the condition more fine-grained
by taking into consideration the number of immediate successors of the root;
this makes our results applicable to logics with bounded branching at
almost no additional cost.
The real reason why we need to consider extension properties with proper clusters will be seen in the proof of
\th\ref{thm:clxprojapx}, but let us give at least a partial motivation now.

Recall that Ghilardi's Assumption~1.2 demands
that if $\{F_i:i<n\}$ is a finite sequence of finite rooted $L$-frames, and ${*}\in\{\I,\R\}$ is a
one-element cluster whose type is compatible with~$L$ (i.e., it occurs in \emph{some} finite $L$-frame), then the frame
$\bigl(\sum_iF_i\bigr)^*$ obtained from the disjoint union $\sum_iF_i$ by attaching $*$ as a new root is
an $L$-frame.

Due to the way it is actually used, this condition should be thought of as a stand-in for the following seemingly more general
property: if $\{F_i:i<n\}$ is a finite sequence of finite rooted $L$-frames, and $C$ is a
finite cluster whose type is compatible with~$L$, then there is a p-morphism from $\bigl(\sum_iF_i\bigr)^C$ to an
$L$-frame which is identical on $\sum_iF_i$. This is equivalent to the original condition, as we can contract the root
cluster~$C$ to a single point by a p-morphism.

Now, when we are dealing with parameters, we actually need the latter property to hold for \emph{parametric} frames. In
this case we cannot contract~$C$ as easily, because the p-morphism has to respect the valuation of parameters; the
minimal p-morphic image we can use has a root cluster containing one point for each valuation of parameters realized
in~$C$. Passing back to ordinary frames, this means we have to require that $\bigl(\sum_iF_i\bigr)^C$ be an $L$-frame
whenever $C$ is small enough so that we can endow it with a valuation of parameters that distinguishes all its points:
that is, $\lh C\le2^{\lh\Par}$ if $\Par$ is finite, and arbitrary finite~$C$ if $\Par$ is infinite.

The official definition below avoids explicit reference to disjoint sums, and incorporates stratification according to the number
of immediate successors of the root cluster mentioned earlier.

\begin{Def}\th\label{def:clx}
A \emph{cluster type} is an isomorphism type of a finite cluster. We
will denote the irreflexive cluster type by~$\I$, and the $k$-point
reflexive cluster type by~$\nr k$. If $C$ is a cluster type
and~$n\in\omega$, a finite rooted frame~$F$ is of \emph{type
$\p{C,n}$} if $\rcl(F)$ has type~$C$ and $n$ immediate successor
clusters.

A logic $L\Sset\lgc{K4}$ is \emph{$\p{C,n}$-extensible}\footnote{The somewhat similar-looking \emph{extension property
up to~$n$} from~\cite{iem:char} is not directly related; it is a restriction of the model extension property
from \th\ref{def:proj}.}
if whenever
$F$ is a type-$\p{C,n}$ frame such that $F\bez\rcl(F)$ is an
$L$-frame, then $F$ is an $L$-frame itself, unless $\Par=\nul$, $n=1$,
and $F\bez\rcl(F)$ has a reflexive root cluster.
$L$ is a \emph{$\Par$-extensible logic} if it has fmp,
and it is $\p{C,n}$-extensible for every $n$ and~$C$ such that there
exists at least one $L$-frame of type~$\p{C,n}$, and $\lh
C\le2^{\lh\Par}$. Note that the
condition $\lh C\le2^{\lh\Par}$ is satisfied automatically if $\Par$
is infinite. Logics $\Par$-extensible with respect to infinite~$\Par$
will also be called \emph{cluster-extensible (clx) logics.}
\end{Def}

The purpose of the seemingly odd exception in the case $\Par=\nul$ and
$n=1$ is to make \th\ref{thm:admext} hold. The underlying reason is
that if $F'$ is a frame with a reflexive root, and $F$ the frame
obtained from~$F'$ by adding a new root cluster~$C$, then the mapping
that contracts~$C$ to a fixed element of~$\rcl(F')$ is a p-morphism
from~$F$ to~$F'$. (This is no longer true for parametric frames
when~$\Par\ne\nul$, as $C$ and~$\rcl(F')$ may have incompatible
valuation of parameters.)

\begin{table}
\centering
\begin{tabular}{|l|l|l|}
\hline
logic&axiomatization over $\lgc{K4}$&frame condition\\
\hline
$\lgc{S4}$&$\Box x\to x$&reflexive\\
$\lgc{D4}$&$\dia\top$&final clusters reflexive\\
$\lgc{GL}$&$\Box(\Box x\to x)\to\Box x$&irreflexive\\
$\lgc{K4Grz}$&$\Box\bigl(\Box(x\to\Box x)\to x\bigr)\to\Box x$&no proper clusters\\
$\lgc{K4.1}$&$\boxdot\dia x\to\dia\Box x$&no proper final clusters\\
$\lgc{K4.3}$&$\Box(\boxdot x\to y)\lor\Box(\Box y\to x)$&width $1$\\
$\lgc{K4B}$&$x\to\Box\dia x$&depth $1$\\
$\lgc{S5}$&$\lgc{S4}\oplus\lgc{K4B}$&reflexive, depth $1$\\
$\lgc{K4BB}_k$
&$\displaystyle
 \Box\Bigl(\LOR_{i\le k}\Box\Bigl(\boxdot x_i\to\LOR_{j\ne i}x_j\Bigr)
 \to\LOR_{i\le k}\boxdot x_i\Bigr)$
&branching at most $k$\\
&\hfill$\displaystyle\to\LOR_{i\le k}\Box\LOR_{j\ne i}x_j$&\\
$\lgc{K4BC}_k$
&$\displaystyle\ET_{i=1}^k\Box\Bigl(\Box\Bigl(
  \ET_{j<i}x_j\to x_i\lor\Box x_0\Bigr)\to x_0\Bigr)\to\Box x_0$
&cluster size at most $k$\\
$\lgc{S4.1.4}$
&$\Box\bigl(\Box(x\to\Box x)\to x\bigr)\to(\Box\dia\Box x\to x)$
&reflexive,\\
&&no inner proper clusters\\
\hline
\end{tabular}
\caption{Some cluster-extensible logics}
\label{tab:axi}
\end{table}
\begin{Exm}\th\label{exm:wellknown}
Table~\ref{tab:axi} lists some important extensions of~$\lgc{K4}$,
along with conditions characterizing their finite  frames.
All these logics as well as $\lgc{K4}$ itself are cluster-extensible,
as can be readily seen from their frame conditions. We will prove
later (\th\ref{cor:subsemi}) that joins of clx logics are themselves
clx, hence arbitrary combinations of logics from the table are also
cluster-extensible. We will denote joins of logics by stacking axiom
labels on top of a name of a base logic, so that
$\lgc{D4.3}=\lgc{D4}\oplus\lgc{K4.3}$,
$\lgc{S4GrzBB}_k=\lgc{S4}\oplus\lgc{K4Grz}\oplus\lgc{K4BB}_k$, etc. (The logic $\lgc{S4.1.4}$,
whose name we take from Zeman~\cite{zeman}, is an exception: this is
not a systematic name, but a meaningless numerical label.) The
axiomatization of the bounded branching logics $\lgc{K4BB}_k$ comes
from~\cite{ej:sfef}, and it is only valid for~$k>0$. We can put
$\lgc{K4BB}_0=\lgc{K4B}$ to maintain the frame condition. Notice that
also $\lgc{K4.3}=\lgc{K4BB}_1$ and~$\lgc{K4Grz}=\lgc{K4BC}_1$, but we still
prefer to call $\lgc{K4B}$, $\lgc{K4.3}$, and~$\lgc{K4Grz}$ by their
more common names.

The trivial logics $\lgc{Triv}=\lgc{K4}\oplus x\eq\Box x=\lgc{S5Grz}$,
$\lgc{Verum}=\lgc{K4}\oplus\Box\bot=\lgc{GLBB}_0$, and
$\lgc{Form}=\lgc{K4}\oplus\bot=\lgc{S4}\oplus\lgc{GL}$ are also cluster-extensible.

We note that some of the axioms have robust definitions
only over~$\lgc{S4}$, whereas their definitions over~$\lgc{K4}$ vary
in the literature. In particular, the Grzegorczyk axiom is often stated
ending with $\cdots\to x$ (which defines over~$\lgc{K4}$, or even
over~$\lgc K$, the same logic as our $\lgc{S4Grz}$), and $\lgc{K4.1}$
is often defined as $\lgc{K4}\oplus\Box\dia x\to\dia\Box x$ (which is
our~$\lgc{D4.1}$). We chose the definitions given
because they seem to be most natural in a potentially irreflexive context, and as
just mentioned, the alternative definitions are covered under other
names.

A notable example of a well-known logic that is not clx
is~$\lgc{S4.2}=\lgc{S4}\oplus\dia\Box x\to\Box\dia x$; this and other
logics whose rooted finite frames have a single top cluster
will be dealt with separately in the sequel. For other examples, the
logics of bounded depth ($\lgc{K4BD}_k$) or width ($\lgc{K4BW}_k$) are
not clx for~$k>1$.

Let $m\ge1$, and $L$ be the logic of finite $\lgc{S4}$-frames such that
clusters of depth $3$ or more have size at most~$m$. Then $L$ is not
cluster-extensible, but it is $\Par$-extensible if $\lh\Par\le\log_2m$.
\end{Exm}

\subsection{Projective approximations}\label{sec:proj-apx}

\begin{Def}\th\label{def:projapx}
Let $L\Sset\lgc{K4}$ and $\fii$ be a formula. A \emph{projective
approximation of~$\fii$} is a finite set of $L$-projective formulas
$\Pi_\fii$ such that
\begin{enumerate}
\item $\fii\adm_L\Pi_\fii$,
\item $\psi\vdash_L\fii$ for every $\psi\in\Pi_\fii$.
\end{enumerate}
\end{Def}

Notice that if $\sigma_\psi$ is a projective unifier of~$\psi$ for
every~$\psi\in\Pi_\fii$, then $\{\sigma_\psi:\psi\in\Pi_\fii\}$ is a
complete set of unifiers of~$\fii$.

The definition of $\Par$-extensible logics is motivated by the
following result, generalizing the parameter-free case proved by
Ghilardi~\cite{ghil}.
\begin{Thm}\th\label{thm:clxprojapx}
If $L$ is a $\Par$-extensible logic, then any formula~$\fii$ has a
projective approximation $\Pi_\fii$ such that every $\psi\in\Pi_\fii$
is a Boolean combination of subformulas of~$\fii$.
\end{Thm}

First we need a simple lemma on preservation of formulas when attaching new root clusters to models, which we will also
use later on. The lemma is easier to prove than to formulate. The most basic
situation is that we have two rooted models whose root clusters $C_0$ and~$C_1$
are isomorphic (i.e., they have the same reflexivity, and their elements can be matched so that
the corresponding elements satisfy the same atoms), the set of proper successors of~$C_i$ is generated by a set~$X_i$,
and we can match $X_0$ with~$X_1$ so that the corresponding elements satisfy the same formulas from a
set~$\Sigma$ closed under subformulas. Then the matching elements of $C_0$ and~$C_1$ also satisfy the same
$\Sigma$-formulas. However, the lemma also applies to cases where the elements only ``behave'' as in the basic
situation: the matchings do not need to be 1--1, $C_i$ need not be disjoint from~$X_i$, and ``irreflexive'' clusters
may be actually reflexive.
\begin{Def}\th\label{def:equivsigma}
Let $\Sigma$ be a set of formulas closed under subformulas.
If $W_0,W_1$ are models and $u_i\in W_i$, $i=0,1$, we will write
$W_0,u_0\equiv_\Sigma W_1,u_1$ if $W_0,u_0\model\psi\EQ
W_1,u_1\model\psi$ for every $\psi\in\Sigma$ (i.e., $\sat_\Sigma(W_0,u_0)=\sat_\Sigma(W_1,u_1)$).
\end{Def}

\begin{Lem}\th\label{lem:tpsigma}
For $i=0,1$, let $W_i$ be a Kripke model, $C_i=\{u_{i,j}:j\in
J\}\sset W_i$, and $X_i=\{w_{i,k}:k\in K\}\sset W_i$, where either
\begin{equation}\label{eq:10}
u_{i,j}\up=C_i\cup X_i\Up
\end{equation}
for every $i=0,1$, $j\in J$, or
\begin{equation}\label{eq:9}
u_{i,j}\up=X_i\Up
\end{equation}
for every $i=0,1$, $j\in J$. Let $\Sigma$ be a set of
formulas closed under subformulas, and assume that
$W_0,w_{0,k}\equiv_\Sigma W_1,w_{1,k}$ for every~$k\in K$, and that
$W_0,u_{0,j}$ satisfies the
same atoms from~$\Sigma$ as $W_1,u_{1,j}$ for every~$j\in J$.

Then $W_0,u_{0,j}\equiv_\Sigma W_1,u_{1,j}$.
\end{Lem}
\begin{Pf}
We will show $u_{0,j}\model\psi\EQ u_{1,j}\model\psi$ for every~$j\in
J$ by induction on the complexity of~$\psi\in\Sigma$. The statement
holds for atoms by assumption, and steps for Boolean connectives are
obvious. Assume $u_{0,j}\nmodel\Box\psi$. If \eqref{eq:10} holds, and
$u_{0,j'}\nmodel\psi$ for some~$j'\in J$, then $u_{1,j'}\nmodel\psi$
by the induction hypothesis, hence $u_{1,j}\nmodel\Box\psi$. Otherwise
there is $k\in K$ such that $w_{0,k}\nmodel\psi$ or
$w_{0,k}\nmodel\Box\psi$. Since $w_{0,k}\equiv_\Sigma w_{1,k}$, this
implies $w_{1,k}\nmodel\psi$ or $w_{1,k}\nmodel\Box\psi$, hence
$u_{1,j}\nmodel\Box\psi$. The reverse direction is symmetric.
\end{Pf}

\begin{Pf}[of \th\ref{thm:clxprojapx}]
Let $\Sigma=\Sub(\fii)$, $B(\Sigma)$ be its Boolean closure,
and $\Pi_\fii$ the set of all $\psi\in B(\Sigma)$ such that
$\psi$ is projective and $\psi\vdash_L\fii$. It suffices to show that
every unifier $\sigma$ of~$\fii$ also unifies some $\psi\in\Pi_\fii$.
Define
\[\psi=\LOR\bigl\{\Sigma^{\sat_\Sigma(\sigma(G),v)}:G\in\Mod_L,v\in G\bigr\}.\]
Clearly, $\psi\in B(\Sigma)$, and as $\vdash_L\sigma(\fii)$, we have
$\vdash_L\psi\to\fii$. The fmp of~$L$ also implies
$\vdash_L\sigma(\psi)$, thus the only thing left to prove is that
$\psi$ is projective. Using \th\ref{thm:proj}, it suffices to show
that $\psi$ has the
model extension property. Notice that
\[\Mod_L(\psi)=\{F\in\Mod_L:\forall u\in F\,\exists G\in\Mod_L\,
  \exists v\in G\,
    (F,u\equiv_\Sigma\sigma(G),v)\}.\]
Let $F_0\in\Mod_L$ be such that $F_0\bez\rcl(F_0)\model\psi$. If
$\Par=\nul$ and $F_0\bez\rcl(F_0)$ has a reflexive root cluster,
the mapping contracting $\rcl(F_0)$ to a fixed
point~$r\in\rcl(F_0\bez\rcl(F_0))$ is a p-morphism, hence we can define a
variant of~$F_0$ satisfying $\psi$ by copying the valuation of
variables from~$r$ to $\rcl(F_0)$.

Otherwise, let $F$ be the model obtained from~$F_0$ by identifying
points of~$\rcl(F_0)$ with the same valuation of parameters (valuation
of variables in~$\rcl(F)$ is immaterial), and let $\p{C,n}$ be its
type. Notice that $\lh C\le2^{\lh\Par}$. Pick elements
$\{u_i:i<n\}\sset F$, one in every immediate successor cluster
of~$\rcl(F)$. Since $F_{u_i}\model\psi$, there exists $G_i\in\Mod_L$
and $v_i\in G_i$ such that $\sigma(G_i),v_i\equiv_\Sigma F,u_i$. We
may assume $v_i\in\rcl(G_i)$. We define a model~$G$ as follows: we
take the disjoint union of $G_i$, $i<n$, and attach a new root cluster
of type~$C$. We may identify elements of~$\rcl(G)$ with elements
of~$\rcl(F)$; we define the valuation of parameters in~$\rcl(G)$
identically to~$F$, the valuation of variables is arbitrary.

Since $L$ is $\p{C,n}$-extensible, $G$ is based on an $L$-frame. Let
$F'$ be the variant of~$F$ such that
\[F',w\model x\iff \sigma(G),w\model x\]
for every variable~$x$ and $w\in\rcl(F)$. Then
$F',w\equiv_\Sigma\sigma(G),w$ for every~$w\in\rcl(F)$ by
\th\ref{lem:tpsigma}, which means $F'\model\psi$. Finally, we can define a
variant~$F'_0\model\psi$ of the original~$F_0$ by lifting the
valuation from~$F'$ by the parameter-preserving p-morphism of $F_0$ onto~$F$.
\end{Pf}

\begin{Cor}\th\label{cor:projapxsize}
If $L$ is a $\Par$-extensible logic, every formula $\fii$ of length~$n$ has a
projective approximation consisting of at most $2^{2^n}$ formulas of
length~$O(n2^n)$.
\noproof\end{Cor}

\begin{Cor}\th\label{cor:admdec}
Let $L$ be a $\Par$-extensible logic, and $\Gamma\ru\Delta$ a rule.
\begin{enumerate}
\item\label{item:16}
$\Gamma\adm_L\Delta$ iff for every projective formula $\psi$ that is
a Boolean combination of subformulas of~$\Gamma$, if
$\psi\vdash_L\fii$ for every $\fii\in\Gamma$, then $\psi\vdash_L\fii$
for some~$\fii\in\Delta$.
\item\label{item:17}
If $L$ is decidable, $\adm_L$ is decidable.
\end{enumerate}
\end{Cor}
\begin{Pf}
\ref{item:16}: This follows from the facts that the set of projective
unifiers of formulas from~$\Pi_{\ET\Gamma}$ is a complete set of
unifiers of~$\Gamma$, and if $\sigma$ is a projective unifier
of~$\psi$, then $\vdash_L\sigma(\fii)$ iff $\psi\vdash_L\fii$.

\ref{item:17}: Projectivity, hence the criterion from~\ref{item:16},
is decidable by condition~\ref{item:theta} of \th\ref{thm:proj}.
\end{Pf}
\begin{Cor}\th\label{cor:clxfinitary}
Every $\Par$-extensible logic~$L$ has at most finitary unification type.
If $L$ is decidable, we can compute a complete set of unifiers
for any given formula.
\noproof\end{Cor}

Corollaries \ref{cor:admdec}~\ref{item:17} and~\ref{cor:clxfinitary}
were proved for a class of transitive modal logics by
Rybakov~\cite{ryb:bk,ryb:modunifcoef} using a different approach.

In \th\ref{cor:clxdec}, we will see that the assumption of
decidability of~$L$ in Corollaries \ref{cor:admdec} and~\ref{cor:clxfinitary} is
redundant when $\Par$ is infinite (i.e., for clx logics).
We will prove more precise estimates on the computational complexity of~$\adm_L$ in
the sequel.

\begin{Exm}\th\label{exm:projapxee}
The bounds in \th\ref{cor:projapxsize} cannot be substantially
improved, even in the parameter-free case.

If $L$ is a $\p{\I,2}$-extensible logic (e.g., $\lgc{K4}$ or
$\lgc{GL}$), consider the formulas
\[\fii_m=\ET_{i<m}(\Box x_i\lor\Box\neg x_i)\to\Box y\lor\Box\neg y\]
of length $n=O(m)$.
We claim that $\fii_m$ has a projective approximation $\Pi_{\fii_m}$
consisting of the formulas
\[\psi_f=\ET_{i<m}(\boxdot x_i\lor\boxdot\neg x_i)\to(y\eq f(\vec x)),\]
where $f\colon\two^X\to\two$ is any Boolean function in the~$m$
variables $X=\{x_i:i<m\}$. The formulas~$\psi_f$ have the model
extension property (we can modify
valuation of~$y$ to match $f(\vec x)$), and
$\psi_f\vdash_L\fii_m$ follows from
$\vdash_\lgc{K4}\ET_{i<m}(\Box x_i\lor\Box\neg
x_i)\to\Box f(\vec x)\lor\Box\neg f(\vec x)$. In order to show
$\fii_m\adm_L\Pi_{\fii_m}$, let $\sigma$ be any unifier of~$\fii_m$.
For every $e\in\two^X$, there is $f(e)\in\two$ such that
\begin{equation}\label{eq:8}
\vdash_L\sigma\bigl(\boxdot X^e\to y^{f(e)}\bigr):
\end{equation}
If not, we could find models $F_0,F_1\in\Mod_L$ with roots $u_0,u_1$
(resp.) such that $\sigma(F_i)\model X^e$, $\sigma(F_0),u_0\model\neg y$,
and $\sigma(F_1),u_1\model y$. Let $F\in\Mod_L$ be the disjoint union
of $F_0$ and $F_1$, endowed with a new irreflexive root~$u$. Then
there is no way to define valuation in~$u$ so
that~$\sigma(F),u\model\fii_m$, contradicting
$\vdash_L\sigma(\fii_m)$.

This defines a function $f\colon\two^X\to\two$, and \eqref{eq:8}
implies $\vdash_L\sigma(\psi_f)$. Thus, $\Pi_{\fii_m}$ is indeed a
projective approximation of~$\fii_m$. Since $\psi_f\nvdash_L\psi_g$
for $f\ne g$, \emph{every} projective approximation $\Pi$ of~$\fii_m$
must contain at least $2^{2^m}=2^{2^{\Omega(n)}}$ formulas, and by a
counting argument, most of these formulas must have length
$\Omega(\log\lh\Pi)=2^{\Omega(n)}$.

If $L$ is a $\p{\nr1,2}$-extensible logic (such as $\lgc{S4}$
or~$\lgc{S4Grz}$), we can use in a similar way the slightly more
complicated formulas
\[\fii'_m=\ET_{i<m}(\Box\dia x_i\lor\Box\dia\neg x_i)
  \to\Box y\lor\Box\neg y,\]
whose projective approximation consists of the $2^{2^m}$ formulas
\[\psi'_f=\ET_{e\in\two^X}
    \Bigl(\ET_{i<m}\boxdot\dia x_i^{e(x_i)}\to y^{f(e)}\Bigr),\]
where again $f\colon\two^X\to\two$.
\end{Exm}

\subsection{Directed unification}\label{sec:directed-unification}

\th\ref{cor:clxfinitary} tells us that the parametric unification type
of any $\Par$-extensible logic is at most finitary, but it does not specify
whether it is of type $1$ or~$\omega$. We will resolve this with the
help of the following concept.
\begin{Def}\th\label{def:directed}
A logic~$L$ has \emph{directed} (or \emph{filtering})
\emph{unification} if for any formula~$\fii$, the preorder of
$L$-unifiers of~$\fii$ is directed, i.e., for every unifiers $\sigma_0$
and~$\sigma_1$ of~$\fii$, there exists a unifier of~$\fii$ more
general than either of~$\sigma_0,\sigma_1$.
\end{Def}
Clearly, if a formula has a mgu, then its preorder of unifiers is
directed, whereas if it has at least two incomparable maximal
unifiers, it is not directed. Thus, if $L$ has non-nullary
unification type, then it has unitary unification if and only if it
has directed unification.

Ghilardi and Sacchetti~\cite{ghi-sacc} discovered a criterion for
directedness of parameter-free unification in transitive modal logics:
namely, $L\Sset\lgc{K4}$ has directed unification if and only if it
includes the logic
\[\lgc{K4.2}:=\lgc{K4}\oplus\dia\boxdot x\to\Box\diadot x.\]
We will give a simple syntactic proof of this result that also applies to
unification with parameters, as well as a much more general class of
logics.

We are temporarily leaving the realm of modal logics, the theorem
below works for logics given by arbitrary structural consequence
relations satisfying the stated conditions. In this context, we will
write $\Gamma\vdash\bot$ as a short-hand for ``$\Gamma$ is
inconsistent'', i.e., $\Gamma\vdash\fii$ for every formula~$\fii$.
Note that we are working with single-conclusion systems here, we will
use $\Gamma\vdash\Delta$ as an abbreviation for $\Gamma\vdash\fii$
for every~$\fii\in\Delta$.
\begin{Thm}\th\label{thm:dirunif}
Let $L$ be a logic such that:
\begin{enumerate}
\def\theenumi{(\alph{enumi})}
\item\label{item:equi} $L$ is equivalential with respect to a set of formulas~$E(x,y)$.
\item\label{item:disj} There is a finite set of formulas~$D(x,y)$ such that
\[\Gamma,D(\fii,\psi)\vdash_L\chi\iff
  \Gamma,\fii\vdash_L\chi\text{ and }\Gamma,\psi\vdash_L\chi\]
for every finite set of formulas~$\Gamma$, and formulas $\fii$,
$\psi$, and~$\chi$.
\item\label{item:swi} There are formulas $S(x,y_0,y_1)$, $C_0(x)$,
and~$C_1(x)$ such that for~$i=0,1$,
\[C_i(x)\vdash_LE(S(x,y_0,y_1),y_i).\]
\item\label{item:neg} There is a formula~$B(x)$ such that for every
$\Gamma$ and~$\fii$,
\begin{gather*}
x\vdash_LC_1(B(x)),\\
\Gamma,\fii\vdash_L\bot\ \TO\ \Gamma\vdash_LC_0(B(\fii)).
\end{gather*}
\end{enumerate}
Then the following are equivalent:
\begin{enumerate}
\item\label{item:dir} $L$ has directed (parametric) unification.
\item\label{item:spec} There is a formula~$\alpha$ such that
$\vdash_LD(C_0(\alpha),C_1(\alpha))$, and $C_0(\alpha)$ and~$C_1(\alpha)$
are $L$-unifiable.
\item\label{item:2} $\vdash_LD(C_0(B(C_0(x))),C_0(B(C_1(x))))$.
\end{enumerate}
Moreover, \ref{item:dir} is equivalent to~\ref{item:spec} for any logic~$L$
that satisfies \ref{item:equi}, \ref{item:disj}, \ref{item:swi}, and
\begin{enumerate}
\def\theenumi{(d')}
\item\label{item:47} $C_0(x)$ and~$C_1(x)$ are $L$-unifiable,
\end{enumerate}
where we can allow $C_0$ and~$C_1$ to be finite sets instead of single formulas.
\end{Thm}
\begin{Rem}
The mnemonics for the letters are Equivalence, Disjunction, Switch,
truth Constant, and Box.

In general, we allow parameters (not indicated by the notation) to
appear in all formulas and unifiers mentioned in \th\ref{thm:dirunif}.
However, if the assumptions are satisfied with parameter-free
$E,D,C_i,B$ (which is the common case), then we can assume without loss of generality that
$\alpha$ and the unifiers in~\ref{item:spec},~\ref{item:47} are
also parameter-free. Consequently, $L$ has directed parametric
unification iff it has directed parameter-free unification.

We could allow the variable $x$ in~$C_i$ and~$S$ to be a list
$x_1,\dots,x_k$ of variables instead, with obvious modifications (we
would have $\alpha_1,\dots,\alpha_k$ and $B_1,\dots,B_k$ to go with these).
\end{Rem}
\begin{Cor}\th\label{cor:dirk42}
A logic $L\Sset\lgc{K4}$ has directed unification if and
only if $L\Sset\lgc{K4.2}$.

More generally, let $L$ be an $n$-transitive multimodal logic (i.e.,
$L$ has finitely many boxes $\Box_1,\dots,\Box_k$, and the combined
modality $\Box x:=\Box_1x\land\dots\land\Box_kx$ satisfies
$\vdash_L\Box^{\le n}x\to\Box^{n+1}x$). Then $L$ has directed
unification iff it proves $\dia^{\le n}\Box^{\le
n}x\to\Box^{\le n}\dia^{\le n}x$.
\end{Cor}
\begin{Pf}
Apply \th\ref{thm:dirunif} with $E(x,y)=\{x\eq y\}$, $D(x,y)=\{\Box^{\le
n}x\lor\Box^{\le n}y\}$, $C_1(x)=x$, $C_0(x)=\neg x$, $S(x,y_0,y_1)=(x\land y_1)\lor(\neg x\land
y_0)$, $B(x)=\Box^{\le n}x$.
\end{Pf}

\begin{Cor}\th\label{cor:extunitary}
Let $L\Sset\lgc{K4}$ be a $\Par$-extensible logic. Then $L$ has
unification of type~$1$ if $L$ is linear (see
\S\ref{sec:single-concl-bases}), and type~$\omega$ otherwise.
\noproof\end{Cor}

\begin{Pf}[of \th\ref{thm:dirunif}]
Using \ref{item:swi} and~\ref{item:equi}, we have
$C_0(x),C_1(x)\vdash_LE(y_0,y_1)$, hence
\begin{equation}\label{eq:incon}
C_0(x),C_1(x)\vdash_L\bot.
\end{equation}
Notice also that \ref{item:neg} implies~\ref{item:47}: by~\ref{item:neg}, we
have $\vdash_LC_1(B(\top))$ for any tautology~$\top$, which implies
$\vdash_LC_0(B(C_0(B(\top))))$ by \ref{item:neg}
and~\eqref{eq:incon}.

\ref{item:dir}${}\to{}$\ref{item:spec}: Let $\sigma_i$ be a unifier
of~$C_i$, $i=0,1$. Both~$\sigma_i$ are unifiers of
$D(C_0(x),C_1(x))$, hence by~\ref{item:dir}, this formula has a
unifier $\tau$ such that $\sigma_0,\sigma_1\le_L\tau$. Then
$\alpha:=\tau(x)$ is as desired.

\ref{item:spec}${}\to{}$\ref{item:dir}: Let $\tau_0$, $\tau_1$ be
unifiers of~$\Gamma$, and define
\[\tau(x_j)=S(\alpha,\tau_0(x_j),\tau_1(x_j))\]
for every variable~$x_j$ occurring in~$\fii$. We may assume that
$\alpha$ (hence $C_i(\alpha)$) shares no variables with~$\tau_i(x_j)$. We have
\begin{equation}\label{eq:tau}
C_i(\alpha)\vdash_LE(\tau(x_j),\tau_i(x_j))
\end{equation}
by \ref{item:swi}, hence
\[C_i(\alpha)\vdash_L\tau(\Gamma)\]
by \ref{item:equi}, which means
\[\vdash_LD(C_0(\alpha),C_1(\alpha))\vdash_L\tau(\Gamma)\]
using \ref{item:disj}. Moreover, if $\sigma_i$ is a unifier
of~$C_i(\alpha)$ identical on variables not occurring in~$\alpha$,
then \eqref{eq:tau} gives
\[\vdash_LE(\sigma_i(\tau(x_j)),\tau_i(x_j)),\]
i.e., $\tau_i\le_L\tau$ via~$\sigma_i$.

\ref{item:spec}${}\to{}$\ref{item:2}: Let $\sigma_i$ be a unifier
of~$C_i(\alpha)$, and define
\[\tau(x_j)=S(x,\sigma_0(x_j),\sigma_1(x_j)).\]
We have
\begin{align*}
C_i(x)&\vdash_LE(\tau(x_j),\sigma_i(x_j))&&\text{by \ref{item:swi},}\\
C_i(x)&\vdash_L\tau(C_i(\alpha))&&\text{by \ref{item:equi},}\\
C_i(x),\tau(C_{1-i}(\alpha))&\vdash_L\bot&&\text{by \eqref{eq:incon},}\\
\tau(C_{1-i}(\alpha))&\vdash_LC_0(B(C_i(x)))&&\text{by \ref{item:neg},}\\
\tau(D(C_0(\alpha),C_1(\alpha)))&\vdash_LD(C_0(B(C_0(x))),C_0(B(C_1(x))))
&&\text{by \ref{item:disj},}\\
&\vdash_LD(C_0(B(C_0(x))),C_0(B(C_1(x))))&&\text{by \ref{item:spec}.}
\end{align*}

\ref{item:2}${}\to{}$\ref{item:spec}: Put $\alpha=B(C_0(B(C_1(x))))$. 
We have
\begin{align*}
C_1(x)&\vdash_LC_1(B(C_1(B(C_1(x)))))&&\text{by \ref{item:neg},}\\
C_1(x),C_0(B(C_1(B(C_1(x)))))&\vdash_L\bot&&\text{by \eqref{eq:incon},}\\
C_0(B(C_1(B(C_1(x)))))&\vdash_LC_0(B(C_1(x)))&&\text{by \ref{item:neg},}\\
C_0(B(C_1(B(C_1(x)))))&\vdash_LC_1(\alpha)&&\text{by \ref{item:neg},}\\
C_0(B(C_0(B(C_1(x)))))&\vdash_LC_0(\alpha)&&\text{by definition,}\\
&\vdash_LD(C_0(\alpha),C_1(\alpha))&&\text{by \ref{item:disj} and \ref{item:2}.}
\end{align*}
Let $\sigma_i$ be a unifier of~$C_i(x)$, $i=0,1$. We have
$\sigma_0(C_1(x))\vdash_L\bot$ by~\eqref{eq:incon}, hence
\[\vdash_L\sigma_0(C_0(B(C_1(x))))\vdash_L\sigma_0(C_1(B(C_0(B(C_1(x))))))
=\sigma_0(C_1(\alpha))\]
by \ref{item:neg}. Similarly,
\begin{align*}
&\vdash_L\sigma_1(C_1(B(C_1(x))))&&\text{by \ref{item:neg},}\\
\sigma_1(C_0(B(C_1(x))))&\vdash_L\bot&&\text{by \eqref{eq:incon},}\\
&\vdash_L\sigma_1(C_0(B(C_0(B(C_1(x))))))=\sigma_1(C_0(\alpha))&&\text{by \ref{item:neg}.}
\end{align*}
\end{Pf}

\begin{Rem}\th\label{cor:flo}
For readers familiar with substructural logics (see~\cite{reslat}): \th\ref{thm:dirunif} can be applied to a large class
of logics as follows.

Let $L$ be an extension (not necessarily simple, i.e., $L$ may have a richer
language) of the $\{\to,\land,\lor,0,1\}$-fragment of $\FL
o$, where $\to$ is either of the two residua. Assume that $L$ is
equivalential with respect to the formula
$E(x,y)=(x\to y)\land(y\to x)$ (note that this holds automatically for
simple axiomatic extensions of fragments of~$\FL o$), and it has the
deduction-detachment theorem in the form
\begin{equation}\label{eq:ddt}
\Gamma,\fii\vdash_L\psi\iff\Gamma\vdash_L\Delta\fii\to\psi
\end{equation}
for some formula $\Delta(x)$. (In systems with Baaz delta, one can usually
take it for~$\Delta$.) Put $\neg\fii:=\fii\to0$. Then $L$
satisfies the assumptions of \th\ref{thm:dirunif} with $D(x,y)=\Delta
x\lor\Delta y$, $S(x,y_0,y_1)=(1\land x\to y_1)\land(1\land\neg x\to
y_0)$, $C_1(x)=x$, $C_0(x)=\neg x$, $B(x)=\Delta x$.
Thus, the following are equivalent:
\begin{enumerate}
\item $L$ has directed unification.
\item There is a formula~$\alpha$ such that
$\vdash_L\Delta\alpha\lor\Delta\neg\alpha$, and $\alpha$ and $\neg\alpha$
are unifiable.
\item $\vdash_L\Delta\neg\Delta x\lor\Delta\neg\Delta\neg x$.
\end{enumerate}

For example, this subsumes \th\ref{cor:dirk42} by taking $\Delta
x=\Box^{\le n}x$. 
For another example, let $L$ be an $n$-contractive simple axiomatic
extension of $\FL{ew}$. (Note that the case~$n=1$ covers
superintuitionistic logics.) Then we can take $\Delta x=x^n$, hence
$L$ has directed unification iff it proves $\bigl(\neg
x^n\bigr)^n\lor\bigl(\neg(\neg x)^n\bigr)^n$. In fact, unification in
some of these logics has been proved unitary by Dzik~\cite{dzik}.
\end{Rem}

\subsection{Structure of cluster-extensible logics}\label{sec:struct-clx}

While there are continuum many extensible logics in the
parameter-free case (see e.g.~\cite{ej:frege}), extensibility is a
much tighter constraint if there are infinitely many parameters. This
is to be expected: we defined cluster-extensible logics for the
purpose that their admissible rules have bases consisting of subsets
of certain explicitly defined rules (\th\ref{thm:clxadmchar}).
Unlike the parameter-free case, it is impossible for a logic to
inherit the admissible rules of its proper sublogic if we have infinitely
many parameters: for any consistent logic $L$ and a parameter-free
formula~$\fii(\vec x)$, we have
\begin{equation}\label{eq:27}
\begin{aligned}
 \vdash_L\fii(\vec x)&\iff\adm_L\fii(\vec p),\\
 \nvdash_L\fii(\vec x)&\iff\fii(\vec p)\adm_L\bot,
\end{aligned}
\end{equation}
where $\vec p$ are pairwise distinct parameters. Thus, each clx logic
is uniquely determined by a set of extension rules, and the
relatively simple structure of these rules carries over to the
corresponding class of logics.

In this section, we
will show that all clx logics have various nice properties that will
be helpful for description of their admissible rules and their
complexity: in particular, clx logics are finitely axiomatizable, have
the exponential-size model property, and are first-order
($\forall\exists$) definable on finite frames. Moreover, clx logics are
closed under joins (hence they form a complete lattice). On the other
hand, the class of clx logics includes most of the best known
particular transitive monomodal logics (a notable exception being
logics with a single top cluster such as~$\lgc{S4.2}$, which require
special treatment).

We assume $\Par$ is infinite for the rest of this section.

\begin{Def}\th\label{def:extcond}
An \emph{extension condition} is a pair~$\p{C,n}$, where
$n\in\omega\cup\{\infty\}$, and $C$ is a cluster type or~$\nr\infty$. 
An extension condition $\p{C,n}$ is \emph{finite} if $C\ne\nr\infty$
and $n\ne\infty$. The set of all extension conditions is denoted
by~$\ECI$, and the set of finite extension conditions
by~$\EC$.

We generalize the notion of a $\p{C,n}$-extensible logic to arbitrary
extension conditions so that $L$ is $\p{\nr\infty,n}$-extensible iff
it is $\p{\nr k,n}$-extensible for every $0<k\in\omega$, and $L$ is
$\p{\nr k,\infty}$-extensible iff it is $\p{\nr k,n}$-extensible for
every $0<n\in\omega$. If $T$ is a set of extension conditions, $L$
is $T$-extensible if it is $t$-extensible for every~$t\in T$.

Let $\le_0$ be the partial order on~$\omega\cup\{\infty\}$ such that
$n\le_0m\le_0\infty$ for every $0<n\le m\in\omega$, and $0$ is
incomparable to any other element. If $C$ and~$D$ are cluster types,
we define $D\preceq C$ iff both $C$, $D$ are irreflexive, or both are
reflexive and $\lh D\le\lh C$. We also put $\nr k\preceq\nr\infty$ for
every $0<k\in\omega$. (Notice that if we identify $\I$ with~$0$, and
$\nr k$ with~$k$ for $0<k\le\infty$, then $\preceq$ is the same order
as~$\le_0$.) 
If $\p{C,n}$ and $\p{D,m}$ are extension conditions, we put
$\p{C,n}\preceq\p{D,m}$ iff $C\preceq D$ and~$n\le_0m$.

The \emph{closure} of a set~$T$ of extension conditions is the
smallest set~$\ob T\Sset T$ downward closed under~$\preceq$, and
closed under the rules
\begin{itemize}
\item if $\p{C,n}\in\ob T$ for every $0<n\in\omega$, then
$\p{C,\infty}\in\ob T$,
\item if $\p{\nr k,n}\in\ob T$ for every $0<k\in\omega$, then
$\p{\nr\infty,n}\in\ob T$.
\end{itemize}
Two sets of extension conditions are \emph{equivalent} if they
have the same closure.

Recall that a \emph{well partial order (wpo)} is a partial order~$\le$
on a set~$X$ satisfying any of the following equivalent conditions:
\begin{itemize}
\item Every subset~$Y\sset X$ has a finite \emph{basis}: a finite set
$B\sset Y$ such that $B\Up\Sset Y$.
\item $<$ is well founded, and there are no infinite antichains.
\item For every sequence $\{x_i:i\in\omega\}\sset X$, there are $i<j$
such that~$x_i\le x_j$.
\end{itemize}
It is easily seen that the class of wpo contains all well-ordered
sets, and it is closed under subsets, finite unions and Cartesian
products, and homomorphic images~\cite{kruskal}.
\end{Def}
\begin{Lem}\th\label{lem:extcond}\
\begin{enumerate}
\item\label{item:18}
If $T$ and~$T'$ are equivalent sets of extension conditions, then
a logic is
$T$-extensible iff it is $T'$-extensible.
\item\label{item:20}
$\preceq$ is a well partial order on~$\ECI$.
\item\label{item:19}
Every set of extension conditions is equivalent to a unique finite
set of extension conditions that is an antichain wrt~$\preceq$.
\end{enumerate}
\end{Lem}
\begin{Pf}

\ref{item:18}: The cases involving infinite conditions are clear from the definition. If $L$ is $\p{\nr k,n}$-extensible
and $0<l\le k$, then $L$ is $\p{\nr l,n}$-extensible as every
type-$\p{\nr l,n}$ frame is a p-morphic image of a type-$\p{\nr k,n}$
frame with the same $F\bez\rcl(F)$. Finally, assume that $L$ is
$\p{C,n}$-extensible, and $F$ is a type-$\p{C,m}$ frame such that
$F\bez\rcl(F)$ is an $L$-frame, where $m\le_0n$. Choose $\{w_i:i<m\}$
such that $F\bez\rcl(F)=\bigcup_{i<m}w_i\Up$, let $f$ be a surjection
of $\{0,\dots,n-1\}$ onto~$\{0,\dots,m-1\}$, and let $G$ be the frame
consisting of~$\bigcupd_{i<n}F_{w_{f(i)}}$ together with
a copy of~$\rcl(F)$ as its root cluster. Then $G$ is an $L$-frame as
it has type~$\p{C,n}$, and $F$ is a p-morphic image of~$G$.

\ref{item:20}: $\preceq$ is the product of two partial orders, each
of which is a disjoint union of a singleton and a well order of type
$\omega+1$, and as such it is a wpo.

\ref{item:19}:
Every set is equivalent to its closure, hence we may assume that $T$
is a closed set of extension conditions. Since any chain in~$T$ has a
supremum in~$T$, the set $M$ of maximal elements of~$T$ is cofinal
in~$T$ by Zorn's lemma, and therefore equivalent to~$T$. Clearly, $M$ is an
antichain, hence it is finite by~\ref{item:20}.
The closure of any finite set of conditions is its downward closure,
hence distinct antichains have distinct closures.
\end{Pf}

\begin{Obs}\th\label{obs:dualec}
There is a bijective correspondence between closed subsets
$T\sset\ECI$, and upward closed subsets $U\sset\EC$, given by
$U=\EC\bez T$ and $T=\ob{\EC\bez U}$.
\noproof\end{Obs}

\begin{Def}\th\label{def:clxchar}
If $L$ is a clx logic, its \emph{type} $\tp(L)$ is the set of all
extension conditions $\p{C,n}$ such that $L$ is $\p{C,n}$-extensible.

Its \emph{basis} $\base(L)$ consists of maximal elements of~$\tp(L)$.
Notice that $\tp(L)=\ob{\tp(L)}$ by
\th\ref{lem:extcond}~\ref{item:18}, hence $\base(L)$ is the unique
finite antichain equivalent to~$\tp(L)$ by the proof
of~\ref{item:19}, and $\p{D,m}\in\tp(L)$ iff $\p{D,m}\preceq\p{C,n}$
for some $\p{C,n}\in\base(L)$.

The \emph{exclusion type of~$L$} is $\xcl(L)=\EC\bez\tp(L)$, and its
\emph{exclusion basis} $\xcb(L)$ is the set of all minimal
elements of~$\xcl(L)$. By \th\ref{lem:extcond}~\ref{item:20},
$\xcb(L)$ is finite, and $\xcl(L)$ is its upward closure, hence
$\p{D,m}\notin\tp(L)$ iff $\p{C,n}\preceq\p{D,m}$ for
some~$\p{C,n}\in\xcb(L)$.

If $U\sset\EC$ is upward closed, then $\frx_U$ is the
class of all finite frames~$F$ such that there is no~$u\in
F$ for which the type of~$F_u$ belongs to~$U$, and $\clx_U$ is the
logic of~$\frx_U$.
\end{Def}
\begin{table}
\centering
\begin{tabular}{|l|l|l|}
\hline
logic $L$&$\base(L)$&$\xcb(L)$\\
\hline
$\lgc{K4}$&$\p{\I/\nr\infty,0/\infty}$&\\
$\lgc{S4}$&$\p{\nr\infty,0/\infty}$&$\p{\I,0/1}$\\
$\lgc{K4Grz}$&$\p{\I/\nr1,0/\infty}$&$\p{\nr2,0/1}$\\
$\lgc{S4Grz}$&$\p{\nr1,0/\infty}$&$\p{\I/\nr2,0/1}$\\
$\lgc{K4.3}$&$\p{\I/\nr\infty,0/1}$&$\p{\I/\nr1,2}$\\
$\lgc{K4BB}_k$&$\p{\I/\nr\infty,0/k}$&$\p{\I/\nr1,k+1}$\\
$\lgc{K4BC}_k$&$\p{\I/\nr k,0/\infty}$&$\p{({\scriptstyle k+1}),0/1}$\\
$\lgc{S4.1.4}$&$\p{\nr\infty,0},\p{\nr1,\infty}$&$\p{\I,0/1},\p{\nr2,1}$\\
\hline
\end{tabular}
\begin{tabular}{|l|l|l|}
\hline
logic $L$&$\base(L)$&$\xcb(L)$\\
\hline
$\lgc{K4B}$&$\p{\I/\nr\infty,0}$&$\p{\I/\nr1,1}$\\
$\lgc{S5}$&$\p{\nr\infty,0}$&$\p{\I,0/1},\p{\nr1,1}$\\
$\lgc{GL}$&$\p{\I,0/\infty}$&$\p{\nr1,0/1}$\\
$\lgc{GL.3}$&$\p{\I,0/1}$&$\p{\nr1,0/1},\p{\I,2}$\\
$\lgc{S4.3}$&$\p{\nr\infty,0/1}$&$\p{\I,0/1},\p{\nr1,2}$\\
$\lgc{Triv}$&$\p{\nr1,0}$&$\p{\I/\nr2,0},\p{\I/\nr1,1}$\\
$\lgc{Verum}$&$\p{\I,0}$&$\p{\nr1,0},\p{\I/\nr1,1}$\\
$\lgc{Form}$&&$\p{\I/\nr1,0/1}$\\
\hline
\end{tabular}
\\[\medskipamount]
\begin{tabular}{|l|l|l|}
\hline
logic $L$&$\base(L)$&$\xcb(L)$\\
\hline
$\lgc{D4}$&$\p{\nr\infty,0},\p{\I/\nr\infty,\infty}$&$\p{\I,0}$\\
$\lgc{K4.1}$&$\p{\I/\nr1,0},\p{\I/\nr\infty,\infty}$&$\p{\nr2,0}$\\
$\lgc{S4.1}$&$\p{\nr1,0},\p{\nr\infty,\infty}$&$\p{\I,0/1},\p{\nr2,0}$\\
\hline
\end{tabular}
\caption{Extension characteristics of some clx logics}
\label{tab:clxbas}
\end{table}
\begin{Exm}\th\label{exm:extcondadm}
Bases and exclusion bases of some concrete clx logics are listed in
Table~\ref{tab:clxbas}. The table employs abbreviations to save
space: for example, the line for~$\lgc{K4.3}$ means that
$\base(\lgc{K4.3})=\{\p{\I,0},\p{\I,1},\p{\nr\infty,0},\p{\nr\infty,1}\}$
and $\xcb(\lgc{K4.3})=\{\p{\I,2},\p{\nr1,2}\}$.
\end{Exm}

\begin{Thm}\th\label{prop:unique}
If $U\sset\EC$ is upward closed, then $\clx_U$ is the
unique clx logic of exclusion type~$U$.
In particular, every clx logic is uniquely determined by either of
$\tp(L)$, $\base(L)$, $\xcl(L)$, or~$\xcb(L)$.
\end{Thm}
\begin{Pf}
Since $\frx_U$ is closed under generated subframes, finite disjoint
unions, and (due to the upward closure of~$U$) under p-morphic
images, it is the class of all finite $\clx_U$-frames (\cite[Exercise
9.34]{cha-zax}). By the definition,
$\clx_U$ has fmp, and no finite rooted $\clx_U$-frame has type~$t\in U$.
On the other hand, $\clx_U$ is $t$-extensible for every~$t\in\EC\bez
U$ by the definition of~$\frx_U$. Thus, $\clx_U$ is
cluster-extensible, and $\xcl(\clx_U)=U$.

If $L$ is any clx logic of exclusion type~$U$, then every finite
$L$-frame belongs to~$\frx_U$ by the definition of type. On the other
hand, if $F\in\frx_U$, we can show that $F$ is an $L$-frame by
induction on~$\lh F$, using the fact that $L$ is $(\EC\bez U)$-extensible. Thus,
$\frx_U$ is the class of all finite $L$-frames, and as $L$ has fmp,
$L=\clx_U$.
\end{Pf}
\begin{Cor}\th\label{cor:ctbl}
There are countably many clx logics.
\end{Cor}
\begin{Pf}
There are countably many choices for $\base(L)$ or~$\xcb(L)$.
\end{Pf}
\begin{Cor}\th\label{cor:lattice}
The set $\CLX$ of all clx logics is a complete lattice under
inclusion, and the
mappings $L\mapsto\xcl(L)$ and $U\mapsto\clx_U$ are mutually inverse
isomorphisms of~$\CLX$ to the lattice of all upward closed sets of
finite extension conditions. Alternatively, $L\mapsto\tp(L)$ and
$T\mapsto\clx_{\EC\bez T}$ are mutually inverse dual isomorphisms
of~$\CLX$ to the lattice of closed sets of extension conditions.
\end{Cor}
\begin{Pf}
Upward closed sets of finite extension conditions are closed under
arbitrary intersections and unions, hence they form a complete
lattice. The rest is clear from \th\ref{prop:unique,obs:dualec} and
the definitions.
\end{Pf}
\begin{Cor}\th\label{cor:nochain}
There is no strictly increasing infinite sequence of clx logics.
\end{Cor}
\begin{Pf}
Assume that $L_0\sset L_1\sset L_2\sset\cdots$ are clx logics, and let
$L$ be the join $\LOR_nL_n$ in~$\CLX$. We have
$\xcl(L)=\bigcup_n\xcl(L_n)$. As $\xcb(L)$ is finite, we must have
$\xcb(L)\sset\xcl(L_n)$ for some~$n\in\omega$, hence
$\xcl(L)=\xcl(L_n)$, and~$L=L_n$.
\end{Pf}

\begin{Rem}\th\label{rem:join}
Let $S$ be a set of clx logics, and $L$ its join in~$\NExt\lgc{K4}$.
Since a finite frame is an $L$-frame iff it is an $L'$-frame for
every~$L'\in S$, the logic determined by finite $L$-frames is the clx
logic of exclusion type $\bigcup\{\xcl(L'):L'\in S\}$, i.e., the join
of~$S$ in~$\CLX$. However, there is no a priori reason why $L$ itself should
have the finite model property.

Nevertheless, we will establish later that this is indeed the
case, hence $\CLX$ is a complete join-subsemilattice
of~$\NExt\lgc{K4}$ (\th\ref{cor:subsemi}).
\end{Rem}

\begin{Exm}\th\label{exm:meet}
Cluster-extensible logics are not closed under intersections,
hence $\CLX$ is not a sublattice of~$\NExt\lgc{K4}$. For example,
consider the logics $\lgc{S4Grz},\lgc{S5}\in\CLX$. Finite rooted
frames of~$L=\lgc{S4Grz}\cap\lgc{S5}$ are exactly those that are either
$\lgc{S4Grz}$-frames or~$\lgc{S5}$-frames. In particular, $L$ has a
type-$\p{\nr1,2}$ frame, but it is not $\p{\nr1,2}$-extensible: for instance, 
the two-element cluster is an $L$-frame, while the frame
\begin{center}
\magicparoff
\unitlength=.1em
\begin{picture}(40,30)
\cput(20,0){$\R$}
\put(18,5){\vector(-1,1){10}}
\put(22,5){\vector(1,1){10}}
\cput(6,15){$\R$}
\put(33,20){\oval(15,8)}
\cput(33,17.5){$\R\ \R$}
\end{picture}
\end{center}
is not. The meet $L'=\lgc{S4Grz}\wedge\lgc{S5}$ in~$\CLX$ is in fact the clx
logic satisfying $\tp(L')=\tp(\lgc{S5})\cup\tp(\lgc{S4Grz})$ and
$\xcl(L')=\xcl(\lgc{S5})\cap\xcl(\lgc{S4Grz})$, namely~$L'=\lgc{S4.1.4}$.

In contrast, we have:
\end{Exm}
\begin{Prop}\th\label{prop:dirmeet}
If $S$ is a chain (or more generally, a downward directed
set) of clx logics, then $\bigcap S$ is a clx logic, and $\xcl\bigl(\bigcap
S\bigr)=\bigcap\{\xcl(L):L\in S\}$.
\end{Prop}
\begin{Pf}
The logic $L_0=\bigcap S$ has fmp, and a finite rooted frame~$F$ is an
$L_0$-frame iff it is an $L$-frame for some $L\in S$: for the left-to-right direction, the Fine--Jankov frame formula
of~$F$ is not in~$L_0$, hence it is not in~$L$ for some $L\in S$, in which case $F$ is an $L$-frame.

Let $F$ be a
finite frame of type $\p{C,n}$ such that $F\bez\rcl(F)$ is an
$L_0$-frame. For every $u\in F\bez\rcl(F)$, there is $L_u\in S$ such
that $F_u$ is an $L_u$-frame. If $L_0$ has a type-$\p{C,n}$ frame,
then so does some $L'\in S$. Let $L\in S$ be such that $L\sset L'$ and
$L\sset L_u$ for every~$u$. Then $F\bez\rcl(F)$ is an $L$-frame and
$\p{C,n}\in\tp(L)$, hence $F$ is an $L$-frame, and a fortiori an
$L_0$-frame.
\end{Pf}

\begin{Thm}\th\label{thm:clxae}
Every clx logic~$L$ is $\forall\exists$-definable on finite frames.
\end{Thm}
\begin{Pf}
Let $U=\xcb(L)$. We know that $U$ is a finite set of finite extension conditions, and a finite 
frame is an $L$-frame iff it has no rooted generated subframe of type
$\p{D,m}\succeq\p{C,n}$, where $\p{C,n}\in U$, hence it suffices to
express the latter property by a $\forall\exists$ formula
$\xi_{C,n}$. (Notice that transitivity is defined by a universal
formula.) This is easy, we can take, e.g.,
\begin{align*}
\xi_{\I,0}&=\forall u\,\exists v\,(u<v),\\
\xi_{\I,n}&=\forall u,w_0,\dots,w_{n-1}\,\exists v\,
 \bigl(\neg(u<u)\land\alpha_n(u,\vec w)\to\beta_n(u,v,\vec w)\bigr),\\
\xi_{\nr k,0}&=\forall u_0,\dots,u_{k-1}\,\exists v\,
  \bigl(\gamma_k(\vec u)\to u_0<v\land\neg(v<u_0)\bigr),\\
\xi_{\nr k,n}&=\forall u_0,\dots,u_{k-1},w_0,\dots,w_{n-1}\,\exists v\,
  \bigl(\gamma_k(\vec u)\land\alpha_n(u_0,\vec w)
  \to\beta_n(u_0,v,\vec w)\bigr)
\end{align*}
for any $k,n\in\omega\bez\{0\}$, where
\begin{align*}
\alpha_n(u,w_0,\dots,w_{n-1})
&=\ET_{i<n}\bigl(u<w_i\land\neg(w_i<u)\bigr)
   \land\ET_{i<j<n}\neg(w_i<w_j\lor w_i=w_j\lor w_j<w_i),\\
\beta_n(u,v,w_0,\dots,w_{n-1})
&=\LOR_{i<n}\bigl(u<v\land\neg(v<u)
     \land v<w_i\land\neg(w_i<v)\bigr),\\
\gamma_k(u_0,\dots,u_{k-1})
&=\ET_{i,j<k}u_i<u_j\land\ET_{i<j<k}\neg(u_i=u_j).
\end{align*}
(Note that the last conjunction in
the definition of $\alpha_1$ and~$\gamma_1$ is empty and
represents~$\top$.)
\end{Pf}

Having established all we could say about clx logics using more-or-less
trivial methods, we now turn to the problem of their finite
axiomatizability. We will use an indirect approach: we will define
a kind of frame formulas semantically corresponding to extension
conditions, and we will show that every logic axiomatized by these formulas
has the finite model property. It will follow easily that any clx
logic is axiomatizable by a set of these formulas, which can be taken
finite due to \th\ref{lem:extcond}. As a byproduct of our proof of the
fmp we obtain an exponential bound on the size of countermodels, and
the form of canonical axiom sets we provide for clx logics also shows
that the class of clx logics is closed under joins, as alluded to in
\th\ref{rem:join}. Last but not least, being a finitely axiomatizable
logic with the fmp, every clx logic is decidable.

\begin{Def}\th\label{def:frfl}
Let $\p{C,n}\in\EC$. The frame $F^\I_{C,n}=\p{F_{C,n},<_\I}$ consists
of a root cluster $\{c_e:e<k\}$ of type~$C$ (where $k=\lh C$), and its
$n$ immediate irreflexive successors $\{s_i:i<n\}$. The frame
$F^\R_{C,n}=\p{F_{C,n},<_\R}$ is defined similarly, but the
$s_i$'s are reflexive.

If $\p{W,<,A}$ is a general frame and~$n>0$, a \emph{weak morphism
from~$W$ to~$F_{C,n}$} is a partial mapping $f$ from~$W$
onto~$F_{C,n}$ such that for every $u\in\dom(f)$ and~$v\in F_{C,n}$,
\begin{enumerate}
\item\label{item:24}
$f^{-1}(v)\in A$,
\item\label{item:25}
$u'<u$ implies $u'\in\dom(f)$ and~$f(u')<_\R f(u)$,
\item\label{item:26}
if $f(u)<_\I v$, there is $u'\in\dom(f)$ such that $u<u'$
and~$f(u')=v$.
\end{enumerate}
(That is, $f$ is
essentially a p-morphism of a downward closed definable subframe
of~$W$ onto~$F_{C,n}$, except that the $s_i$'s are hermaphroditic.)
For~$n=0$, a weak morphism
from~$W$ to~$F_{C,0}$ is defined to be a p-morphism from~$W$
to~$F_{C,0}$ (i.e., to~$C$).
We define
\begin{align*}
\alpha_{\I,0}&=\dia\top,\\
\alpha_{\I,1}&=\Box y\to y\lor\Box\bot,\\
\alpha_{\I,n}
&=\ET_{i<j<n}\Box(\boxdot x_i\lor\boxdot x_j)\to\LOR_{i<n}\Box x_i
\qquad(n>1),\\
\alpha_{\nr k,0}&=\LOR_{e<k}\diadot\Box\Bigl(\ET_{d<e}y_d\to y_e\Bigr),\\
\alpha_{\nr k,n}&=\boxdot\beta_{\nr k,n}\to y_0
\qquad(n>0),
\end{align*}
where
\begin{multline*}
\beta_{\nr k,n}=
\ET_{d<e<k}(y_d\lor y_e)
 \land\ET_{d,e<k}(\Box y_d\to y_e)
\land\ET_{i\ne j<n}(x_i\lor\boxdot x_j)\\
\land\ET_{\substack{e<k\\i<n}}\bigl((\Box x_i\to y_e)\land(x_i\lor\boxdot y_e)\bigr)
\land\ET_{i<n}\Bigl(x_i\land\ET_{e<k}y_e\to\Box x_i)\Bigr).
\end{multline*}
\end{Def}

\begin{Lem}\th\label{lem:refut}
For any $t\in\EC$ and a general frame~$W$, the following are
equivalent.
\begin{enumerate}
\item\label{item:28}
$\alpha_t$ is not valid in~$W$.
\item\label{item:29}
There is $w\in W$ and a weak morphism from the generated
subframe~$W_w$ to~$F_t$.
\end{enumerate}
\end{Lem}
\begin{Pf}
We consider $t=\p{\nr k,n}$ with~$n>0$, the other cases are left to
the reader.

\ref{item:28}${}\to{}$\ref{item:29}:
Assume that $w\in W$, and $\model$ is an admissible valuation that
makes $W_w\model\beta_{\nr k,n}$ and~$w\nmodel y_0$. Define a partial
function $f$ from $W_w$ to~$F_{\nr k,n}$ by
\begin{equation}\label{eq:17}
\begin{split}
f(u)=c_e&\iff u\nmodel y_e,\\
f(u)=s_i&\iff u\nmodel x_i.
\end{split}
\end{equation}
The truth of~$\beta_{\nr k,n}$ (namely, the clauses $y_d\lor y_e$,
$x_i\lor x_j$, and $x_i\lor y_e$) ensures that $f$ is well-defined,
and clearly $f^{-1}(v)$ is admissible in~$W_w$ for every~$v\in F_{\nr
k,n}$, so condition~\ref{item:24} from \th\ref{def:frfl} is
satisfied. Every $u$ mapped to~$c_e$ sees some points mapped to each
element of~$F_{\nr k,n}$ (using the conjuncts $\Box y_d\to y_e$ and $\Box
x_i\to y_e$), which ensures condition~\ref{item:26}, and in
view of~$f(w)=c_0$, also the surjectivity of~$f$.
Points mapped to~$s_i$ can only see points mapped to~$s_i$ (due to
$x_i\lor\Box x_j$ and $x_i\lor\Box y_e$), hence
condition~\ref{item:25} holds whenever $u'\in\dom(f)$. The last
conjunct of~$\beta_{\nr k,n}$ implies that $u'\in\dom(f)$ whenever
$u'<u$ is such that $f(u)=s_i$. Finally, if $u'<u$ and $f(u)=c_e$,
then there is $u<u''$ such that $f(u'')=s_0$, hence $u'\in\dom(f)$ as
well.

\ref{item:29}${}\to{}$\ref{item:28}:
Let $f$ be a weak morphism from $W_w$ to~$F_{\nr k,n}$. Since $f$ is onto,
we may assume that $f(w)=c_0$. Define an
admissible valuation in~$W_w$ by~\eqref{eq:17}. Then by inspection
$W_w\model\beta_{\nr k,n}$, and $w\nmodel y_0$, hence $\alpha_{\nr
k,n}$ is not valid in~$W_w$. Since $W_w$ is a generated subframe
of~$W$, it is not valid in~$W$ either.
\end{Pf}

\begin{Cor}\th\label{cor:canorder}
If $t\preceq t'$, then $\lgc{K4}\oplus\alpha_t$ proves $\alpha_{t'}$.
\noproof\end{Cor}

\begin{Rem}\th\label{rem:zax}
The axioms~$\alpha_{C,n}$ are variants of Zakharyaschev's canonical
formulas~\cite{cha-zax}. Using the refutation criterion from
\th\ref{lem:refut}, one can show
that $\alpha_{C,0}$ generates over~$\lgc{K4}$ the same logic
as Zakharyaschev's~$\alpha(C,\bot)$, and for $n>0$,
$\lgc{K4}\oplus\alpha_{C,n}$ can be axiomatized by the $n+1$ canonical
formulas $\alpha(F_i,D)$ ($i\le n$), where $F_i$ is the version
of~$F_{C,n}$ with $i$~irreflexive and $n-i$~reflexive leaves, and $D$
consists of all sets of leaves of size at least~$2$.
\end{Rem}
\begin{Lem}\th\label{lem:canfin}
If $t\in\EC$, a finite frame $F$ validates $\alpha_t$ iff it has no
rooted generated subframe of type $t'\succeq t$.
\end{Lem}
\begin{Pf}
If $F_u$ has type $\p{D,m}\succeq\p{C,n}=t$, we can define a weak
morphism from~$F_u$ to~$F_{C,n}$ by fixing a surjection of~$\cls(u)$
to $\rcl(F_{C,n})$, and picking $n$ distinct immediate successor
clusters of~$u$, each of which is mapped to one leaf of~$F_{C,n}$.

Conversely, let $f$ be a weak morphism from~$F_u$ to~$F_{C,n}$.
Without loss of generality, we may assume that $\cls(u)$ is a
$<$-maximal cluster intersecting $f^{-1}(c_0)$. Let $\p{D,m}$ be the
type of~$F_u$. Since $f$ is a p-morphism of~$\cls(u)$ to~$C$, we have
$C\preceq D$. If $n=0$, then $F_u=\dom(f)=\cls(u)$, hence~$m=0$.
Otherwise, choose a $<$-minimal point $w_i>u$ such that $f(w_i)=s_i$
for each~$i<n$. Since $\dom(f)$ is downward closed, $w_i$ must be an
immediate successor of~$u$, and we have $w_i\nleq w_j$ for~$i\ne j$,
hence~$m\ge n$.
\end{Pf}

We will now prove the crucial lemma that logics axiomatized by the $\alpha_t$ formulas have the finite model property.
The argument is a variant of selective filtration. We have to be somewhat careful, as a blind selection of witnesses
for false boxed formulas might result in nodes with too many immediate successors, violating the requisite~$\alpha_t$
(note that for example, an arbitrary finite tree can be embedded in a binary tree).
On the other hand, we also want to ensure a rather tight upper bound on the size of the extracted model, though for
clarity we formulate this separately only in \th\ref{thm:clxfinax}. To this end, we will use a combinatorial principle
stated in \th\ref{lem:treesize}, which is implicit in the proof of the exponential model property for cofinal-subframe
logics by Zakharyaschev~\cite[Thm.~4.3]{zakh-k4-ii}.
\begin{Lem}\th\label{lem:canfmp}
If $U\sset\EC$, then $\lgc{K4}\oplus\{\alpha_t:t\in U\}$ has the finite
model property.
\end{Lem}
\begin{Pf}
Put $L=\lgc{K4}\oplus\{\alpha_t:t\in U\}$, and assume
$\Gamma\nvdash_L\fii$, hence there exists a descriptive $L$-frame~$W$
and an admissible valuation~$\model$ in~$W$ such that $W\model\Gamma$
and $W\nmodel\fii$. We will construct a finite $L$-model~$F$ separating $\fii$ from~$\Gamma$.

Let $\Sigma$ be the set of all subformulas
of~$\Gamma\cup\{\fii\}$, and
$B=\{\fii\}\cup\{\psi:\Box\psi\in\Sigma\}$. For any~$u\in W$, we put
\[\bdt(u)=\{\psi\in B:u\model\boxdot\psi\}.\]
Notice that $u\le v$ implies $\bdt(u)\sset\bdt(v)$, and in particular,
$u\sim v$ implies $\bdt(u)=\bdt(v)$.
The set of \emph{critical formulas of~$u$} is
\[\crit(u)=\bigcap_{v\gnsim u}\bdt(v)\bez\bdt(u).\]
Clearly, $\crit(u)=\crit(v)$ if~$u\sim v$ (hence we can also write
$\crit(C)$ when $C$ is a cluster), and
$\crit(u)\cap\crit(v)=\nul$ if~$u\lnsim v$. By
\th\ref{lem:descrmax},
\begin{equation}\label{eq:18}
B\bez\bdt(u)=\bigcup_{v\ge u}\crit(v).
\end{equation}
We are going to construct a finite subtree $T\sset\omega^{<\omega}$,
a labelling of~$T$ by finite clusters $\{C_\sigma:\sigma\in T\}$ (including
valuation of atoms) that together define a finite rooted model
$F=\bigcup_\sigma C_\sigma$, and a mapping $f\colon F\to W$ such that
\begin{enumerate}
\item\label{item:33} $a$ and~$f(a)$ satisfy the same atoms,
\item\label{item:27} $a<b$ implies $f(a)<f(b)$ (hence $a\sim b$
implies $f(a)\sim f(b)$),
\item\label{item:32} $a\lnsim b$ implies $f(a)\lnsim f(b)$,
\item\label{item:34} $f(a)\lnsim u$ implies $\bdt(f(a))\ssset\bdt(u)$,
\item\label{item:35}
either $C_\sigma=\{a_\sigma\}$ and $\crit(f(a_\sigma))=\nul$, or
$C_\sigma=\{a_{\sigma,\psi}:\psi\in\crit(f(C_\sigma))\}$ (hence
$\crit(f(C_\sigma))\ne\nul$) and $f(a_{\sigma,\psi})\nmodel\psi$ for
every $\psi\in\crit(f(C_\sigma))$
\end{enumerate}
for every $a,b\in F$ and $u\in W$. Note that the
elements~$a_{\sigma,\psi}$ are not necessarily distinct. We denote
by~$D_\sigma$ the cluster of~$W$ including~$f(C_\sigma)$.

We build $T$ and~$F$ from bottom up. First, we put the root~$\nulstr$
in~$T$, and we find a cluster~$D_\nulstr\sset W$ such that
$\fii\in\crit(D_\nulstr)$ using~\eqref{eq:18}.

Assume that $\sigma\in T$ and a cluster~$D_\sigma\sset W$ has been
defined in such a way that
\begin{equation}\label{eq:19}
\bdt(D_\sigma)\ssset\bdt(u)\text{ for every }u\gnsim D_\sigma.
\end{equation}
If $\crit(D_\sigma)=\nul$, we pick any point~$u_\sigma\in D_\sigma$,
and we let $C_\sigma=\{a_\sigma\}$ be a copy of~$u_\sigma$,
putting~$f(a_\sigma)=u_\sigma$. Otherwise, we find an $\sset$-minimal
subset $D'_\sigma\sset D_\sigma$ such that every
$\psi\in\crit(D_\sigma)$ is refuted in some $u\in D'_\sigma$. We put $k_\sigma=\lh{D'_\sigma}$, enumerate
$D'_\sigma=\{u_{\sigma,i}:i<k_\sigma\}$, and for each $\psi\in\crit(D_\sigma)$, fix $i(\psi)<k_\sigma$ such that
$u_{\sigma,i(\psi)}\nmodel\psi$. We make
$C_\sigma=\{a_{\sigma,i}:i<k_\sigma\}$ a copy
of~$\{u_{\sigma,i}:i<k_\sigma\}$, and we put
$f(a_{\sigma,i})=u_{\sigma,i}$ and
$a_{\sigma,\psi}=a_{\sigma,i(\psi)}$. Notice that the minimality
of~$D'_\sigma$ implies
$C_\sigma=\{a_{\sigma,\psi}:\psi\in\crit(D_\sigma)\}$.

If $D_\sigma$ is a maximal cluster of~$W$, $\sigma$ will be a leaf
of~$T$, and we put~$n_\sigma=0$. Otherwise, let $S_\sigma\ne\nul$ be
the collection of all $\sset$-minimal sets in $\{\bdt(u):u\gnsim
D_\sigma\}$. If $\lh{S_\sigma}\ge2$ and some $d\in S_\sigma$ includes
$\bigcap\{d'\in S_\sigma:d'\ne d\}$, we remove~$d$ from~$S_\sigma$.
Continuing in the same way, we eventually obtain a subset
$R_\sigma\sset S_\sigma$ such that either $\lh{R_\sigma}=1$, or
\begin{equation}\label{eq:20}
d\nSset\bigcap_{\substack{d'\in R_\sigma\\d'\ne d}}d'
\end{equation}
for every $d\in R_\sigma$. We fix an enumeration
$R_\sigma=\{d_{\sigma,i}:i<n_\sigma\}$, where $n_\sigma=\lh{R_\sigma}$. For every~$i<n_\sigma$, we add
$\sigma^\cat i$ into~$T$, and we choose a $<$-maximal cluster
$D_{\sigma^\cat i}\sset W$ such that $D_\sigma\lnsim D_{\sigma^\cat
i}$ and $\bdt(D_{\sigma^\cat i})=d_{\sigma,i}$ using
\th\ref{lem:descrmax}. Notice that \eqref{eq:19} holds
for~$D_{\sigma^\cat i}$, hence we can carry on with the construction.

Assume that the construction has been completed. Since
$\bdt(D_\sigma)\ssset\bdt(D_{\sigma^\cat i})$ for every $\sigma^\cat
i\in T$, each $\sigma\in T$ has $\lh \sigma\le\lh B$, and in
particular, $T$ and~$F$ are finite.
Properties \ref{item:33}--\ref{item:35} are clearly
satisfied. We claim $a\equiv_\Sigma f(a)$ for every~$a\in F$.
We will prove
\begin{equation}\label{eq:21}
F,a\model\psi\iff W,f(a)\model\psi\qquad(a\in C_\sigma)
\end{equation}
by outer top-down induction on~$\sigma\in T$, and inner induction on
the complexity of~$\psi\in\Sigma$. The steps for atoms and Boolean
connectives are trivial. If $f(a)\model\Box\psi$, then
$f(b)\model\psi$ for every $b>a$ by~\ref{item:27}, hence
$b\model\psi$ by the induction hypothesis. Thus, $a\model\Box\psi$.
Conversely, assume that $f(a)\nmodel\Box\psi$. If there is $u\gnsim
D_\sigma$ such that $u\nmodel\boxdot\psi$, i.e., $\psi\notin\bdt(u)$,
then $\psi\notin d$ for some $d\in S_\sigma$, hence also $\psi\notin
d$ for some~$d\in R_\sigma$. Putting $d_{\sigma,i}=d$, we have
$D_{\sigma^\cat i}\nmodel\boxdot\psi$. If we pick any $b\in
C_{\sigma^\cat i}$, then $f(b)\nmodel\psi$ or $f(b)\nmodel\Box\psi$,
hence $b\nmodel\psi$ or~$b\nmodel\Box\psi$ by the induction hypothesis
for~$\sigma^\cat i$, hence $a\nmodel\Box\psi$, as $a<b$. On the other
hand, if $\psi\in\bdt(u)$ for every $u\gnsim D_\sigma$, then $D_\sigma$
(and $C_\sigma$) must be reflexive, and $\psi\in\crit(D_\sigma)$.
By~\ref{item:35}, we have $f(a_{\sigma,\psi})\nmodel\psi$, hence
$a_{\sigma,\psi}\nmodel\psi$ by the induction hypothesis for~$\psi$,
which implies $a\nmodel\Box\psi$ as~$a\sim a_{\sigma,\psi}$.

In particular, \eqref{eq:21} implies that $F\model\Gamma$ and
$F,a_{\nulstr,\fii}\nmodel\fii$. It remains to show that $F$ is an
$L$-frame. Assume that $a\in C_\sigma$ has type
$\p{C,n}$, and let $W'=W_{f(a)}$. Notice that $\lh C=k_\sigma$, and
$n=n_\sigma$. By~\eqref{eq:19}, $D_\sigma$ is definable in~$W'$. If
$k_\sigma=1$, we put~$E_0=D_\sigma$. Otherwise
$D_\sigma\Sset\{u_{\sigma,i}:i<k_\sigma\}$, and each $u_{\sigma,i}$
refutes a formula~$\psi\in\crit(D_\sigma)$ that holds in
all~$u_{\sigma,j}$, $j\ne i$. Thus, we can partition~$D_\sigma$ into
$k_\sigma$ nonempty subsets $E_i$, $i<k_\sigma$, definable
in~$W'$.

If $n_\sigma=0$, then $W'=D_\sigma$. Otherwise, let $A_i=\{u\in
W':\bdt(u)=d_{\sigma,i}\}$ for every~$i<n_\sigma$. Clearly, $A_i$ is
definable in~$W'$, disjoint from~$D_\sigma$, and as $d_{\sigma,i}$
and~$d_{\sigma,j}$ are incomparable for~$i\ne j$, we have $u\nleq v$
for any $u\in A_i$, $v\in A_j$. Moreover, $A_i\cup D_\sigma$ is
downward closed: if $u\in W'$, $u<v\in A_i$, then
$\bdt(D_\sigma)\sset\bdt(u)\sset d_{\sigma,i}$. Since $d_{\sigma,i}\in
S_\sigma$, this means that either $\bdt(u)=d_{\sigma,i}$, i.e., $u\in
A_i$, or $\bdt(u)=\bdt(D_\sigma)$, i.e., $u\in D_\sigma$
by~\eqref{eq:19}. It follows that the partial mapping
\begin{align*}
g(u)=c_i&\iff u\in E_i,\\
g(u)=s_i&\iff u\in A_i
\end{align*}
is a weak morphism of~$W'$ to $F_{C,n}$. Since $W$, and
therefore~$W'$, is an $L$-frame, we must have $\p{C,n}\nsucceq t$ for
every~$t\in U$. Thus, $F$ is an $L$-frame.
\end{Pf}

\begin{Lem}\th\label{lem:treesize}
Let $T$ be a tree with root~$\roo$, labelled by finite sets
$\{X_\sigma:\sigma\in T\}$. If $\sigma\in T$, and $\{\sigma_i:i<n\}$
is the set of all immediate successors of~$\sigma$, we assume that
\begin{enumerate}
\item\label{item:36} $X_{\sigma_i}\ssset X_\sigma,$
\item\label{item:37} if $n>1$, then $X_{\sigma_i}\nsset\bigcup_{j\ne
i}X_{\sigma_j}$
\end{enumerate}
for every~$i<n$. Then $\lh T<3\cdot2^{\lh{X_\roo}-1}$.
\end{Lem}
\begin{Pf}
For every~$\sigma\in T$, let $T_\sigma$ be the subtree of~$T$ rooted
at~$\sigma$, and $m_\sigma=\lh{X_\sigma}$. We will prove
\begin{equation}\label{eq:22}
\lh{T_\sigma}<3\cdot2^{m_\sigma-1}
\end{equation}
by induction on~$m_\sigma$. Let $\{\sigma_i:i<n\}$ be the set of
immediate successors of~$\sigma$, and assume that \eqref{eq:22} holds
for every~$\tau$ such that $m_\tau<m_\sigma$, and in particular, for
every~$\sigma_i$ in view of~\ref{item:36}.

If $n=0$, then $\lh{T_\sigma}=1<3\cdot2^{m_\sigma-1}$. If $n=1$, then
\[\lh{T_\sigma}=1+\lh{T_{\sigma_0}}<1+3\cdot2^{m_{\sigma_0}-1}
  <3\cdot2^{m_{\sigma_0}}\le3\cdot2^{m_\sigma-1}\]
by the induction hypothesis, as~$1<3/2$.

If $n\ge2$, then for every~$i<n$, there exists $x_i\in X_{\sigma_i}\sset
X_\sigma$ such that $x_i\notin X_{\sigma_j}$ for every~$j\ne i$
by~\ref{item:37}. This means $X_{\sigma_i}\sset
X_\sigma\bez\{x_j:j\ne i\}$, hence $m_{\sigma_i}\le m_\sigma-n+1$.
Also $m_{\sigma_i}>0$, hence by the induction hypothesis, we obtain
\[\lh{T_\sigma}=1+\sum_{i<n}\lh{T_{\sigma_i}}
  \le1+n\bigl(3\cdot2^{m_\sigma-n}-1\bigr)
  \le1+2\bigl(3\cdot2^{m_\sigma-2}-1\bigr)
  =3\cdot2^{m_\sigma-1}-1,\]
using the fact that $n(3\cdot2^{m_\sigma-n}-1)$ is nonincreasing as a
function of~$n$.
\end{Pf}

\begin{Exm}\th\label{exm:bintree}
Let $m\ge1$, and $T'$ be the full binary tree of height~$m$. Label the
root by an $m$-element set, and for every inner node, label its two
immediate successors by two distinct subsets of its label of size
smaller by one. Notice that nodes of depth~$d$ have labels of
size~$d$; in particular, the labels of the leaves are singletons. Let
$T$ be obtained from~$T'$ by adding a new leaf with empty label over
every leaf of~$T'$. Then $T$ satisfies the assumptions of
\th\ref{lem:treesize} with $\lh{X_\roo}=m$, and $\lh
T=2^m-1+2^{m-1}=3\cdot2^{m-1}-1$.
\end{Exm}

\begin{Thm}\th\label{thm:clxfinax}
Let $L$ be a clx logic.
\begin{enumerate}
\item\label{item:30}
$L$ is finitely axiomatizable. Specifically,
\[L=\lgc{K4}\oplus\{\alpha_t:t\in\xcb(L)\}.\]
\item\label{item:31}
$L$ has an exponential-size model property: if $\Gamma\nvdash_L\fii$,
where $\Gamma\cup\{\Box\fii\}$ has $b$ boxed subformulas, then
$\Gamma\ru\fii$ can be refuted in a rooted $L$-frame that is a tree
of clusters of total size $<3\cdot2^{b-1}$, depth $\le b+1$, cluster
size $\le b$, and branching $\le\max\{b-1,1\}$.
\end{enumerate}
\end{Thm}
\begin{Pf}

\ref{item:30}:
The logic $L'=\lgc{K4}\oplus\{\alpha_t:t\in\xcb(L)\}$ has the same finite
frames as~$L$ by \th\ref{lem:canfin}, and enjoys fmp by
\th\ref{lem:canfmp}, hence $L=L'$.

\ref{item:31}:
If $\Gamma\nvdash_L\fii$, then $\Gamma\nvdash_{L'}\fii$. Consider the
finite $L'$-model $F\model\Gamma$, $F\nmodel\fii$, constructed in the
proof of \th\ref{lem:canfmp}. As already observed there, $F$ has depth
at most $\lh B+1=b+1$. By property~\ref{item:35} of~$f$, $F$ has
cluster size at most~$b$. If $C_\sigma$ has $n_\sigma>1$ immediate
successor clusters,
then \eqref{eq:20} implies that for every~$i<n_\sigma$, we can choose
$\psi_i\in B$ such that $\psi_i\notin d_{\sigma,i}$,
and $\psi_i\in d_{\sigma,j}$ for every~$j\ne i$. Moreover, since
$\fii\in\crit(D_\nulstr)$, we have $\psi_i\in B\bez\{\fii\}$ for
every~$i$. Thus, $n_\sigma\le b-1$.

We will bound $\lh F$ using \th\ref{lem:treesize}. If
$C_\sigma=\{a_\sigma\}$, we label~$\sigma$ with
$X_\sigma:=B\bez\bdt(D_\sigma)$. Otherwise,
$k_\sigma=\lh{C_\sigma}\le\lh{\crit(D_\sigma)}$. We first ``linearize
$C_\sigma$'' by replacing $\sigma$ with a chain
$\{\sigma^i:i<k_\sigma\}$, where $\sigma^{i+1}$ is a successor
of~$\sigma^i$, and we choose labels $X_{\sigma^i}$
so that $B\bez\bdt(D_\sigma)=X_{\sigma^0}\sSset
X_{\sigma^1}\sSset\dots\sSset X_{\sigma^{k_\sigma-1}}\sSset
B\bez\bigcap_{u\gnsim
D_\sigma}\bdt(u)=B\bez(\bdt(D_\sigma)\cup\crit(D_\sigma))$. In this
way, we obtain a tree~$T'$ of the same size as~$F$, and its labelling
satisfies the assumptions of \th\ref{lem:treesize} due
to~\eqref{eq:20}. Thus, $\lh F=\lh{T'}<3\cdot2^{\lh{X_\nulstr}-1}\le3\cdot2^{b-1}$.
\end{Pf}

The fact that the bounds in \th\ref{thm:clxfinax} depend only on the
number of boxed subformulas and not on the overall size
of~$\Gamma\cup\{\fii\}$ will be important in the proof of
\th\ref{thm:clxadm}.

\begin{Cor}\th\label{cor:subsemi}
If $S$ is a set of clx logics, the join of~$S$ in~$\NExt\lgc{K4}$ is
also a clx logic.
\end{Cor}
\begin{Pf}
By \th\ref{thm:clxfinax}, $L'=\LOR S$ can be axiomatized by
$\{\alpha_t:t\in U\}$, where $U=\bigcup_{L\in S}\xcb(L)$. Thus,
$L'=\clx_{U'}$ using \th\ref{thm:clxfinax} again, where $U'$ is the
upward closure of~$U$.
\end{Pf}
\begin{Cor}\th\label{cor:clxdec}
If $L$ is a clx logic, then $L$ and~$\adm_L$ are decidable, and given
any formula, we can compute its projective approximation and a
complete set of unifiers.
\noproof\end{Cor}

\section{Extension rules}\label{sec:extension-rules}
We now proceed to the main part of this paper, namely the construction
of bases of admissible rules for $\Par$-extensible logics, and their
semantic description. We introduce certain rules related to extension
conditions, called extension rules, and we show that their validity in
nice (i.e., descriptive or Kripke) parametric frames corresponds to
the existence of (a suitable version of) tight predecessors for finite
sets of points in the frame (\th\ref{thm:irrextchar,thm:reflextchar}).
As a consequence, we obtain a characterization of $\Par$-extensible
logics as those that admit appropriate sets of extension rules
(\th\ref{thm:admext}). We prove that consequence relations axiomatized
by extension rules are complete with respect to locally finite Kripke
frames (\th\ref{thm:extlocfin}), and with the help of the description
of projective formulas from Section~\ref{sec:projective-formulas}, we
derive our main result (\th\ref{thm:clxadmchar}) stating that
$\Par$-extensible logics have bases of admissible rules consisting of
extension rules, and that the consequence relation~$\adm_L$ is
sound and complete with respect to frames having enough tight
predecessors. Such frames are nearly always infinite (cf.~\cite{rko}), which may be
sometimes inconvenient; for this reason, we also give a description of
admissible rules in terms of suitable finite (in fact, exponentially
bounded) models (\th\ref{thm:pext,thm:clxadm}).

While the natural form of extension rules has multiple conclusions, we
indicate in Section~\ref{sec:single-concl-bases} how to turn them into
single-conclusion rules providing bases of single-conclusion rules for
$\Par$-extensible logics. Further properties of bases are investigated
in Section~\ref{sec:finite-indep-bases}: we show how to modify
extension rules so as to obtain independent bases of admissible rules
for finite sets of parameters, and we characterize which
$\Par$-extensible logics have finite bases.

The semantics of extension rules will be given in terms of the notion
of tight predecessor\footnote{Tight predecessors are also called co-covers in~\cite[\S6.2]{ryb:bk}.} defined below. The main difference from the
parameter-free case is that the predecessors are no longer just
singletons: we also need to take care of proper clusters whose
individual points are distinguishable by a valuation of parameters, following the discussion at the beginning of Section~\ref{sec:clx-logics}.
\begin{Def}\th\label{def:tp}
Let $W$ be a parametric general frame, $P\sset\Par$ finite,
$n\in\omega$, and $X=\{w_i:i<n\}\sset W$ (where the $w_i$ are
not necessarily distinct).

If $e\in\two^P$, a \emph{tight $\p{\I,\{e\}}$-predecessor ($\p{\I,\{e\}}$-tp)}
of~$X$ is $\{u\}\sset W$ such that
\[W,u\model P^e,\qquad u\up=X\Up.\]
Tight $\p{\I,\{e\}}$-predecessors are also collectively called
\emph{irreflexive tight predecessors.}
(However, notice that when $v$ is a reflexive smallest element of~$X$
satisfying $P^e$, then $\{v\}$ is a $\p{\I,\{e\}}$-tp of~$X$, despite not
being irreflexive.) $W$ is \emph{$\p{\I,n,\{e\}}$-extensible} if every
$\{w_i:i<n\}\sset W$ has a $\p{\I,\{e\}}$-tp, and it is
\emph{$\p{\I,n}$-extensible} if it is $\p{\I,n,\{e\}}$-extensible for
every finite~$P$ 
and~$e\in\two^P$.

Similarly, if $E\sset\two^P$, $E\ne\nul$, then a \emph{tight
$\p{\R,E}$-predecessor ($\p{\R,E}$-tp)} of~$X$ is $\{u_e:e\in E\}\sset
W$ such that
\[W,u_e\model P^e,\qquad u_e\up=X\Up\cup\{u_f:f\in E\}.\]
Any tight $\p{\R,E}$-predecessor is also called a
\emph{reflexive tight predecessor.}
(Again, a $\p{\R,E}$-tp may be included in $X\up$ when $X$ has a
reflexive smallest element whose cluster realizes every $e\in
E$.) $W$ is \emph{$\p{\R,n,E}$-extensible} if every
$\{w_i:i<n\}\sset W$ has a $\p{\R,E}$-tp, and it is
\emph{$\p{\nr k,n}$-extensible} if it is $\p{\R,n,E}$-extensible for
every finite~$P$ 
and $E\sset\two^P$ such that $0<\lh E\le k$.
\end{Def}

\begin{figure}
\centering
\includegraphics{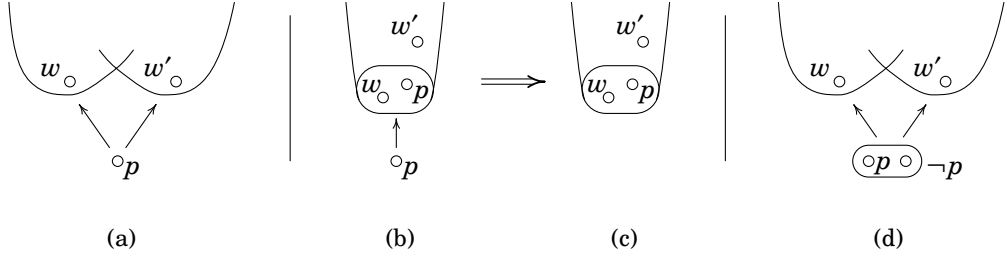}
\caption{Illustrations of tight predecessors (see \th\ref{exm:rules})}
\label{fig:tp}
\end{figure}

Next we define our rules syntactically. The irreflexive case is a
straightforward modification of the parameter-free rules given
in~\cite{ejadm,ej:indep}, but the reflexive case is more peculiar, as we need to enforce the exact composition of the
root cluster in terms of the valuation of parameters.
\begin{Def}\th\label{def:extrules}
Let $n\in\omega$, $P\sset\Par$ be finite, and~$e\in\two^P$. 
(If $\Par$ itself is finite, it suffices to consider $P=\Par$.) The
\emph{irreflexive extension rule} $\Ext_{\I,n,\{e\}}$ is
\[P^e\land\Box y\to\LOR_{i<n}\Box x_i\Ru\{\boxdot y\to x_i:i<n\}.\]
Let $\Ext^\Par_{\I,n}$ denote the set of all rules of the form
$\Ext_{\I,n,\{e\}}$.

If $n$ and~$P$ are as above, $E\sset\two^P$, and $e_0\in E$, the
\emph{reflexive extension rule} $\Ext_{\R,n,E,e_0}$ is
\[P^{e_0}\land\boxdot\Bigl(y\to\LOR_{e\in E}\Box(P^e\to y)\Bigr)
     \land\ET_{e\in E}\boxdot\bigl(\Box(P^e\to\Box y)\to y\bigr)
   \to\LOR_{i<n}\Box x_i
\Ru\{\boxdot y\to x_i:i<n\}.\]
Let $\Ext_{\R,n,E}=\{\Ext_{\R,n,E,e_0}:e_0\in E\}$.
If $k\in\omega\bez\{0\}$, let $\Ext^\Par_{\nr k,n}$ denote the set of all
rules of the form $\Ext_{\R,n,E,e_0}$, where $\lh E\le k$.
\end{Def}
We remark that the $P^{e_0}$ conjunct in the reflexive rules can be
dropped for $n\ne1$, as long as the corresponding rules for~$n=1$
(with the conjunct) are present, cf.\
\th\ref{rem:RnEe,lem:mcindepbas}. However, one can check that other
elements of the definition are essential, using variants of the
construction from the proof of \th\ref{lem:mcindepbas}.
\begin{Exm}\th\label{exm:rules}
For $n=2$ and $P=\nul$ we have one reflexive extension rule:
\[\boxdot(y\to\Box y)\land\boxdot(\Box\Box y\to y)\to\Box x\lor\Box x'\ru\boxdot y\to x,\boxdot y\to x'.\]
This simplifies to the rule
\[\boxdot(y\eq\Box y)\to\Box x\lor\Box x'\ru\boxdot y\to x,\boxdot y\to x'\]
denoted as $(A'_2)$ in~\cite{ej:indep}, and semantically corresponds to the existence of ordinary reflexive tp's for
every pair of points $\{w,w'\}$.

For $P=\{p\}$, we have three kinds of reflexive extension rules. The first rule~$\Ext_{\R,2,\{p\}}$,
\[p\land\boxdot\bigl(y\to\Box(p\to y)\bigr)\land\boxdot\bigl(\Box(p\to\Box y)\to y\bigr)\to\Box x\lor\Box x'
\ru\boxdot y\to x,\boxdot y\to x',\]
expresses the existence of ordinary tp's with the valuation of~$p$ set to true in the tp. Typically these tp's look as
in Fig.~\ref{fig:tp}~(a), but in the exceptional situation in Fig.~\ref{fig:tp}~(c), where $w$ belongs to a
reflexive cluster which contains a (not necessarily distinct) point~$u\model p$, and $w'$ is above~$w$, we can take $u$
for the tp of $\{w,w'\}$. In this case, no rule can force the existence of a ``proper'' tp as in Fig.~\ref{fig:tp}~(b),
because there is a p-morphism from (b) to~(c) which collapses the tp upwards.

The second rule~$\Ext_{\R,2,\{\neg p\}}$,
\[\neg p\land\boxdot\bigl(y\to\Box(\neg p\to y)\bigr)\land\boxdot\bigl(\Box(\neg p\to\Box y)\to y\bigr)\to\Box x\lor\Box x'
\ru\boxdot y\to x,\boxdot y\to x',\]
is completely analogous to~$\Ext_{\R,2,\{p\}}$, except that $p$ is made false at the tp.

Finally, we have the pair of rules~$\Ext_{\R,2,\{p,\neg p\}}=\{\Ext_{\R,2,\{p,\neg p\},p},\Ext_{\R,2,\{p,\neg p\},\neg p}\}$:
\begin{gather*}
\frac{\phantom\neg p\land\boxdot\bigl(y\to\Box(p\to y)\lor\Box(\neg p\to y)\bigr)
   \land\boxdot\bigl(\Box(p\to\Box y)\lor\Box(\neg p\to\Box y)\to y\bigr)\to\Box x\lor\Box x'}
{\boxdot y\to x,\boxdot y\to x'}\\[\medskipamount]
\frac{\neg p\land\boxdot\bigl(y\to\Box(p\to y)\lor\Box(\neg p\to y)\bigr)
   \land\boxdot\bigl(\Box(p\to\Box y)\lor\Box(\neg p\to\Box y)\to y\bigr)\to\Box x\lor\Box x'}
{\boxdot y\to x,\boxdot y\to x'}
\end{gather*}
They express the existence of two-element tp clusters as in Fig.~\ref{fig:tp}~(d), where one point in the tp
satisfies~$p$, and the other one does not. If $w\nmodel\boxdot y\to x$, and~$w'\nmodel\boxdot y\to x'$, then the
premise of the first rule is false in the element of the tp that satisfies~$p$, and the premise of the second rule
is false in its mate. Either rule alone is sufficient to guarantee tp's in the generic situation depicted in~(d). To
see that we actually need both rules, assume that $w,w'$ are arranged as in Fig.~\ref{fig:tp}~(c), and 
that all points in~$\cls(w)$ satisfy~$p$. Then the premise of the first rule is false in~$w$, and only the second rule
calls for an honest tp of $\{w,w'\}$.
\end{Exm}

The correspondence of extension rules to extensible parametric frames will be our main technical tool, thus we proceed
to state it formally and prove it. While the correspondence is straightforward to show for irreflexive rules where
parameters do not have much effect, the reflexive case is more intricate, hence we start with the former.
\begin{Thm}\th\label{thm:irrextchar}
Let $P\sset\Par$ be finite, $W$ a parametric general
frame, $n\in\omega$, and $e\in\two^P$.
\begin{enumerate}
\item\label{item:irr-tp-rule}
If $W$ is $\p{\I,n,\{e\}}$-extensible, then $W\model\Ext_{\I,n,\{e\}}$.
\item\label{item:irr-rule-tp}
If $W$ is a descriptive or Kripke frame, and
$W\model\Ext_{\I,n,\{e\}}$, then $W$ is $\p{\I,n,\{e\}}$-extensible.
\end{enumerate}
\end{Thm}
\begin{Pf}
\ref{item:irr-tp-rule}: Let $\model$ be an admissible valuation
in~$W$ that refutes the conclusion of~$\Ext_{\I,n,\{e\}}$. Pick $w_i\in
W$ such that $w_i\model\boxdot y\land\neg x_i$, and let $\{u\}$ be a
$\p{\I,\{e\}}$-tp of $\{w_i:i<n\}$. Then $u$ refutes the premise
of~$\Ext_{\I,n,\{e\}}$.

\ref{item:irr-rule-tp}:
Assume first that $W\model\Ext_{\I,n,\{e\}}$ is a Kripke frame. Given
$X=\{w_i:i<n\}\sset W$, define a valuation in~$W$ by
\begin{align*}
u&\model x_i\iff u\ne w_i,\\
u&\model y\iff u\in X\Up.
\end{align*}
This valuation refutes the conclusion of~$\Ext_{\I,n,\{e\}}$ as
$w_i\nmodel\boxdot y\to x_i$, hence there exists $u\in W$ such that
\[u\model P^e\land\Box y\land\ET_{i<n}\dia\neg x_i.\]
Thus, $\{u\}$ is a $\p{\I,\{e\}}$-tp of~$X$.

Now, let $\p{W,<,A,\model_p}\model\Ext_{\I,n,\{e\}}$ be a descriptive
parametric frame, and $X=\{w_i:i<n\}\sset W$. We will write $P^e$ for
the set of points of~$W$ where the formula~$P^e$ holds. We claim that
the set 
\[U=\{P^e\}\cup\{\Box B:B\in A,X\sset\boxdot B\}
  \cup\{\dia C:C\in A,C\cap X\ne\nul\}\]
has fip. Indeed, assume $X\sset\boxdot B_j$, $w_i\in C_{i,j}$, $j<m$.
Let $\model$ be the valuation that makes $x_i$ true on
$W\bez\bigcap_jC_{i,j}$, and $y$ true on $\bigcap_jB_j$. We have
$w_i\nmodel\boxdot y\to x_i$, hence by $\Ext_{\I,n,\{e\}}$, there is $u$
such that $u\model P^e\land\Box y\land\ET_{i<n}\dia\neg x_i$. Then
$u\in P^e\cap\bigcap_j\Box B_j\cap\bigcap_{i,j}\dia C_{i,j}$.

Since $W$ is compact, there is a point $u\in\bigcap U$. Clearly,
$u\model P^e$. Since $u\in\dia C$ for every $C\in A$ such that $w_i\in
C$, and $W$ is refined, we have $u<w_i$, thus $u\up\Sset X\Up$. On the
other hand, if $v\notin X\Up$, we can find $B_i,B'_i\in A$ such that
$w_i\in B_i\cap\Box B'_i$, $v\notin B_i\cup B'_i$ using the
refinedness of~$W$. Putting $B=\bigcup_i(B_i\cup B'_i)$, we have
$X\sset\boxdot B$, hence $u\in\Box B$. However, $v\notin B$, hence
$u\nless v$. Thus, $\{u\}$ is a $\p{\I,\{e\}}$-tp of~$X$.
\end{Pf}

\begin{Thm}\th\label{thm:reflextchar}
Let $P\sset\Par$ be finite, $W$ a parametric general
frame, $n\in\omega$, and $E\sset\two^P$, $E\ne\nul$.
\begin{enumerate}
\item\label{item:refl-tp-rule}
If $W$ is $\p{\R,n,E}$-extensible, then $W\model\Ext_{\R,n,E}$.
\item\label{item:refl-rule-tp}
If $W$ is a descriptive or Kripke frame, and
$W\model\Ext_{\R,n,E}$, then $W$ is $\p{\R,n,E}$-extensible.
\end{enumerate}
\end{Thm}
\begin{Pf}
\ref{item:refl-tp-rule}: Let $e_0\in E$, and $\model$ be an
admissible valuation in~$W$ that refutes the conclusion
of~$\Ext_{\R,n,E,e_0}$. Pick $w_i\in W$ such that $w_i\model\boxdot
y\land\neg x_i$, and let $\{u_e:e\in E\}$ be a $\p{\R,E}$-tp of
$X=\{w_i:i<n\}$. Then $u_{e_0}$ refutes the premise
of~$\Ext_{\R,n,E,e_0}$: clearly $u_{e_0}\model P^{e_0}$, and
$u_{e_0}\model\ET_i\neg\Box x_i$ as $u_{e_0}<w_i$. If $u_{e_0}\le
v\model y$, then either $v\in X\Up$, in which case $v\model\Box y$, or
$v=u_e$ for some~$e\in E$, in which case $v\model\Box(P^e\to y)$. If
$v\ge u$ and~$v\nmodel y$, then $v\notin X\Up$, thus for every~$e\in
E$, $v<u_e\model P^e\land\neg\Box y$ (as $u_e<v$), hence
$v\nmodel\Box(P^e\to\Box y)$.

\ref{item:refl-rule-tp}: The proof is a bit involved, hence we defer
it to \th\ref{lem:reflextkrip,lem:reflextdesc} below.
\end{Pf}

\begin{Cor}\th\label{cor:extchar}
Let $W$ be a parametric frame, $n\in\omega$, and
$C$ a cluster type. Then
$W\model\Ext^\Par_{C,n}$ if $W$ is $\p{C,n}$-extensible. If $W$ is a
descriptive or Kripke frame, the converse implication also holds.
\noproof\end{Cor}

The overall strategy for proving \th\ref{thm:reflextchar}~\ref{item:refl-rule-tp} will be the same as in
\th\ref{thm:irrextchar}: assuming $X=\{w_i:i<n\}$ has no $\p{\R,E}$-tp, we want to refute $\Ext_{\R,n,E}$ using a
valuation which makes $x_i$ true everywhere except~$w_i$ (or its small neighbourhood in the descriptive case). The
involvement of parameters and proper clusters leads to difficulties in defining a valuation of~$y$ to make this work.
There are many different ways in which a given cluster may fail to be a $\p{\R,E}$-tp of~$X$ (it may see too many or too
few points; it may realize a valuation of parameters not present in~$E$, or fail to realize some $e\in E$; and it
may include multiple points realizing the same $e\in E$), and in each case we will need a different way of defining the
valuation of~$y$ so that it detects the failure, but on the other hand, we need to do this coherently for all
clusters at once. As a result, we will end up with an unsightly definition by cases.

We begin the proof of with the case of Kripke frames. We will use the following variant of
Kat\v etov's lemma on three sets~\cite{katet}, whose proof
we include for completeness.
\begin{Lem}\th\label{lem:kat}
Let $f\colon A\to A$ be a function such that $f^k(x)\ne x$ for every
$x\in A$ and odd~$k$. Then we can partition $A$ into disjoint sets $A_0$
and~$A_1$ such that $f(A_i)\sset A_{1-i}$, $i=0,1$.
\end{Lem}
\begin{Pf}
For $x,y\in A$, we write $x\approx y$ if $f^n(x)=f^m(y)$ for some
$n,m\in\omega$, and $x\sim y$ if in addition $n\equiv m\pmod2$. It is
easy to see that $\approx$ and $\sim$ are equivalence relations, and
$x\approx y$ iff $x\sim y$ or $f(x)\sim y$. On the other hand,
$x\nsim f(x)$, as otherwise $f^n(x)=f^k(f^n(x))$ for some~$n$ and an
odd~$k$. Let $B\sset A$ be a set containing one point in each
equivalence class of~$\approx$, and put $A_0=\{x\in A:\exists y\in
B\,(x\sim y)\}$, $A_1=A\bez A_0=\{x\in A:\exists y\in B\,(f(x)\sim
y)\}$. Then the properties of $\approx$ and~$\sim$ ensure $f(A_0)\sset
A_1$ and~$f(A_1)\sset A_0$.
\end{Pf}

\begin{Lem}\th\label{lem:reflextkrip}
Let $P\sset\Par$ be finite, $n\in\omega$, and $E\sset\two^P$,
$E\ne\nul$. If $W$ is a Kripke frame such that
$W\model\Ext_{\R,n,E}$, then $W$ is $\p{\R,n,E}$-extensible.
\end{Lem}
\begin{Pf}
Assume that $X=\{w_i:i<n\}\sset W$ does not have a $\p{\R,E}$-tp, we
will show that $W\nmodel\Ext_{\R,n,E,e_0}$ for some $e_0\in E$. If
there is $i_0<n$ such that $w_{i_0}<w_i$ for every~$i<n$ (including
itself), then $\cls(w_{i_0})$ contains a $\p{\R,E}$-tp of~$X$ unless
some~$e\in E$ is not realized in~$\cls(w_{i_0})$, and we choose~$e_0$
to be one such~$e$. Otherwise, we can take an arbitrary~$e_0\in E$.

For every $i<n$ and~$u\in W$, we define
\[u\model x_i\iff u\ne w_i.\]
Put $S=W\bez X\Up$. Let us say that an \emph{$E$-cluster} is a
reflexive cluster $C\sset W$ such that $\sat_P(u)\in E$ for every~$u\in C$,
and conversely, for every~$e\in E$ there is a unique $u\in C$ such
that $\sat_P(u)=e$. We also consider the condition
\begin{equation}\label{eq:7}
\forall e\in E\,\forall v\in u\Up\cap S\,\exists w\in v\up\cap S\:
  w\model P^e
\end{equation}
on $u\in W$. Notice that a point~$u$ in a $<$-maximal cluster of~$S$
satisfies~\eqref{eq:7} iff it is reflexive and for every~$e\in E$, $\cls(u)$ includes a point
realizing~$P^e$  (in particular, this holds if it is
an $E$-cluster). We define a valuation of~$y$ by cases:
\begin{itemize}
\item If $u\in X\Up$, we put $u\model y$.
\item If $u\in S$, and \eqref{eq:7} does not hold, then $u\nmodel y$.
\item If $C$ is an $E$-cluster that is a $<$-maximal cluster of~$S$,
we put $u\nmodel y$ for every $u\in C$.
\item Let $C$ be a $<$-maximal cluster of~$S$ that
satisfies~\eqref{eq:7}, but is not an $E$-cluster. Then there is a point
$v\in C$ such that $\sat_P(v)\notin E$, or a point $v\in C$ such that
$v\equiv_Pv'$ for some $v'\in C$, $v'\ne v$. We pick one such~$v$, and
make $v\model y$, $u\nmodel y$ for $u\in C\bez\{v\}$.
\item If $u\in S$ satisfies~\eqref{eq:7}, sees a maximal cluster
in~$S$, but is not itself in such a maximal cluster, then $u\model y$.
\item Let $T$ be the set of all $u\in S$ that satisfy~\eqref{eq:7},
but do not see any maximal cluster of~$S$, and for every~$e\in E$, let
$T^e=\{u\in T:u\model P^e\}$. Since condition~\eqref{eq:7} is
preserved upwards in~$S$, for every $u\in T$ there is $v\gnsim u$,
$v\in T$, which in turn sees some $w\in T^e$ for any~$e\in E$. Thus,
we can choose a function $f_e\colon T^e\to T^e$ such that $u\lnsim 
f_e(u)$ for every~$u\in T^e$.
Since $f_e$ is strictly increasing, it is cycle-free, hence by
\th\ref{lem:kat}, we can write $T^e$ as a disjoint union $T^e_0\cupd T^e_1$
such that $f_e(T^e_i)\sset T^e_{1-i}$, $i=0,1$. We put $u\model y$ for $u\in
T^e_0$, and $u\nmodel y$ for $u\in T^e_1$. Satisfaction of~$y$ in
$T\bez\bigcup_{e\in E}T^e$ is arbitrary.
\end{itemize}
\begin{Cl}\th\label{cl:vtrwevtr}
Let $u\in S$ satisfy~\eqref{eq:7}.
\begin{enumerate}
\item\label{item:3}
For every $e\in E$, there is $v\in S$ such that $u<v\model
P^e\land\neg y$.
\item\label{item:4}
If $\cls(u)$ is not an $E$-cluster maximal in~$S$, then there is $v\in
S$ such that $u\le v\model y$.
\end{enumerate}
\end{Cl}
\begin{Pf*}
\ref{item:3}: If $u$ sees a maximal cluster~$C$ of~$S$, then $C$ is
reflexive and contains $v\model P^e$ by~\eqref{eq:7}. We have $v\nmodel
y$, unless $C$ contains another point $v'\model P^e$, which then does
not satisfy~$y$.

If $u\in T$, there is $v\in S$ such that $u<v\model P^e$
by~\eqref{eq:7}. We have $v\in T^e$, hence either $v'=v$ or
$v'=f_e(v)>v$ belongs to~$T^e_1$, i.e., $v'\model P^e\land\neg y$.

\ref{item:4}: If $\cls(u)$ is a maximal cluster of~$S$, then
$\cls(u)$ is not an $E$-cluster, hence there is $v\in\cls(u)$ such
that $v\model y$.

If $\cls(u)$ sees a maximal cluster of~$S$, but is not a maximal
cluster itself, then $u\model y$.

Otherwise, $u\in T$. Fix $e\in E$. As above, there is $v\in T^e$,
$v>u$, and either $v'=v$ or $v'=f_e(v)>v$ is an element of~$T^e\sset S$
satisfying~$y$.
\end{Pf*}
Clearly, $w_i\nmodel\boxdot y\to x_i$; we will show
\[u\model P^{e_0}\land\boxdot\Bigl(y\to\LOR_{e\in E}\Box(P^e\to y)\Bigr)
     \land\ET_{e\in E}\boxdot\bigl(\Box(P^e\to\Box y)\to y\bigr)
   \to\LOR_{i<n}\Box x_i\]
for every $u\in W$. We distinguish several cases:
\begin{itemize}
\item Let $u\in X\Up$. If $u\nless w_i$ for some $i<n$, then
$u\model\Box x_i$. Otherwise $u\sim w_{i_0}$, and $u\nmodel P^{e_0}$
by the choice of~$e_0$.
\item If $u\in S$ does not satisfy~\eqref{eq:7}, let $e\in E$
and~$v\ge u$, $v\in S$ be such that $w\nmodel P^e$ for every $w>v$,
$w\in S$. Then $v\model\Box(P^e\to\Box y)$ as $y$ holds outside~$S$,
but $v\nmodel y$, hence $u\nmodel\boxdot\bigl(\Box(P^e\to\Box y)\to
y\bigr)$.
\item If $\cls(u)$ is an $E$-cluster maximal in~$S$, then
$u\nless w_i$ for some $i<n$, as otherwise $\cls(u)$ would be a
$\p{\R,E}$-tp of~$X$. Consequently, $u\model\Box x_i$.
\item In other cases, \th\ref{cl:vtrwevtr} gives a $v\in S$ such that
$u\le v\model y$. Also, for every~$e\in E$, there is $w\in S$ such
that $v<w\model P^e\land\neg y$, hence $v\model\ET_{e\in
E}\neg\Box(P^e\to y)$, and $u\nmodel\boxdot\bigl(y\to\LOR_{e\in
E}\Box(P^e\to y)\bigr)$.
\end{itemize}
Thus, $\Ext_{\R,n,E,e_0}$ is refuted in~$W$.
\end{Pf}

We cannot adapt the proof of \th\ref{lem:reflextkrip} directly to descriptive frames, as we made many non-definable
choices in the construction of the valuation (even appealing to the axiom of choice). We need to approach the problem in a
different way. Observe that the non-constructiveness in the proof would be substantially alleviated if we were allowed to
use many different variables $y_u$ to handle each $u\in S$ separately. Of course, we cannot afford infinitely many
variables as our rules are finitary, but let us assume that the compactness of descriptive frames will take care
of that, and focus on variants of the $\Ext_{\R,n,E,e_0}$ rules with finitely many copies of the $y$~variable:
\[\fracd{P^{e_0}\land
  \ET_{j<m}\boxdot\Bigl(y_j\to\LOR_{e\in E}\Box(P^e\to y_j)\Bigr)
  \land\ET_{\substack{j<m\\e\in E}}\boxdot\bigl(\Box(P^e\to\Box y_j)\to y_j\bigr)
 \to\LOR_{i<n}\Box x_i}
{\Bigl\{\ET_{j<m}\boxdot y_j\to x_i:i<n\Bigr\}}.\]
Let us denote this rule as $\Ext_{\R,n,E,e_0}^m$.

It is easy to see that $\Ext_{\R,n,E,e_0}^m$ is valid
in all $\p{\R,n,E}$-extensible frames, which suggests that it should follow from $\Ext_{\R,n,E}$, but we cannot infer
this directly from \th\ref{lem:reflextkrip}, as we do not a priori know that
$\lgc{K4}+\Ext_{\R,n,E}$ is Kripke-complete. Instead, we will show that $\Ext_{\R,n,E}$ derives $\Ext_{\R,n,E,e_0}^m$
by means of the following lemma, whose parameter-free special case was already used in~\cite{ej:indep} for a similar
purpose.

\begin{Lem}\th\label{lem:funnyfla}
Let $m\in\omega$, $P\sset\Par$ be finite, and $E\sset\two^P$. Then there
exists a formula $\alpha(P,y_0,\dots,y_{m-1})$ such that $\lgc{K4}$
proves
\[\ET_{j<m}\boxdot y_j\to\alpha,\]
and
\begin{multline*}
\boxdot\Bigl(\alpha\to\LOR_{e\in E}\Box(P^e\to \alpha)\Bigr)
   \land\ET_{e\in E}\boxdot\bigl(\Box(P^e\to\Box \alpha)\to\alpha\bigr)\\
\to
\boxdot\Bigl(y_j\to\LOR_{e\in E}\Box(P^e\to y_j)\Bigr)
     \land\ET_{e\in E}\boxdot\bigl(\Box(P^e\to\Box y_j)\to y_j\bigr)
\end{multline*}
for every $j<m$.
\end{Lem}
\begin{Pf}
If $E=\nul$, we can take $\alpha=\top$, hence we may assume $E\ne\nul$.
Put
\begin{align*}
\beta&=\ET_{j<m}\boxdot y_j,\\
\gamma^e_j&=\boxdot(P^e\to y_j)\lor\boxdot(P^e\land y_j\to\beta),
   \qquad j<m,e\in E,\\
\alpha&=\beta\lor\!\ET_{e\in E}\biggl[
  \boxdot\bigl(\Box(P^e\to\beta)\to\beta\bigr)\land
  \biggl(P^e\to\!\LOR_{j<m}\Bigl[\ET_{i<j}\gamma^e_i\land\neg\gamma^e_j
          \land\bigl(\Box(\gamma^e_j\to\beta)\to y_j
  \bigr)\Bigr]\biggr)\biggr].
\end{align*}
Clearly, $\vdash\beta\to\alpha$.
\begin{Cl}\th\label{cl:1}
$\lgc{K4}$ proves
\begin{enumerate}
\item\label{item:1} $\neg\beta\land\ET_{e\in E}\boxdot\bigl(\Box(P^e\to\beta)\to\beta\bigr)
       \to\ET_{e\in E}\dia(P^e\land\neg\alpha)$,
\item\label{item:2x} $\boxdot\alpha\to\beta$.
\end{enumerate}
\end{Cl}
\begin{Pf*}
\ref{item:1}:
Assume $F\in\Mod_\lgc{K4}$, $u\in F$, and
$u\model\neg\beta\land\ET_e\boxdot\bigl(\Box(P^e\to\beta)\to\beta\bigr)$.
Fix~$e\in E$, and let $v\ge u$ be maximal such that $v\nmodel\beta$. Since
$v\model\Box(P^e\to\beta)\to\beta$, we have
$v\model\dia(P^e\land\neg\beta)$, hence $v$ is reflexive, and
$P^e$ is realized in some $v_e\sim v$. If $\cls(v)\model\gamma^e_j$ for
every $j<m$, then $v_e\nmodel\alpha$. Otherwise let $j$ be minimal such
that $\cls(v)\nmodel\gamma^e_j$. Since $v\nmodel\boxdot(P^e\to y_j)$,
we may assume without loss of generality $v_e\nmodel y_j$. Since
also $\cls(v)\model\Box(\gamma^e_j\to\beta)$, we have
$v_e\nmodel\alpha$. Either way, $u\model\dia(P^e\land\neg\alpha)$.

\ref{item:2x}:
Let $u\model\alpha\land\neg\beta$. Then $u\model\ET_{e\in
E}\boxdot\bigl(\Box(P^e\to\beta)\to\beta\bigr)$, hence
$u\model\dia(P^e\land\neg\alpha)$ for any $e\in E$ by~\ref{item:1},
in particular $u\nmodel\Box\alpha$.
\end{Pf*}
Now, let $u\in F\in\Mod_L$, and assume
\begin{equation}\label{eq:1}
u\model\boxdot\Bigl(\alpha\to\LOR_{e\in E}\Box(P^e\to \alpha)\Bigr)
   \land\ET_{e\in E}\boxdot\bigl(\Box(P^e\to\Box \alpha)\to\alpha\bigr),
\end{equation}
we have to show
\begin{equation}\label{eq:2}
u\model\boxdot\Bigl(y_j\to\LOR_{e\in E}\Box(P^e\to y_j)\Bigr)
     \land\ET_{e\in E}\boxdot\bigl(\Box(P^e\to\Box y_j)\to y_j\bigr)
\end{equation}
for every $j<m$. If $u\model\beta$, then \eqref{eq:2} follows, hence
we may assume $u\nmodel\beta$. If
$v\model\Box(P^e\to\beta)\land\neg\beta$ for some $v\ge u$ and $e\in
E$, then
$v\model\Box(P^e\to\Box\alpha)$. On the other hand, \th\ref{cl:1}
implies $v\nmodel\boxdot\alpha$, contradicting~\eqref{eq:1}. Thus,
\begin{equation}\label{eq:3}
u\model\boxdot\bigl(\Box(P^e\to\beta)\to\beta\bigr)\qquad(e\in E).
\end{equation}
Consequently, if $v\ge u$ is such that $v\model\alpha\land\neg\beta$,
then $v\model\ET_{e\in E}\dia(P^e\land\neg\alpha)$ by \th\ref{cl:1},
which again contradicts~\eqref{eq:1}. Thus,
\begin{equation}\label{eq:4}
u\model\boxdot(\alpha\eq\beta).
\end{equation}
Also, \eqref{eq:3} implies that any $v\ge u$, $v\model\ET_{e\in E}\neg
P^e$, satisfies $\alpha$, hence
\begin{equation}\label{eq:5}
u\model\boxdot\Bigl(\beta\lor\LOR_{e\in E}P^e\Bigr).
\end{equation}
We claim
\begin{equation}\label{eq:6}
u\model\boxdot(P^e\to\gamma^e_j)\qquad(e\in E,j<m).
\end{equation}
Assume for contradiction that $v\ge u$ satisfies
$P^e\land\neg\gamma^e_j$ for some~$j$. W.l.o.g., $v$ is maximal with
this property, and $v\model\ET_{i<j}\gamma^e_i$. In particular,
$v\model\diadot(P^e\land\neg y_j)$, hence $v\nmodel\beta$. We obtain
$v\nmodel\alpha$ by~\eqref{eq:4}, which is only possible if
$v\model\Box(\gamma^e_j\to\beta)\land\neg y_j$. By the maximality
of~$v$ and~\eqref{eq:3}, this implies $w\model\beta$ for every
$w\gnsim v$. Since $v\nmodel\gamma^e_j$, there is $w\ge v$, $w\model
P^e\land\neg\beta\land y_j$. We must have $w\sim v$. But then
$w\model\alpha$, hence $w\model\beta$, a contradiction.

Now we can complete the proof of~\eqref{eq:2}. Let $v\ge u$ be such
that $v\model y_j$. If $v\model\beta$, then $v\model\Box(P^e\to y_j)$
for any~$e\in E$, hence we can assume $v\nmodel\beta$. Then $v\model
P^e$ for some~$e\in E$ by~\eqref{eq:5}, hence $v\model\gamma^e_j$
by~\eqref{eq:6}. Since $v\nmodel P^e\land y_j\to\beta$, we
must have $v\model\boxdot(P^e\to y_j)$.

Let $e\in E$ and $v\ge u$ be such that $v\nmodel y_j$.
By~\eqref{eq:5}, there is $f\in E$ such that $v\model P^f$, and we
have $v\model\gamma^f_j$ by~\eqref{eq:6}, which means
$v\model\boxdot(P^f\land y_j\to\beta)$. By~\eqref{eq:3}, there is
$w>v$ such that $w\model P^e\land\neg\beta$, and $z>w$ such that
$z\model P^f\land\neg\beta$. Then $z\model\neg y_j$, hence
$w\nmodel\Box y_j$, and $v\nmodel\Box(P^e\to\Box y_j)$.
\end{Pf}

\begin{Cor}\th\label{cor:reflextmultiy}
If $n,m\in\omega$, $P\sset\Par$ is finite, and $e_0\in
E\sset\two^P$, then
$\lgc{K4}+\Ext_{\R,n,E,e_0}$ proves the rule $\Ext^m_{\R,n,E,e_0}$.
\noproof\end{Cor}

\begin{Lem}\th\label{lem:reflextdesc}
Let $P\sset\Par$ be finite, $n\in\omega$, and $E\sset\two^P$,
$E\ne\nul$. If $W$ is a descriptive frame such that
$W\model\Ext_{\R,n,E}$, then $W$ is $\p{\R,n,E}$-extensible.
\end{Lem}
\begin{Pf}
Let $A$ be the algebra of admissible sets of~$W$. Let
$X=\{w_i:i<n\}\sset W$, we have to find a $\p{\R,E}$-tp of~$X$.

Assume there is $i_0<n$ such that $w_{i_0}<w_i$ for every~$i$. If
$\cls(w_{i_0})$ realizes every~$e\in E$, then there is a $\p{\R,E}$-tp
of~$X$ included in~$\cls(w_{i_0})$. Otherwise, we can fix $e_0\in E$
not realized in~$\cls(w_{i_0})$. If there is no such~$i_0$,
we let $e_0\in E$ be arbitrary.

As in \th\ref{thm:irrextchar}, we identify $P^{e_0}$ with the set of
points of~$W$ where it is satisfied. We also use connectives to denote
the corresponding operations on sets from~$A$. By \th\ref{cor:reflextmultiy},
$W\model\Ext^m_{\R,n,E,e_0}$ for every $m\in\omega$, hence the set
\begin{multline*}
U=\{P^{e_0}\}\cup\{\dia B:B\in A,B\cap X\ne\nul\}\\\cup
\Bigl\{\boxdot\Bigl(C\to\LOR_{e\in E}\Box(P^e\to C)\Bigr)
  \land\ET_{e\in E}\boxdot\bigl(\Box(P^e\to\Box C)\to C\bigr):
  C\in A,C\Sset X\Up\Bigr\}
\end{multline*}
has fip, and there is $u_{e_0}\in\bigcap U$ as $W$ is compact. We have
$u_{e_0}\model P^{e_0}$ and $u_{e_0}<w_i$ for every~$i<n$. The choice
of~$e_0$ ensures that $u_{e_0}\notin X\Up$.
\pagebreak[2]
\begin{Cl}\th\label{cl:2}
Let $u_{e_0}\le u\notin X\Up$.
\begin{enumerate}
\item\label{item:5}
For every~$e\in E$, there is
$u<v<u$ such that $v\model P^e$. In particular, $u$ is reflexive.
\item\label{item:6}
Putting $e=\sat_P(u)$, we have $e\in E$, and there is no~$u\le v\notin
X\Up$ such that $v\model P^e$, other than $u$ itself.
\end{enumerate}
\end{Cl}
\begin{Pf*}
Since $W$ is refined, for every~$i<n$ there exists $C_i\in A$ such that
$w_i\in\boxdot C_i$, and $u\notin C_i$. If we put
$C=\bigcup_{i<n}C_i$, we have $C\Sset X\Up$, and $u\notin C$.

\ref{item:5}: Using the definition of~$U$,
$u\in\Box(P^e\to\Box(C\lor\neg D))\to\neg D$ for every~$D\in A$. Thus,
the set
\[\{P^e\}\cup\{B\in A:u\in\Box B\}\cup\{\dia D:D\in A,u\in D\}\]
has fip, and consequently its intersection contains an element~$v$.
Clearly, $v\model P^e$, and the refinedness of~$W$ implies $u<v<u$.

\ref{item:6}: Since $u\in P^e\lor\Box C\Sset X\Up$, the definition
of~$U$ implies that there is $e'\in E$ such that
$u\in\Box\bigl(P^{e'}\to P^e\lor\Box C\bigr)$. If $e'\ne e$, this
would in fact mean $u\in\Box(P^{e'}\to\Box C)$,
contradicting~\ref{item:5}. Thus, $e'=e$, which implies $e\in E$.

Assume $u<v\notin X\Up$, $v\model P^e$. By reducing~$C$ if
necessary, we may assume $v\notin C$. For every $D\in A$ such that
$u\in D$, there is $e'\in E$ such that
$u\in\Box\bigl(P^{e'}\to(P^e\land D)\lor\boxdot C\bigr)$. As above, we
must have $e'=e$, hence $u\in\Box(P^e\to D\lor C)$, and $v\in D$. As
$D$ was arbitrary, and $W$ is refined, we obtain~$v=u$.
\end{Pf*}
Part \ref{item:5} of \th\ref{cl:2} implies that $\cls(u_{e_0})$ is
reflexive, and for every $e\in E$, there is $u_e\in\cls(u_{e_0})$ such
that $u_e\model P^e$. By \ref{item:6}, this $u_e$ is a unique point in
$u_{e_0}\Up\bez X\Up$ satisfying~$P^e$, and every point of
$u_{e_0}\Up\bez X\Up$ satisfies some~$P^e$, $e\in E$, hence it
equals~$u_e$. Thus, $u_e\up=\{u_{e'}:e'\in E\}\cup X\Up$, and
$\{u_e:e\in E\}$ is a $\p{\R,E}$-tp of~$X$.
\end{Pf}

This completes the proof of \th\ref{thm:reflextchar}.
\begin{Rem}\th\label{rem:RnEe}
The proof of \th\ref{lem:reflextdesc} shows a bit more: if $n$, $P$,
and $E$ are as in the lemma, $e\in E$,
$W\model\Ext_{\R,n,E,e}$ is descriptive, and $X=\{w_i:i<n\}\sset W$
either does not have a reflexive root, or its root cluster
avoids~$P^e$, then $X$ has a $\p{\R,E}$-tp in~$W$.
\end{Rem}
\begin{Rem}\th\label{rem:infext}
Exploiting compactness, one can show that descriptive frames
validating certain extension rules are also extensible wrt
infinite subsets.

First, if $W$ is a descriptive frame such that
$W\model\Ext_{*,\infty,E}:=\{\Ext_{*,n,E}:0<n\in\omega\}$ (where
$*\in\{\I,\R\}$, and $\lh E=1$ if $*=\I$), and if $\nul\ne
X\sset W$ is \emph{closed,} then $X$ has a $\p{*,E}$-tp
in~$W$.

Second, if $P$ is infinite, $n\in\omega$, $e\in\two^P$,
and $W$ validates
\[\Ext_{\I,n,\{e\}}:=\{\Ext_{\I,n,\{e\res P'\}}:P'\sset P\text{ finite}\},\]
then every $X=\{w_i:i<n\}\sset W$ has a $\p{\I,\{e\}}$-tp. The
reflexive case is slightly more complicated: if $E\sset\two^P$ is
closed (in the product topology on~$\two^P$), $E_0$ is a set
of isolated points of~$E$, and $W$ satisfies appropriate instances of
the extension rules, then any $X$ as above has a tp cluster consisting
of one point realizing~$e$ for each~$e\in E_0$, and
one or more points realizing~$e$ for each~$e\in E\bez E_0$. This can
be generalized to closed infinite~$X$ as above. We leave the details to
the interested reader, as we have no further use for these properties.
\end{Rem}

We are now going to show that the admissibility of $\Ext_{C,n}^\Par$
in a logic~$L$ is equivalent to $\p{C,n}$-extensibility of~$L$. The basic idea
is that $\Ext_{C,n}^\Par$ is admissible iff it holds in canonical
frames~$C_L(P,V)$ by \th\ref{lem:admcanon}, which is equivalent to
extensibility of~$C_L(P,V)$ by
\th\ref{thm:irrextchar,thm:reflextchar}. Since every finite frame can
be embedded in a canonical frame, the existence of tight predecessors
in~$C_L(P,V)$ is equivalent to extensibility of the logic. (This is
not quite true as not all rooted subframes of~$C_L(P,V)$ are finite,
but one can make it work anyway.)

Recall \th\ref{def:extcond}. If $D\preceq C$, then $\Ext^\Par_{D,n}\sset\Ext^\Par_{C,n}$ by
definition. Also, since we can identify some of the variables~$x_i$ in
extension rules by a substitution, we have:
\begin{Obs}\th\label{obs:order}
If $\p{D,m}\preceq\p{C,n}$, then $\lgc{K4}+\Ext^\Par_{C,n}$ proves
$\Ext^\Par_{D,m}$, $\lgc{K4}+\Ext_{*,n,E}$ proves $\Ext_{*,m,E}$, and
$\lgc{K4}+\Ext_{\R,n,E,e}$ proves $\Ext_{\R,m,E,e}$.
\noproof\end{Obs}

\begin{Thm}\th\label{thm:admext}
Let $L\Sset\lgc{K4}$ have fmp, $\p{C,n}\in\EC$, and $e\in
E\sset\two^P$ for some finite $P\sset\Par$, where $\lh E=\lh C$ (note
that if $\Par$ is finite, this is only possible when $\lh
C\le2^{\lh\Par}$). Then the following are equivalent.
\begin{enumerate}
\item\label{item:7}
$L$ is $\p{C,n}$-extensible.
\item\label{item:8}
$L$ admits $\Ext_{\I,n,E}$ if $C=\I$, and $\Ext_{\R,n,E,e}$ if $C$ is
reflexive.
\item\label{item:9}
$L$ admits $\Ext^\Par_{D,m}$ for every $\p{D,m}\preceq\p{C,n}$.
\end{enumerate}
\end{Thm}
\begin{Pf}
\ref{item:9}${}\to{}$\ref{item:8} is trivial.

\ref{item:8}${}\to{}$\ref{item:7}:
Let $F$ be a finite rooted frame of type $\p{C,n}$ such that
$F\bez\rcl(F)$ is an $L$-frame, and if $\Par=\nul$ and $n=1$, then $F$
does not have a reflexive root. We can endow~$F$ with a valuation
of~$P$ such that if $F\bez\rcl(F)$ has a reflexive root~$r$, then
$\cls(r)\model\neg P^e$. By \th\ref{lem:descrcanon},
there is a finite set~$V$ of variables
such that $F\bez\rcl(F)$ can be identified with a generated subframe
of the canonical frame~$C_L(P,V)$, including the valuation of~$P$.
Choose $X=\{w_i:i<n\}\sset F$ 
such that $F\bez\rcl(F)=X\Up$. We have $C_L(P,V)\model\Ext_{\I,n,E}$
($\Ext_{\R,n,E,e}$, respectively) by \th\ref{lem:admcanon}, hence $X$ has
a $\p{*,E}$-tp $U$ by
\th\ref{thm:irrextchar,rem:RnEe}. The choice of the valuation of~$P$
in $F\bez\rcl(F)$ ensures that $U$ is disjoint from~$F\bez\rcl(F)$,
hence $F$ is isomorphic to the generated subframe
$U\cup(F\bez\rcl(F))$ of~$C_L(P,V)$ (minus its valuation), and as such
it is an $L$-frame.

\ref{item:7}${}\to{}$\ref{item:9}:
In view of \th\ref{obs:order}, we may assume $m=n$. Let $P'\sset\Par$
be finite, $e'\in E'\sset\two^{P'}$, $\lh{E'}\le\lh C$. Let $\sigma$
be a substitution
such that $\nvdash_L\sigma(\boxdot y\to x_i)$ for every~$i<n$. Since
$L$ has fmp, we 
can find $F_i\in\Mod_L$ with root $w_i$ such that $\sigma(F_i)\model
y$ and 
$\sigma(F_i),w_i\nmodel x_i$. If $\Par=\nul$, $n=1$, and $w_0$ is
reflexive, 
we put $F=F_0$ and $u_{e'}=w_0$. Otherwise,
let $F'$ be the disjoint union of $\{F_i:i<n\}$
extended by a new root cluster of type~$C$, and $F$ a similar model
where the root cluster is shrunk to size~$\lh{E'}$. $F'$ is an
$L$-frame as $L$ is $\p{C,n}$-extensible, and $F$ is its p-morphic
image. We enumerate elements of~$\rcl(F)$ as $\{u_f:f\in E'\}$, and we
define $F,u_f\model P^f$; the valuation of variables in~$u_f$ is
arbitrary.

Either way, $\{u_f:f\in E'\}$ is a $\p{*,E'}$-tp of~$\{w_i:i<n\}$, hence
$\sigma(F),u_{e'}$ refutes the premise of~$\sigma(\Ext_{\I,n,e'})$
($\sigma(\Ext_{\R,n,E',e'})$, resp.) by the proof of \th\ref{thm:irrextchar}
(\ref{thm:reflextchar}).
\end{Pf}

\begin{Def}\th\label{def:locfin}
A frame $W$ is \emph{locally finite} if $u\Up$ is finite for every
$u\in W$.
\end{Def}

\begin{Thm}\th\label{thm:extlocfin}
Let $T$ be a set of conditions of the form $\p{*,n,E}$, where
$*\in\{\I,\R\}$, $n\in\omega$, $\nul\ne E\sset\two^P$ for a finite
$P\sset\Par$ (not necessarily the same for each~$t\in T$), and if
$*=\I$, $\lh E=1$. Assume that
$L\Sset\lgc{K4}$ has fmp, and is $\p{\I,n}$-extensible ($\p{\nr k,n}$-extensible) whenever $\p{\I,n,E}\in T$ ($\p{\R,n,E}\in T$ with
$k=\lh E$, respectively). The following are equivalent for any
rule~$\Gamma\ru\Delta$.
\begin{enumerate}
\item\label{item:10}
$L+\{\Ext_t:t\in T\}$ proves $\Gamma\ru\Delta$.
\item\label{item:11}
$\Gamma\ru\Delta$ holds in every parametric general $L$-frame that is
$t$-extensible for every~$t\in T$.
\item\label{item:12}
$\Gamma\ru\Delta$ holds in every parametric countable locally finite Kripke
$L$-frame that is $t$-extensible for every~$t\in T$.
\end{enumerate}
\end{Thm}
\begin{Pf}
\ref{item:11}${}\to{}$\ref{item:12} is trivial, and
\ref{item:10}${}\eq{}$\ref{item:11}
follows from \th\ref{thm:irrextchar,thm:reflextchar}, as any
parametric consequence relation is complete with respect to parametric
descriptive frames.

\ref{item:12}${}\to{}$\ref{item:11}:
Let $W$ be an $L$-frame $t$-extensible for every~$t\in T$, and let
$\model$ be a valuation on~$W$ such that $W\nmodel\Gamma\ru\Delta$.
Let $\Sigma$ be the set of subformulas of~$\Gamma$, and put
\[\psi=\LOR\bigl\{\Sigma^{\sat_\Sigma(W,w)}:w\in W\bigr\}.\]
We have $\vdash_\lgc{K4}\psi\to\fii$ for every $\fii\in\Gamma$, hence
it suffices to find a model of the requested form refuting
$\psi\ru\Delta$.

For every $\fii\in\Delta$, $W$ witnesses that $\psi\nvdash_L\fii$.
Since $L$ has fmp, we can find a finite $L$-model whose root refutes
$\boxdot\psi\to\fii$, and by taking a disjoint union of these, we
obtain a finite $L$-model~$F_0$ such that $F_0\model\psi$, and
$F_0\nmodel\fii$ for every~$\fii\in\Delta$. We will construct a
sequence $F_0\sset F_1\sset F_2\sset\cdots$ of finite $L$-models such
that $F_k\model\psi$, $F_k$ is a generated submodel of~$F_{k+1}$, and
$F:=\bigcup_kF_k$ is $t$-extensible for every $t\in T$. Then $F$ is an
$L$-frame, and $F\nmodel\psi\ru\Delta$, hence completing the proof.

We may assume that $F_0$ is included in a countable set~$Z$,
and we will choose all the models~$F_k$ so that their underlying set
is also included in~$Z$. Let $\{\p{t_k,X_k}:k\in\omega\}$ be an
enumeration of all pairs of $t_k=\p{*_k,n_k,E_k}\in T$ and
$X_k=\{z_{k,0},\dots,z_{k,n_k-1}\}\sset Z$, where each pair occurs
infinitely many times in the enumeration.

Starting with~$F_0$, we define the models~$F_k\model\psi$ by induction
on~$k$. Assume that $F_k$ has already been defined. If $X_k\nsset
F_k$, or $X_k$ has a $\p{*_k,E_k}$-tp in~$F_k$, we put $F_{k+1}:=F_k$.
Otherwise, we have $F_k\model\psi$ by the induction hypothesis, hence
for every~$i<n_k$, we can find $w_i\in W$ such that
$F_k,z_{k,i}\equiv_\Sigma W,w_i$. Since $W$ is $t_k$-extensible, we
can find a $\p{*_k,E_k}$-tp $\{u_e:e\in E_k\}\sset W$
of~$\{w_i:i<n_k\}$. Choose distinct elements $\{v_e:e\in E_k\}\sset
Z\bez F_k$, and put $F_{k+1}=F_k\cup\{v_e:e\in E_k\}$, where the
accessibility relation and valuation of parameters is defined so that
$\{v_e:e\in E_k\}$ is a $\p{*_k,E_k}$-tp of~$X_k$, and the valuation
of variables in~$v_e$ is the same as in~$u_e$. By
\th\ref{lem:tpsigma}, we have $v_e\equiv_\Sigma u_e$, hence
$F_{k+1}\model\psi$. Also, $F_{k+1}$ is based on an $L$-frame: it
suffices to show this for the rooted subframe generated by the new
elements, which is indeed an $L$-frame by the extensibility
assumptions on~$L$ (note that the exceptional case when $\Par=\nul$,
$n_k=1$ cannot happen: if $z_{k,0}$ is reflexive, it already has the
requisite tight predecessor in~$F_k$, namely itself).
\end{Pf}

A more constructive proof of \th\ref{thm:extlocfin} will be given in
the course of proving \th\ref{thm:pext}.

We can now give the main result of this section. The only part left to
prove is that a failure of a rule in an extensible frame implies its
nonadmissibility; we do this by ``approximating'' the frame by a set of
finite models with the model extension property, and using the
characterization from Section~\ref{sec:projective-formulas} to
extract a projective formula whose mgu witnesses nonadmissibility of
the rule.
\begin{Thm}\th\label{thm:clxadmchar}
Let $L$ be a $\Par$-extensible logic, and
\[T=\bigl\{\p{C,n}:L\text{ has a type-$\p{C,n}$ frame, and }\lh
  C\le2^{\lh\Par}\bigr\}.\]
Then the following are equivalent for any rule~$\Gamma\ru\Delta$.
\begin{enumerate}
\item\label{item:13}
$\Gamma\adm_L\Delta$.
\item\label{item:14}
$\Gamma\ru\Delta$ holds in every parametric $L$-frame, $\p{C,n}$-extensible for
every $\p{C,n}\in T$.
\item\label{item:15}
$\Gamma\ru\Delta$ is derivable in $L+\{\Ext^\Par_{C,n}:\p{C,n}\in T\}$.
\end{enumerate}
Moreover, it suffices to consider only countable, locally finite
Kripke frames in~\ref{item:14}.

In particular, $\{\Ext^\Par_{C,n}:\p{C,n}\in T\}$ is a basis of
$L$-admissible rules.
\end{Thm}
\begin{Pf}
\ref{item:14}${}\to{}$\ref{item:15}${}\to{}$\ref{item:13}
follow from \th\ref{thm:extlocfin,thm:admext}, respectively.

\ref{item:13}${}\to{}$\ref{item:14}:
Let $W$ be a parametric $L$-frame, $\p{C,n}$-extensible
for every $\p{C,n}\in T$, and fix an admissible valuation
in~$W$ that refutes~$\Gamma\ru\Delta$. Let $\Sigma$ be the set of all
subformulas of formulas occurring in~$\Gamma$, and define
\[\psi=\LOR\bigl\{\Sigma^{\sat_\Sigma(W,v)}:v\in W\}.\]
Clearly, $\psi\vdash_L\fii$ for every $\fii\in\Gamma$, and
$\psi\nvdash_L\fii$ for every $\fii\in\Delta$, as $W\model\psi$. The
same argument as in the proof of
\th\ref{thm:clxprojapx} shows that
\[\Mod_L(\psi)=\{F\in\Mod_L:\forall u\in F\,\exists v\in
W\,(F,u\equiv_\Sigma W,v)\}\]
has the model extension property, hence $\psi$ is projective by
\th\ref{thm:proj}. Thus, if $\sigma$ is the projective unifier
of~$\psi$, we have $\vdash_L\sigma(\fii)$ for every~$\fii\in\Gamma$,
but $\nvdash_L\sigma(\fii)$ for every~$\fii\in\Delta$, which implies
$\Gamma\nadm_L\Delta$.
\end{Pf}
As a sort of converse to \th\ref{thm:clxadmchar}, one can show that if
$\Par$ is infinite, and a logic $L\Sset\lgc{K4}$ has a basis of
admissible rules consisting of 
a set of extension rules, then $L$ is a clx logic.

We will also give a characterization of consequences of extension
rules using finite models, which will be helpful in the sequel for
determination of the computational complexity of admissibility in clx
logics. The characterization is similar in spirit to criteria for
admissibility in various modal logics presented by
Rybakov~\cite{ryb:s4int,ryb:s4con,ryb:provlog,ryb:grz}, and generalized in \cite[\S6.1]{ryb:bk}.

Our assumptions are somewhat different:
on the one hand, clx logics have the generalized property of branching
below~$1$ in Rybakov's terminology, on the other hand, we do not need
to assume
any analogue of the effective $m$-drop point property. (In fact, one
can use the proof of \th\ref{lem:canfmp} to show that clx logics
satisfy this property automatically. We suspect that the property
actually holds for \emph{all} logics with the generalized property
of branching below~$m$.) We also obtain better bounds: the models we
construct in \th\ref{thm:clxadm} have size exponential in the size of
the rule, whereas the bounds in~\cite{ryb:bk} are at least doubly
exponential.

In order to keep the notation manageable, we will only state the
result for combinations of~$\Ext^\Par_{C,n}$, rather than individual
$\Ext_{*,n,E}$ rules; this is of course enough for the application to
admissibility.

\begin{Def}\th\label{def:extruleinfty}
We generalize the $\Ext^\Par_{C,n}$ notation to infinite extension
conditions (\th\ref{def:extcond}) by putting
$\Ext^\Par_{\nr\infty,n}:=\bigcup_{k=1}^\infty\Ext^\Par_{\nr k,n}$, and
$\Ext^\Par_{C,\infty}:=\bigcup_{n=1}^\infty\Ext^\Par_{C,n}$. We generalize in a
similar way the notion of $\p{C,n}$-extensible frames.

Moreover, if $T\sset\ECI$, we put $\Ext^\Par_T:=\bigcup_{t\in 
T}\Ext^\Par_t$, and a frame is $T$-extensible if it is
$t$-extensible for every~$t\in T$.
\end{Def}

Recall that every set of extension conditions is equivalent to a
finite one by \th\ref{lem:extcond}.
\begin{Obs}
If $T,T'$ are equivalent sets of extension conditions, then
$\lgc{K4}+\Ext^\Par_T=\lgc{K4}+\Ext^\Par_{T'}$, and a frame is
$T$-extensible iff it is $T'$-extensible.
\noproof\end{Obs}

\begin{Def}\th\label{def:tpp}
Let $\Sigma$ be a finite set of formulas closed under subformulas,
$P=\Sigma\cap\Par$, and $F$ be a model.

If $X\sset F$, and $e\in\two^P$, then a \emph{tight
$\p{\I,\{e\}}$-pseudopredecessor ($\p{\I,\{e\}}$-tpp) of~$X$ wrt~$\Sigma$}
is $\{u\}\sset F$ such that $u\model P^e$, and for every
$\Box\psi\in\Sigma$,
\[u\model\Box\psi\iff w\model\boxdot\psi\text{ for every }w\in X.\]
If $n\in\omega\cup\{\infty\}$, then $F$ is
\emph{$\p{\I,n}$-pseudoextensible wrt~$\Sigma$,} if every finite $X\sset F$ such that $\lh
X\le_0n$ has a $\p{\I,\{e\}}$-tpp wrt~$\Sigma$ for every~$e\in\two^P$.

If $X\sset F$, and $\nul\ne E\sset\two^P$, a \emph{tight
$\p{\R,E}$-pseudopredecessor ($\p{\R,E}$-tpp) of~$X$ wrt~$\Sigma$}
is $\{u_e:e\in E\}\sset F$ such that for every~$e\in
E$ and $\Box\psi\in\Sigma$, we have $u_e\model P^e$, and
\begin{equation}\label{eq:14}
u_e\model\Box\psi\iff w\model\boxdot\psi\text{ for every }w\in X
  \text{ and }u_f\model\psi\text{ for every }f\in E.
\end{equation}
If $n,k\in\omega\cup\{\infty\}$, $k\ne0$, then $F$ is
\emph{$\p{\nr k,n}$-pseudoextensible wrt~$\Sigma$,} if every finite $X\sset F$ such that
$\lh X\le_0n$ has a $\p{\R,E}$-tpp wrt~$\Sigma$ for every~$E\sset\two^P$ such that
$\lh E\le_0k$.

If $T$ is a set of extension conditions, $F$ is
\emph{$T$-pseudoextensible wrt~$\Sigma$} if it is $t$-pseudoextensible
for every~$t\in T$.

Note that every $\p{*,E}$-tp is also a $\p{*,E}$-tpp wrt~$\Sigma$, and a
$T$-extensible frame is $T$-pseudoextensible wrt~$\Sigma$. Essentially,
$\p{*,E}$-tpp's wrt~$\Sigma$ are sets of points that behave as if they
were $\p{*,E}$-tp's as far as formulas from~$\Sigma$ are concerned.

Let $B=\{\fii:\Box\fii\in\Sigma\}$, and let
$\pext_T^\Sigma$ consist of the following rules:
\begin{itemize}
\item If $\p{\I,\infty}\in T$, rules of the form
\begin{equation}\label{eq:11}
P^e\land\ET_{\fii\in B_+}\Box\fii\to\LOR_{\psi\in B_-}\Box\psi
 \Ru\Bigl\{\ET_{\fii\in B_+}\boxdot\fii\to\psi:\psi\in B_-\Bigr\},
\end{equation}
where $e\in\two^P$, $B=B_+\cupd B_-$, $B_-\ne\nul$ (the case
of~$B_-=\nul$ is actually the rule for~$\p{\I,0}$ below).
\item If $\p{\I,n}\in T$, $n\in\omega$, rules of the form
\begin{equation}\label{eq:12}
P^e\land\ET_{\fii\in B_+}\Box\fii\to\LOR_{\psi\in B_-}\Box\psi
 \Ru\Bigl\{\ET_{\fii\in B_+}\boxdot\fii\to\LOR_{\psi\in B_i}\boxdot\psi:i<n\Bigr\},
\end{equation}
where $e\in\two^P$, $B=B_+\cupd B_-$, $B_-=\bigcup_{i<n}B_i$.
\item If $\p{\nr k,\infty}\in T$, rules of the form
\begin{equation}
\fracd{\Bigl\{P^{S(f)}\land\!\!\!\ET_{\fii\in B_+\bez D}\!\!\!\boxdot\fii\to
         \mkern-20mu\LOR_{\substack{\psi\in D\\f(\psi)=S(f)}}\mkern-20mu\psi
         \lor\!\!\!\LOR_{\psi\in B_-\cup D}\!\!\!\Box\psi:
       D\sset B_+,f\colon D\to E\Bigr\}}
{\Bigl\{\ET_{\fii\in B_+}\boxdot\fii\to\psi:\psi\in B_-\Bigr\}},
\end{equation}
where $E\sset\two^P$, $\lh E\le_0k$, $B=B_+\cupd B_-$, $B_-\ne\nul$,
$S\colon\bigcupd_{D\sset B_+}E^D\to E$. (Here, $E^D$ denotes the set
of all functions $f\colon D\to E$.)
\item If $\p{\nr k,n}\in T$, $n\in\omega$, rules of the form
\begin{equation}\label{eq:13}
\fracd{\Bigl\{P^{S(f)}\land\!\!\!\ET_{\fii\in B_+\bez D}\!\!\!\boxdot\fii\to
         \mkern-20mu\LOR_{\substack{\psi\in D\\f(\psi)=S(f)}}\mkern-20mu\psi
         \lor\!\!\!\LOR_{\psi\in B_-\cup D}\!\!\!\Box\psi:
       D\sset B_+,f\colon D\to E\Bigr\}}
{\Bigl\{\ET_{\fii\in B_+}\boxdot\fii\to\LOR_{\psi\in B_i}\boxdot\psi:i<n\Bigr\}},
\end{equation}
where $E\sset\two^P$, $\lh E\le_0k$, $B=B_+\cupd B_-$, $B_-=\bigcup_{i<n}B_i$,
$S\colon\bigcupd_{D\sset B_+}E^D\to E$.
\end{itemize}
Notice that $\pext_T^\Sigma$ is finite, and all formulas occurring
in~$\pext_T^\Sigma$ are Boolean combinations of $\Sigma$-formulas. The
reader should think of~$\pext_T^\Sigma$ as rule \emph{instances}
rather than rule schemata, as we will use them in a context where they
do not get closed under substitution. In fact, the gist of
\th\ref{thm:pext} below is that $\pext_T^\Sigma$ axiomatizes the consequences
of~$\Ext_T^\Par$ involving only (Boolean combinations of)
$\Sigma$-formulas. But first we need a
semantic characterization of~$\pext_T^\Sigma$:
\end{Def}

\begin{Lem}\th\label{lem:pext}
Let $\Sigma$ be a finite set of formulas closed under subformulas, $T$
a set of extension conditions, and $F$ a model. Then $F$ is
$T$-pseudoextensible wrt~$\Sigma$ iff $F\model\pext_T^\Sigma$.

In particular, $\pext_T^\Sigma$ is provable in $\lgc{K4}+\Ext^\Par_T$.
\end{Lem}
\begin{Pf}
We will show the lemma for the most complicated case of $t=\p{\nr
k,n}\in T$, $n\in\omega$, the other cases are similar and left to the
reader.

First, assume that $F$ is $t$-pseudoextensible, $E,B_+,B_-,B_i,S$ are
as in~\eqref{eq:13}, and $w_i\in F$, $i<n$, witness that the
conclusion of~\eqref{eq:13} fails, i.e., $w_i\model\ET_{\fii\in
B_+}\boxdot\fii\land\ET_{\psi\in B_i}\neg\boxdot\psi$. Let $\{u_e:e\in
E\}$ be a $\p{\R,E}$-tpp of $\{w_i:i<n\}$ wrt~$\Sigma$. Put
$D=\{\psi\in B_+:u_e\nmodel\Box\psi\}$, where~$e\in E$. By~\eqref{eq:14}, this
definition is independent of~$e$, and for every~$\psi\in D$, there
exists $f(\psi)\in E$ such that $u_{f(\psi)}\nmodel\psi$. This defines
a function $f\colon D\to E$. Putting
$e=S(f)$, inspection shows that $u_e$ refutes the premise
of~\eqref{eq:13} corresponding to $D$ and~$f$.

Conversely, let
$X=\{w_i:i<n\}\sset F$ and $E\sset\two^P$ be such that $\lh E\le_0k$,
and $X$ has no $\p{\R,E}$-tpp wrt~$\Sigma$.
Put $B_i=\{\psi\in B:w_i\nmodel\boxdot\psi\}$, $B_-=\bigcup_{i<n}B_i$,
$B_+=B\bez B_-$.

Let $D\sset B_+$ and $f\colon D\to E$. If there were $\{u_e:e\in
E\}\sset F$ such that
\[u_e\model P^e\land\ET_{\fii\in B_+\bez D}\boxdot\fii
  \land\ET_{\psi\in B_-\cup D}\neg\Box\psi
  \land\ET_{f(\psi)=e}\neg\psi,\]
then $\{u_e:e\in E\}$ would be a $\p{\R,E}$-tpp of~$X$, with
$\{\psi\in B_+:u_e\nmodel\Box\psi\}=D$. Thus,
there must exist $S(f):=e\in E$ such that
\begin{equation}\label{eq:15}
F\model P^e\land\ET_{\fii\in B_+\bez D}\boxdot\fii
  \to\LOR_{\psi\in B_-\cup D}\Box\psi
  \lor\LOR_{f(\psi)=e}\psi.
\end{equation}
This defines a function
$S\colon\bigcupd_{D\sset B_+}E^D\to E$ for which all premises
of~\eqref{eq:13} hold in~$F$ by~\eqref{eq:15}. However, the
definition of $B_+$ and~$B_i$ ensures that the $i$th conclusion
of~\eqref{eq:13} is false in~$w_i$, hence \eqref{eq:13} does not hold
in~$F$.
\end{Pf}

\begin{Thm}\th\label{thm:pext}
Let $L\Sset\lgc{K4}$ have fmp, $T$ be a set of extension
conditions such that $L$ is $T$-extensible, and $\Sigma$ a finite set
of formulas closed under subformulas. The following are equivalent for
any rule $\Gamma\ru\Delta$ such that $\Gamma\sset\Sigma$.
\begin{enumerate}
\item\label{item:21} $L+\Ext^\Par_T$ proves $\Gamma\ru\Delta$.
\item\label{item:22} $\pext_T^\Sigma\vdash_L\Gamma\ru\Delta$.
\item\label{item:23} $\Gamma\ru\Delta$ holds in every finite
$L$-model, $T$-pseudoextensible wrt~$\Sigma$.
\end{enumerate}
\end{Thm}
\begin{Pf}
\ref{item:23}${}\to{}$\ref{item:22}${}\to{}$\ref{item:21}
follows from \th\ref{lem:pext,lem:fmprules}.

\ref{item:21}${}\to{}$\ref{item:23}:
Let $F_0$ be a finite $T$-pseudoextensible $L$-model such that
$F_0\nmodel\Gamma\ru\Delta$, we will find a locally finite
$T$-extensible $L$-model $F\nmodel\Gamma\ru\Delta$.

Similarly to the proof of \th\ref{thm:extlocfin}, we will construct a
sequence of finite
$L$-models $F_0\sset F_1\sset F_2\sset\cdots$ such that $F_k$ is a
generated submodel of~$F_{k+1}$, while maintaining the property
\begin{equation}\label{eq:16}
\forall v\in F_k\,\exists u\in F_0\,u\equiv_\Sigma v.
\end{equation}
As in \th\ref{thm:extlocfin}, we assume that $F_0$ is included in a
countable set $Z$, and we will define $F_k$ so that their underlying
sets are also included in~$Z$. Let $\{\p{t_k,X_k}:k\in\omega\}$ be an
enumeration of all pairs of $t_k=\p{*_k,n_k,E_k}$ and
$X_k\sset Z$, where $\lh{X_k}=n_k$, $E_k\sset\two^{P_k}$
for some finite $P_k\sset\Par$, $P_k\Sset P:=\Sigma\cap\Par$,
$\lh{E_k}=1$ if $*_k=\I$,
$\p{C_k,n_k}\preceq\p{C,n}$ for some $\p{C,n}\in T$, where $C_k=\I$ if
$*_k=\I$, and $C_k=\nr m$ if $*_k=\R$ and $\lh{E_k}=m$,
and each pair occurs infinitely many times in the enumeration.

Assuming $F_k$ is already defined, let $F_{k+1}=F_k$ if $X_k\nsset
F_k$, or if $X_k$ has a $\p{*_k,E_k}$-tp in~$F_k$. Otherwise, fix
$\p{C,n}\in T$ such that $\p{C_k,n_k}\preceq\p{C,n}$, and let
$E=\{e\res P:e\in E_k\}$. Write $X_k=\{z_i:i<n_k\}$, and for
every~$i<n_k$, find $w_i\in F_0$ such that $z_i\equiv_\Sigma w_i$ by
the induction hypothesis. Since $F_0$ is $\p{C,n}$-pseudoextensible,
there exists a $\p{*_k,E}$-tpp $\{u_e:e\in E\}\sset F_0$
of~$\{w_i:i<n\}$ wrt~$\Sigma$. We choose distinct elements $\{v_e:e\in
E_k\}\sset Z\bez F_k$, and define $F_{k+1}=F_k\cup\{v_e:e\in E_k\}$,
where the accessibility relation and valuation of parameters in~$v_e$
is defined so that $\{v_e:e\in E_k\}$ is a $\p{*_k,E_k}$-tp of~$X_k$,
and valuation of variables is defined by
\[F_{k+1},v_e\model x\EQ F_0,u_{e\res P}\model x.\]
By induction on the complexity of~$\psi$, we can prove $v_e\model\psi$
iff $u_{e\res P}\model\psi$ for every $\psi\in\Sigma$ as in
\th\ref{lem:tpsigma}, which shows that \eqref{eq:16} holds
for~$F_{k+1}$. Moreover, $F_{k+1}$ is based on an $L$-frame, as $L$ is
$\p{C,n}$-extensible, and therefore $\p{C_k,n_k}$-extensible.

When the construction is completed, we put $F=\bigcup_kF_k$. Then $F$
is a locally finite model based on an $L$-frame, it is $T$-extensible
by construction, and \eqref{eq:16} implies $F\model\Gamma$. On the
other hand, $F_0$ is a generated submodel of~$F$, hence $F\nmodel\psi$
for every $\psi\in\Delta$, thus $F\nmodel\Gamma\ru\Delta$.
\end{Pf}

In view of \th\ref{thm:clxadmchar}, \th\ref{thm:pext} provides a
description of admissible rules in $\Par$-extensible logics.
We will state it explicitly for cluster-extensible logics, as we can
give explicit bounds in this case. Recall \th\ref{def:clxchar}.
\begin{Thm}\th\label{thm:clxadm}
Let $L$ be a clx logic, and $\Gamma\ru\Delta$ a rule. The following
are equivalent.
\begin{enumerate}
\item\label{item:38} $\Gamma\adm_L\Delta$.
\item\label{item:40} $L+\Ext^\Par_{\base(L)}$ proves $\Gamma\ru\Delta$.
\item\label{item:39} $\Gamma\ru\Delta$ holds in every (countable,
locally finite, Kripke) $\base(L)$-extensible $L$-frame.
\item\label{item:41} $\pext^\Sigma_{\base(L)}\vdash_L\Gamma\ru\Delta$,
where $\Sigma=\Sub(\Gamma)$.
\item\label{item:42} $\Gamma\ru\Delta$ holds in every
$L$-model that is $\base(L)$-pseudoextensible wrt~$\Sigma$, and
has size at most~$4^n$, where $n=\sum_{\fii\in\Gamma\cup\Delta}\lh\fii$.
\end{enumerate}
More precisely, let $b$ be the number of boxed subformulas
of~$\Gamma\cup\Delta$, and $m$ the cardinality of
\[(\Sigma\cap\Par)\cup\{\psi,\Box\psi:\Box\psi\in\Sigma\}.\]
If $\Gamma\nadm_L\Delta$, there exists an $L$-model
$\base(L)$-pseudoextensible wrt~$\Sigma$ and
refuting~$\Gamma\ru\Delta$, of size at most
\[3\cdot2^b(2^m+\lh\Delta).\]
\end{Thm}
\begin{Pf}
The equivalence of the conditions follows from
\th\ref{thm:clxadmchar,thm:pext}, except for the size bound.
Assume $\Gamma\nadm_L\Delta$, and let us estimate the size of the
countermodel $F$ to~\ref{item:41} constructed using
\th\ref{lem:fmprules}. Let
\[\Theta=(\Sigma\cap\Par)\cup\{\psi,\Box\psi:\Box\psi\in\Sigma\},\]
and denote by~$B(\Theta)$ its Boolean closure. Notice that premises and
conclusions of all rules from~$\pext^\Sigma_{\base(L)}$ are
in~$B(\Theta)$. In the proof of \th\ref{lem:fmprules}, we find a
suitable partition $B(\Theta)\cup\Gamma\cup\Delta=X\cupd Y$ such that
$\Gamma\sset X$ and $\Delta\sset Y$, for every~$\psi\in Y$ we fix an
$L$-model $F_\psi\model X$, $F_\psi\nmodel\psi$, and we define~$F$ as
the disjoint union of all the~$F_\psi$'s.

Since all boxed subformulas of $B(\Theta)\cup\Gamma\cup\Delta$ are
already subformulas of~$\Gamma\cup\Delta$, we can
make~$\lh{F_\psi}<3\cdot2^b$ for every~$\psi\in Y$ by
\th\ref{thm:clxfinax}. We have $\lh\Delta$ models~$F_\psi$
for~$\psi\in\Delta$. As for the rest, the number of nonequivalent
formulas in~$Y\cap B(\Theta)$ may be as large
as~$2^{2^{\lh\Theta}}=2^{2^m}$, however we will not need so many
models. Every $\psi\in B(\Theta)$ can be expressed in full conjunctive
normal form as $\psi=\ET_i\psi_i$, where each~$\psi_i$ is a clause of
the form
\begin{equation}\label{eq:23}
\LOR_{\tet\in\Theta}\tet^{e(\tet)}
\end{equation}
for some $e\in\two^\Theta$. Since $X$ is closed under~$\vdash_L$,
$\psi\in Y$ iff $\psi_i\in Y$ for some~$i$, and if we include in~$F$ a
model $F_{\psi_i}\nmodel\psi_i$, we will automatically have
$F\nmodel\psi$. Thus, it suffices to include in~$F$ models~$F_\psi$
only for $\psi\in\Delta$ or~$\psi\in Y$ of the
form~\eqref{eq:23}, and there are at most $\lh\Delta+2^m$ such
formulas~$\psi$, which gives $\lh F<3\cdot2^b(2^m+\lh\Delta)$. We can
also estimate $2^b(2^m+\lh\Delta)\le2^{2n-2}$, hence $\lh F<4^n$.
\end{Pf}

\subsection{Single-conclusion bases}\label{sec:single-concl-bases}

\th\ref{thm:clxadmchar,thm:pext,thm:clxadm} provide a description of
multiple-conclusion admissible rules of $\Par$-extensible logics.
Clearly, the description also applies to single-conclusion rules as
its special case, however it does not provide \emph{bases} of
single-conclusion admissible rules. We will construct such bases in
this section; it amounts to axiomatization of single-conclusion
fragments of consequence relations generated by extension rules.

We will distinguish two cases, depending on the properties of the
logic. A logic $L\Sset\lgc{K4}$ is called \emph{linear}, if it is
complete wrt a class of general frames $\p{W,<,A}$ such that the
induced relation~$\le$ is a linear preorder; equivalently, a logic is
linear iff it has width~$1$ iff it extends~$\lgc{K4.3}$. Notice that a
$\Par$-extensible logic is \emph{not} linear iff it is
$\p{C,2}$-extensible for some~$C\in\{\I,\nr1\}$.

If $L$ is a linear $\Par$-extensible logic, the multiple-conclusion basis
given in \th\ref{thm:clxadmchar} consists of rules with at most one
conclusion, hence we can easily fix it up to obtain a
single-conclusion basis.
\begin{Def}\th\label{def:emptyconc}
For any rule $\Gamma\ru\Delta$, we define
$(\Gamma\ru\Delta)^\bot=\Gamma\ru\Delta,\bot$. 
If $B$ is a set of rules, let $B^\bot:=\{\roo^\bot:\roo\in B\}$.
\end{Def}

\begin{Lem}\th\label{lem:emptyconc}
Let $L\Sset\lgc{K4}$, $X$ a set of rules, and $\Gamma\ru\Delta$ a
rule. Then
\[\Gamma\vdash_{L+X^\bot}\Delta\iff
\Delta\ne\nul\text{ and }\Gamma\vdash_{L+X}\Delta.\]
\end{Lem}
\begin{Pf}
The set of rules with nonempty conclusion is closed under cut, hence
the right-hand side defines a consequence relation. Let us call it
$\vdash_1$. On the one hand, $\vdash_1$ includes~$L$ and all rules
from~$X^\bot$, hence ${\vdash_{L+X^\bot}}\sset{\vdash_1}$. On the
other hand, $\{\roo:\roo^\bot\in L+X^\bot\}$
defines a consequence relation including~$L+X$, therefore
$\Gamma\vdash_1\Delta$ implies
$\Gamma\vdash_{L+X^\bot}\Delta,\bot$. If $\Delta\ne\nul$, we can use
cut on $\bot\vdash_L\psi$ for any~$\psi\in\Delta$ to obtain
$\Gamma\vdash_{L+X^\bot}\Delta$.
\end{Pf}
\begin{Rem}\th\label{rem:emptyconc}
Obviously, the only property of~$L$ we used in the proof is
$\bot\vdash_Lx$. If $L$ is an arbitrary logic, an analogous
lemma holds where we use a variable not appearing in~$\Gamma\cup\Delta$ instead
of~$\bot$ in \th\ref{def:emptyconc}.
\end{Rem}

\begin{Cor}\th\label{thm:scbasislin}
If $L$ is a linear $\Par$-extensible logic, then
single-conclusion
$L$-admissible rules have a basis consisting of the rules
\[\begin{cases}
\bigl(\Ext^\Par_{C,0}\bigr)^\bot,&n=0,\\
\Ext^\Par_{C,1},&n=1
\end{cases}\]
for every $\p{C,n}\in\EC$ such that $L$ has a type-$\p{C,n}$ frame,
and $\lh C\le2^{\lh\Par}$.
\noproof\end{Cor}

We now turn to non-linear $\Par$-extensible logic. Any such logic
admits the \emph{disjunction property} rules
$\DP=\{\DP_n:n\in\omega\}$, where $\DP_n$ is
\[\LOR_{i<n}\Box x_i\Ru\{x_i:i<n\}.\]
Notice that $\lgc{K4}+\DP=\lgc{K4}+\DP_0+\DP_2$: the rule $\DP_1$ is a
substitution instance of~$\DP_2$, and then we can prove~$\DP_n$ in
$\lgc{K4}+\DP_0+\DP_2$ by induction on~$n$:
\[\LOR_{i<n+1}\Box x_i
\vdash_\lgc{K4}\Box x_n\lor\Box\LOR_{i<n}\Box x_i
\vdash_{\DP_2}x_n,\LOR_{i<n}\Box x_i,\]
hence
\[\LOR_{i<n+1}\Box x_i\vdash_{\lgc{K4}+\DP_2+\DP_n}\{x_i:i<n+1\}\]
by a cut. Since every rule~$\Gamma\ru\Delta$ is equivalent over $\lgc{K4}+\DP$ to a
single-conclusion rule (namely~$\Gamma\ru\LOR_{\psi\in\Delta}\Box\psi$), we are left with the question for which
sets~$B$ of single-conclusion rules is $L+B+\DP$ conservative
over~$L+B$ in the sense that it proves the same single-conclusion rules. We will use the following general result.
\begin{Lem}\th\label{lem:maxmulti}
For every single-conclusion consequence relation~$\vdash$, there
exists a largest multiple\hyph conclusion consequence relation~$\vdash_m$
whose single-conclusion fragment is~$\vdash$, and it can be described
explicitly by
\begin{equation}\label{eq:24}
\Gamma\vdash_m\Delta\iff
\forall\Theta,\fii,\sigma\,\bigl(
  \forall\psi\in\Delta\,(\sigma(\psi),\Theta\vdash\fii)
  \TO\sigma(\Gamma),\Theta\vdash\fii\bigr),
\end{equation}
where the quantification is over all finite sets of formulas~$\Theta$,
formulas~$\fii$, and substitutions~$\sigma$.
\end{Lem}
\begin{Pf}
The right-hand side of~\eqref{eq:24} defines a consequence relation.
For example, we verify that $\vdash_m$ is closed under cut. Assume
that $\Gamma\vdash_m\Delta,\chi$, $\Gamma,\chi\vdash_m\Delta$, and
$\sigma(\psi),\Theta\vdash\fii$ for every $\psi\in\Delta$. Using
$\Gamma,\chi\vdash_m\Delta$, we have
$\sigma(\Gamma),\sigma(\chi),\Theta\vdash\fii$, hence
$\sigma(\psi),\sigma(\Gamma),\Theta\vdash\fii$ for every
$\psi\in\Delta\cup\{\chi\}$. Since $\Gamma\vdash_m\Delta,\chi$, we
obtain $\sigma(\Gamma),\Theta\vdash\fii$.

Since $\vdash$ is closed under cuts and substitutions, $\vdash_m$
extends~$\vdash$. On the other hand, $\Gamma\vdash_m\psi$ implies
$\Gamma\vdash\psi$ by taking $\Theta=\nul$, $\fii=\psi$, and
$\sigma=\id$.

Let $\vdash'$ be another consequence relation whose single-conclusion
fragment is~$\vdash$, and assume $\Gamma\vdash'\Delta$. If
$\Theta$, $\fii$, $\sigma$ are such that
$\sigma(\psi),\Theta\vdash\fii$ for every~$\psi\in\Delta$, we have
$\sigma(\psi),\Theta\vdash'\fii$ for every~$\psi\in\Delta$ as
${\vdash'}\Sset{\vdash}$, hence $\sigma(\Gamma),\Theta\vdash'\fii$ by
a cut with $\sigma(\Gamma)\vdash'\sigma(\Delta)$. This is a
single-conclusion rule, hence $\sigma(\Gamma),\Theta\vdash\fii$. Thus, $\Gamma\vdash_m\Delta$.
\end{Pf}

\begin{Lem}\th\label{lem:dpcons}
Let $L\Sset\lgc{K4}$, and $B$ be a set of single-conclusion rules.
Then the following are equivalent.
\begin{enumerate}
\item\label{item:43}
$L+B$ is the single-conclusion fragment of~$L+B+\DP$.
\item\label{item:44}
For every $\fii$, $\psi$, and~$\chi$: $\psi\vdash_{L+B}\fii$
and $\chi\vdash_{L+B}\fii$ implies
$\Box\psi\lor\Box\chi\vdash_{L+B}\fii$.
\item\label{item:45}
$\Box x\vdash_{L+B}x$, and
for every $\fii$, $\psi$, and~$\xi$: $\psi\vdash_{L+B}\fii$
implies $\Box\xi\lor\Box\psi\vdash_{L+B}\Box\xi\lor\Box\fii$.
\item\label{item:46}
$\Box x\vdash_{L+B}x$, and
for every $(\Gamma\ru\fii)\in B$,
\[\Box x\lor\ET_{\psi\in\Gamma}\Box\psi\vdash_{L+B}\Box x\lor\Box\fii,\]
where $x$ is a variable not occurring in~$\Gamma\cup\{\fii\}$.
\end{enumerate}
\end{Lem}
\begin{Pf}
Let $\vdash$ denote the consequence relation~$L+B$.

\ref{item:44}${}\to{}$\ref{item:43}:
By \th\ref{lem:maxmulti}, it suffices to show that
$\DP\sset{\vdash_m}$. For $\DP_0$, this amounts to
$\bot,\Theta\vdash\fii$. As for $\DP_2$, assume
$\psi,\Theta\vdash\fii$ and $\chi,\Theta\vdash\fii$. We have
$\psi\land\tet\vdash\fii$ and $\chi\land\tet\vdash\fii$, where
$\tet=\ET\Theta$, hence
\[\Box\psi\lor\Box\chi,\Theta\vdash_\lgc{K4}
  \Box(\psi\land\tet)\lor\Box(\chi\land\tet)\vdash\fii\]
using~\ref{item:44}.

\ref{item:45}${}\to{}$\ref{item:44}:
If $\psi\vdash\fii$ and $\chi\vdash\fii$, then
\[\Box\psi\lor\Box\chi\vdash\Box\fii\lor\Box\chi\vdash\Box\fii\lor\Box\fii\vdash\fii\]
by~\ref{item:45}.

\ref{item:46}${}\to{}$\ref{item:45}:
Define
\[\Gamma\vdash_1\fii\iff\Box\xi\lor\Box\ET_{\psi\in\Gamma}\psi\vdash\Box\xi\lor\Box\fii
  \quad\text{for every $\xi$.}\]
Then $\vdash_1$ is a single-conclusion consequence relation
including~$L$. We will verify closure under cut: assume
$\Gamma\vdash_1\chi$ and $\Gamma,\chi\vdash_1\fii$. Put
$\gamma=\ET\Gamma$. We have
\begin{align*}
\Box\xi\lor\Box\gamma&\vdash\Box\xi\lor\Box\chi,\\
\Box\xi\lor\Box\gamma&\vdash\Box\xi\lor\Box\gamma,
\end{align*}
hence
\[\Box\xi\lor\Box\gamma\vdash\Box\xi\lor\Box(\gamma\land\chi)
  \vdash\Box\xi\lor\Box\fii.\]
Clearly, $\vdash_1$ includes $B$ by~\ref{item:46} and substitution,
hence it includes $\vdash$, which gives~\ref{item:45}.

\ref{item:43}${}\to{}$\ref{item:46}:
$\Box x\vdash x$ is a special case of~$\DP$. For any $\Gamma\ru\fii$
in~$B$, we have
\begin{gather*}
\Box x\lor\Box\ET\Gamma
\vdash_\DP x,\ET\Gamma,\\
x\vdash_\lgc{K4}\Box x\lor\Box\fii,\\
\ET\Gamma\vdash_B\fii\vdash_\lgc{K4}\Box x\lor\Box\fii,
\end{gather*}
hence
$$\Box x\lor\Box\ET\Gamma\vdash_{L+B+\DP}\Box x\lor\Box\fii.\qedhere$$
\end{Pf}
\th\ref{lem:dpcons} suggests that we can turn a basis into a
single-conclusion basis by taking the single-conclusion rules
$\Gamma\ru\LOR_{\psi\in\Delta}\Box\psi$ equivalent to rules from the
basis over~$\DP$, and adding ``side variables'' in the spirit
of~\ref{item:46}. The way we do it below (unboxing the side
variable in the conclusion) also ensures the property $\Box x\vdash_{L+B}x$.
\begin{Def}\th\label{def:scsiderule}
For any rule $\Gamma\ru\Delta$, we define
$(\Gamma\ru\Delta)^\lor$ to be the rule
\[\Box x\lor\ET_{\fii\in\Gamma}\Box\fii
  \Ru x\lor\LOR_{\psi\in\Delta}\Box\psi,\]
where $x$ is a variable not occurring in~$\Gamma\cup\Delta$. If $B$ is
a set of rules, we put $B^\lor:=\{\roo^\lor:\roo\in B\}$.

We note that for the specific case of extension rules,
$\Ext^\lor_{\I,n,\{e\}}$ is
\[\Box z\lor\Box\Bigl(P^e\land\Box y\to\LOR_{i<n}\Box x_i\Bigr)
  \Ru z\lor\LOR_{i<n}\Box(\boxdot y\to x_i),\]
and $\Ext^\lor_{\R,n,E,e_0}$ is
\[\fracd{\Box z\lor\Box\Bigl[
  P^{e_0}\land\boxdot\Bigl(y\to\LOR_{e\in E}\Box(P^e\to y)\Bigr)
     \land\ET_{e\in E}\boxdot\bigl(\Box(P^e\to\Box y)\to y\bigr)
   \to\LOR_{i<n}\Box x_i\Bigr]}
{z\lor\LOR_{i<n}\Box(\boxdot y\to x_i)}.\]
\end{Def}

\begin{Thm}\th\label{thm:clxadmsc}
Let $L$ be a non-linear $\Par$-extensible logic, and
\[T=\bigl\{\p{C,n}:L\text{ has a type-$\p{C,n}$ frame, and }\lh
  C\le2^{\lh\Par}\bigr\}.\]
Then $\bigl\{\bigl(\Ext^\Par_{C,n}\bigr)^\lor:\p{C,n}\in T\bigr\}$ is a basis of
$L$-admissible single-conclusion rules.
\end{Thm}
\begin{Pf}
Clearly, $(\Gamma\ru\Delta)^\lor$ is derivable in
$L+(\Gamma\ru\Delta)+\DP$, hence all rules in
\[B:=\bigl\{\bigl(\Ext^\Par_{C,n}\bigr)^\lor:\p{C,n}\in T\bigr\}\]
are $L$-admissible.

On the other hand, since $L$ is consistent, it has a type-$\p{C,0}$
frame for some~$C$,
and $\bigl(\Ext^\Par_{C,0}\bigr)^\lor$ derives $\Box z\ru z$, i.e.,
\[\Box x\vdash_{L+B}x.\]
Moreover, if $\Box z\lor\Box\fii\ru z\lor\LOR_{i<n}\Box\psi_i$ is one
of the rules $\bigl(\Ext^\Par_{C,n}\bigr)^\lor\in B$, we can derive
\begin{align*}
\Box u\lor\Box(\Box z\lor\Box\fii)
&\vdash_\lgc{K4}\Box\bigl(\Box(\Box u\lor\Box z)\lor\Box\fii\bigr)\\
&\vdash_{L+B}\Box(\Box u\lor\Box z)\lor\Box\fii\\
&\vdash_{L+B}\Box u\lor\Box z\lor\LOR_{i<n}\Box\psi_i\\
&\vdash_\lgc{K4}\Box u\lor\Box\Bigl(z\lor\LOR_{i<n}\Box\psi_i\Bigr).
\end{align*}
Thus, $L+B$ derives all single-conclusion rules provable in $L+B+\DP$
by \th\ref{lem:dpcons}. However, if $\fii\ru\{\psi_i:i<n\}$ is again
one of the rules $\Ext^\Par_{C,n}$, $\p{C,n}\in T$, we have
\[\fii\vdash_\lgc{K4}\Box\bot\lor\Box\fii\vdash_B\bot\lor\LOR_{i<n}\Box\psi_i\vdash_\DP\{\psi_i:i<n\},\]
hence $L+B+\DP$ includes the basis of~$\adm_L$ from
\th\ref{thm:clxadmchar}. Thus, $L+B$ derives all single-conclusion
$L$-admissible rules.
\end{Pf}

\subsection{Finite and independent bases}\label{sec:finite-indep-bases}

In this section, we investigate whether $\Par$-extensible logics have
finite or independent bases of admissible rules with parameters.
Recall that a basis of admissible rules~$B$ is \emph{independent} if
for every~$\roo\in B$, $B\bez\{\roo\}$ is not a basis. Clearly, a
finite basis can be made independent by successively omitting
redundant rules, but this may not be possible for infinite bases.

First, we observe that finite bases are out of question in the
presence of infinitely many parameters, apart from the trivial cases
of inconsistent logics (which have the empty set for a basis) and
logics with no tautologies (which have a multiple-conclusion basis
$\{x\ru\nul\}$, and a single-conclusion basis $\{x\ru y\}$; of course,
modal logics always have some tautologies anyway).
\begin{Prop}\th\label{prop:infparbasis}
If $L$ is a consistent logic with at least one tautology, then every
basis of $L$-admissible multiple-conclusion or single-conclusion rules
has to involve all parameters.

In particular, if\/ $\Par$ is infinite, then $L$ has no finite basis of
multiple-conclusion or single-conclusion admissible rules.
\end{Prop}
\begin{Pf}
Assume for contradiction that $B$ is a basis, and $p\in\Par$ does not
appear in~$B$. We have $p\adm_Lx$, hence the rule $p\ru x$ is
derivable in $L+B$. Since $L$ is closed under atomic substitutions, and $p$ does not appear in~$B$, we can substitute
another variable~$y$ for~$p$ in the derivation. Thus, $L+B$ derives
$y\ru x$, i.e., $y\adm_Lx$. If $L$ has a tautology, we can substitute
it for~$y$, and we can substitute an arbitrary formula for~$x$, hence $L$ is
inconsistent.
\end{Pf}
Nevertheless, logics can have independent bases of admissible rules in
infinitely many parameters in a nontrivial way. Consider the simplest
example of the classical logic~$\CPC$. One can check easily that it
has a basis~$B$ consisting of the rules
\begin{equation}\label{eq:25}
\neg P^e\ru\bot,\qquad e\in\two^P,P\sset\Par\text{ finite}
\end{equation}
(together with the rule $\bot\ru\nul$ in the multiple-conclusion
case). This basis is not independent, since $\neg P'^{e'}\vdash_\CPC\neg
P^e$ when $e'$ is the restriction of~$e$ to~$P'\sset P$. However, we
can make it independent by splitting~\eqref{eq:25} into rules that do
away with one parameter at a time: namely, if we bijectively enumerate
$\Par=\{p_n:n\in\omega\}$, and put $P_n=\{p_i:i<n\}$, the set of
rules
\begin{equation}\label{eq:26}
\neg P_{n+1}^e\ru\neg P_n^{e\res P_n},\qquad n\in\omega,e\in\two^{P_{n+1}}
\end{equation}
is equivalent to~$B$, and independent over~$\CPC$ (if we fix a rule~$\roo$ as
in~\eqref{eq:26}, the set of formulas implied by $\neg
P_{n+1}^e$ is closed under the rules of~$\CPC$, and
rules of the form~\eqref{eq:26} other than~$\roo$, but it is not closed under~$\roo$).

This kind of transformation works well for variable-free rules such
as~\eqref{eq:25}, but it is rather unclear how to adapt it to
extension rules. We thus leave it
as an open problem whether clx logics have independent bases of
admissible rules with infinitely many parameters.

We now turn to the case of finitely many parameters. We define a
transitive modal logic~$L$ to have \emph{bounded branching} if it
includes~$\lgc{K4BB}_k$ for some~$k\in\omega$. If $L$ has fmp, this is
equivalent to the condition that all finite $L$-frames have branching
at most~$k$.
\begin{Lem}\th\label{lem:finbas}
If\/ $\Par$ is finite and $L$ is a $\Par$-extensible logic with bounded
branching, then $L$ has a finite basis (and a fortiori an independent
basis) of either multiple-conclusion or single-conclusion admissible rules.
\end{Lem}
\begin{Pf}
There are only finitely many $\p{C,n}$ such that $\lh C\le2^{\lh\Par}$
and $L$ has a type-$\p{C,n}$ frame, as well as finitely many choices
for $e_0\in E\sset\two^\Par$, hence the bases given in
\th\ref{thm:clxadmchar,thm:scbasislin,thm:clxadmsc} are finite.
\end{Pf}

For logics with unbounded branching, the bases we constructed earlier
are infinite, and we will show that in fact no finite bases exist;
nevertheless the logics do have independent bases. The main source of
non-independence in the bases given in
\th\ref{thm:clxadmchar,thm:clxadmsc} is that $L+\Ext_{C,n}$
derives~$\Ext_{C,m}$ when $m\le_0n$, as an $m$-element set~$X$ whose
tight predecessor we are seeking can be (non-injectively) enumerated
as $\{w_i:i<n\}$. We get around this problem in a similar way as
in~\cite{ej:indep} by considering variants of extension rules that
express the existence of tight predecessors of antichains of size
exactly~$n$.
\begin{Def}\th\label{def:extrulesindep}
Assume that $\Par$ is finite. For any~$n\in\omega$
and~$e\in\two^\Par$, $\Ext^=_{\I,n,\{e\}}$ denotes the rule
\[P^e\land\Box y\to\LOR_{i<n}\Box x_i\Ru
   \Bigl\{\boxdot\Bigl(y\land\ET_{j\ne i}x_j\Bigr)\to x_i:i<n\Bigr\}.\]
Note that the big conjunction is empty if~$n\le1$.
If $n\ne1$ and $\nul\ne E\sset\two^\Par$, $\Ext^=_{\R,n,E}$ is
defined as
\[\fracd{\boxdot\Bigl(y\to\LOR_{e\in E}\Box(P^e\to y)\Bigr)
     \land\ET_{e\in E}\boxdot\bigl(\Box(P^e\to\Box y)\to y\bigr)
   \to\LOR_{i<n}\Box x_i}
{\Bigl\{\boxdot\Bigl(y\land\ET_{j\ne i}x_j\Bigr)\to x_i:i<n\Bigr\}}.\]
\end{Def}

\begin{Lem}\th\label{lem:indeprulesem}
Assume that $\Par$ is finite, and let $W$ be a descriptive or Kripke
parametric frame.
\begin{enumerate}
\item\label{item:48}
If $*\in\{\I,\R\}$, $n\in\omega$, and $\nul\ne E\sset\two^\Par$,
where $\lh E=1$ if~$*=\I$, and $n\ne1$ if~$*=\R$, then
$W\model\Ext^=_{*,n,E}$ iff every antichain $X\sset W$ of size~$n$ has
a $\p{*,E}$-tp.
\item\label{item:49}
If $e\in E\sset\two^\Par$, then $W\model\Ext_{\R,1,E,e}$ iff
$\{w\}$ has a $\p{\R,E}$-tp whenever $w\in W$ is irreflexive, or no
point of~$\cls(w)$ satisfies $\Par^e$.
\end{enumerate}
\end{Lem}
\begin{Pf}
By a straightforward modification of the proofs of
\th\ref{thm:irrextchar,thm:reflextchar}, which we leave to the reader.
Recall that the hard part of~\ref{item:49} was already noted in
\th\ref{rem:RnEe}.
\end{Pf}

\begin{Lem}\th\label{lem:mcindepbas}
Assume that $\Par$ is finite. Let $L$ be a $\Par$-extensible logic,
and put
\[T=\bigl\{\p{C,n}\in\EC:\text{$L$ has a type-$\p{C,n}$ frame,
 and $\lh C\le2^{\lh\Par}$}\bigr\}.\]
For every $\p{C,n}\in T$, we include in~$B$ the following rules.
\begin{enumerate}
\item\label{item:50} If $C=\I$: $\Ext^=_{\I,n,E}$ for every $E=\{e\}$,
$e\in\two^\Par$.
\item\label{item:51} If $C=\nr k$ and~$n\ne1$: $\Ext^=_{\R,n,E}$ for
every $E\sset\two^\Par$ such that~$\lh E=k$.
\item\label{item:52} If $C=\nr k$ and~$n=1$: $\Ext_{\R,1,E,e}$ for
every $e\in E\sset\two^\Par$ such that~$\lh E=k$, unless $\Par=\nul$
and~$L\Sset\lgc{S4}$.
\end{enumerate}
Then $B$ is an independent basis of $L$-admissible rules.
\end{Lem}
\begin{Pf}
Let $W$ be a descriptive $L$-frame. For every finite $X\sset W$, there
exists an antichain $Y\sset X$ (nonempty if $X$ is nonempty) such that
$X\Up=Y\Up$. Then it follows from
\th\ref{cor:extchar,lem:indeprulesem} that $W\model B$ iff
$W\model\Ext^\Par_t$ for every~$t\in T$; the only thing to note is
that the rule omitted in the exceptional case of~\ref{item:52} is
valid in~$W$ automatically (in other words, it is derivable
in~$\lgc{S4}$), because in the absence of parameters, a reflexive
point is its own reflexive tight predecessor. Thus, $L+B$ is equivalent
to $L+\{\Ext^\Par_t:t\in T\}$, and $B$ is a basis of $L$-admissible
rules by \th\ref{thm:clxadmchar}.

It remains to show that $B$ is independent. Fix $\roo\in B$
corresponding to $C$, $n$, $E$, and (in case~\ref{item:52}) $e$, we will
construct a parametric Kripke $L$-frame where $B\bez\{\roo\}$ is valid, while
$\roo$ is not. Let $F_0$ be a finite $L$-frame generated by an antichain
$X=\{w_i:i<n\}$ (notice that $\p{C,n}\in T$
implies that $L$ is consistent). If $n=1$, we can
arrange that $w_0$ is irreflexive if $C=\I$ or~$\Par=\nul$, or
otherwise that $\cls(w_0)$ realizes~$\Par^{e'}$ for every $e'\in
E\bez\{e\}$, but not $\Par^e$.

As in the proof of
\th\ref{thm:extlocfin}, we will construct by induction on~$k$ a
sequence of finite 
$L$-frames $F_0\sset F_1\sset F_2\sset\cdots$ whose underlying sets
are included in a countable set~$Z$. Let
$\{\p{C_k,n_k,E_k,X_k}:k\in\omega\}$ be an enumeration of all
quadruples where $\p{C_k,n_k}\in T$, $E_k\sset\two^\Par$,
$\lh{E_k}=\lh{C_k}$, $X_k\sset Z$, $\lh{X_k}=n_k$, such that every
quadruple appears infinitely many times in the enumeration. Assume
that $F_k$ has already been constructed. If $X_k\nsset F_k$, $X_k$
is not an antichain, or $X_k$ has a $\p{*_k,E_k}$-tp in~$F_k$ (where
$*_k\in\{\I,\R\}$ is the reflexivity of~$C_k$), or
if $X_k\Up=X\Up$, $C_k=C$, and~$E_k=E$, we put $F_{k+1}=F_k$.
Otherwise, $F_{k+1}$ consists of~$F_k$ together with a
$\p{*_k,E_k}$-tp of~$X_k$ whose elements are taken from $Z\bez F_k$.
Let $F=\bigcup_{k\in\omega}F_k$.

Since $L$ is $\p{C_k,n_k}$-extensible, we obtain by induction that
every~$F_k$ is an $L$-frame, hence also $F$ is an $L$-frame. If $*$
denotes the reflexivity of~$C$, then $X$ has no $\p{*,E}$-tp in~$F_0$,
and we never added one in the later stages, hence it has no
$\p{*,E}$-tp in~$F$. In case \ref{item:52}, we also ensured that the
root cluster of~$X$ is irreflexive, or avoids~$e$. Thus,
$F\nmodel\roo$ by \th\ref{lem:indeprulesem}. On the other hand, we
have $F\models B\bez\{\roo\}$, since we added all the required tight
predecessors. The only potential problem is with rules
$\Ext_{\R,1,E,e'}$ for~$e'\ne e$ if $n=1$ and~$C$ is reflexive, as
we did not include a $\p{*,E}$-tp of~$X$ in~$F$. However, in this case
$\Par^{e'}$ is satisfied in~$\rcl(X)$.
\end{Pf}

\begin{Lem}\th\label{lem:scindepbas}
Assume that $\Par$ is finite. Let $L$ be a non-linear
$\Par$-extensible logic, and $B$ the basis from
\th\ref{lem:mcindepbas}. Define $B_1$ as the set of rules $\roo^\lor$
for~$\roo\in B$, except that we omit $(\Ext^=_{\R,0,E})^\lor$ if\/
$\Par=\nul$ and~$L\Sset\lgc{D4.1}$.
Then $B_1$ is an independent basis of $L$-admissible single-conclusion
rules.
\end{Lem}
\begin{Pf}
First, assume that $\Par=\nul$ and~$L\Sset\lgc{D4.1}$. If
$L\nSset\lgc{S4}$, $B_1$ includes $(\Ext^=_{\I,1,E})^\lor$
or~$(\Ext_{\R,1,E,e})^\lor$, where $E=\two^\nul$ and $e$ is its unique
element, hence it derives $\Box z\ru z\lor\Box(\boxdot y\to x_0)$.
Substituting $y\mapsto\top$ and~$x_0\mapsto\bot$, we obtain $\Box z\ru
z\lor\Box\bot$, hence $\Box z\ru z$ as $L\Sset\lgc{D4}$. (If
$L\Sset\lgc{S4}$, the rule $\Box z\ru z$ is outright derivable
in~$L$.) Now, the omitted rule~$(\Ext^=_{\R,0,E})^\lor$ is equivalent
to $\Box z\lor\Box\neg\boxdot(y\eq\Box y)\ru z$, and
$\vdash_\lgc{D4.1}\neg\Box\neg\boxdot(y\eq\Box y)$, hence
$(\Ext^=_{\R,0,E})^\lor$ is derivable in~$\lgc{D4.1}+\Box z\ru z\sset
L+B_1$.

Thus, $L+B_1$ is equivalent to $L+B^\lor$. The same argument as in
\th\ref{thm:clxadmsc} shows that $L+B^\lor+\DP=L+B$ is conservative
over~$L+B^\lor$, hence $B^\lor$ and~$B_1$ are bases of $L$-admissible
single-conclusion rules by \th\ref{lem:mcindepbas}.
In order to show that $B_1$ is independent, let $\roo^\lor\in B_1$
with~$\roo\in B$, we need to construct a frame $F\model
L+(B_1\bez\{\roo^\lor\})$ where $\roo^\lor$ fails. Let $C,n,E,e,*$ be
the data associated with~$\roo$ as in the proof of
\th\ref{lem:mcindepbas}.

If $\Par=\nul$, $n=0$, and $\vdash_L\diadot\Box\bot$ (whence $C=\I$),
we can take for~$F$ the irreflexive one-element model: the rule
$\Box\bot\ru\bot$, implied by~$(\Ext^=_{\I,0,E})^\lor$, fails in~$F$, whereas
any other rule in~$B_1$ corresponds to~$n'>0$, thus its conclusion
includes a boxed disjunct, and as such the rule holds in any model
satisfying~$\Box\bot$.

Otherwise, we construct~$F$ as in the proof of
\th\ref{lem:mcindepbas} with the following modification: when defining
$F_{k+1}$ from~$F_k$, we also include as a disjoint part of the model
a fresh copy of a fixed finite $L$-frame~$G$. The original argument remains
valid, hence $F\model L+(B\bez\{\roo\})$ and~$F\nmodel\roo$, as long
as $G$ does not include a $\p{*,E}$-tp of~$X$. This can only happen
if~$n=0$, and we avoid it by choosing~$G$ more carefully in this case:
if $\Par\ne\nul$, we take for~$G$ a one-element model
satisfying~$\Par^{e'}$, where~$E\ne\{e'\}$. If $\Par=\nul$, we
take a one-element model of different reflexivity than~$*$, if
possible. Since we already dealt with the case when $\R\nmodel L$
(i.e., $\vdash_L\diadot\Box\bot$), the only remaining possibility is
that $\I\nmodel L$. If the two-element cluster is an $L$-frame, we
take it for~$G$. If not, we have $L\Sset\lgc{D4.1}$, but then $B_1$ does not include
any rule with~$n=0$, so this case cannot happen.

Since $B_1$ includes at least one rule~$\roo'$ with~$n'=2$, $F$ is
downward directed: if $u,v\in F_k$, there is $w\in F_{k+1}$
incomparable with~$u,v$ in~$F_{k+1}$, hence due to tight predecessors
added for~$\roo'$, there are $w',w''\in F$ such that $w'<u,w$
and~$w''<v,w'$, in particular~$w''<u,v$. Consequently, $F\model\DP$.
Since $\roo$ is equivalent to $\roo^\lor$ over $\lgc{K4}+\DP$, this
implies $F\model B_1\bez\{\roo^\lor\}$ and~$F\nmodel\roo^\lor$.
\end{Pf}

\begin{Thm}\th\label{thm:indepbas}
Assume that $\Par$ is finite, and $L$ is a $\Par$-extensible logic.
Then $L$ has an independent basis of either multiple-conclusion or
single-conclusion admissible rules. Moreover, $L$ has a finite basis
if and only if $L$ has bounded branching.
\end{Thm}
\begin{Pf}
If $L$ has bounded branching, it has finite and independent bases by
\th\ref{lem:finbas}, thus let us assume that $L$ has unbounded branching.
$L$ has an independent basis~$B$ (or $B_1$ in the single-conclusion
case) by \th\ref{lem:mcindepbas} (\th\ref{lem:scindepbas}, resp.).
This basis is infinite, since $T$ includes $\p{*,n}$ for arbitrarily
large~$n$ for some $*\in\{\I,\nr1\}$. If we assume for contradiction
that $L$ has a finite basis, then all its elements are derivable
over~$L$ from a finite fragment of~$B$ ($B_1$, resp.), which is then
also a basis. But this contradicts the independence of~$B$ ($B_1$, resp.).
\end{Pf}

\section{Conclusion}\label{sec:conclusion}

As we have demonstrated in this paper, essential parts of the theory
of admissible rules and unification in transitive modal logics
satisfying extension properties appropriately generalize to
admissibility and unification with parameters: this includes the
semantic description of projective formulas and the existence of
projective approximations (which implies finitary unification type
and decidability of admissibility), the existence of relatively
transparent bases of admissible rules reflecting extension conditions
satisfied by finite frames for the logic, basic properties of bases
(their finite or independent axiomatizability), and  semantic
correspondence of the consequence relation given by admissible rules
to frames having suitable tight predecessors.

On the one hand, the results and methods used are similar to the
parameter-free case on the most basic level. On the other hand, there
are also significant differences that mostly boil down to the fact
that admissibility and unification without parameters is insensitive
to sizes of clusters in the models involved, whereas we have to take
proper clusters seriously when working with parameters since elements
of a finite cluster can be distinguished by a valuation of parameters.
This makes the analysis technically more complicated, but more
importantly, it shows up in the statements of some of the results:
for example, the L\"owenheim substitutions serving as building blocks
of projective unifiers are no longer based on a simple choice between
two formulas $\boxdot\fii\to x$ and~$\boxdot\fii\land x$. (While it was
not formulated that way in \th\ref{def:proj}, we can view the
functions~$D$ parametrizing the substitutions as follows: we fix the
parametric frame consisting of a large cluster with one point
satisfying each possible combination of the parameters, and we
consider all models~$D$ based on this frame.)

Perhaps even more striking is the effect on bases of admissible rules:
in the parameter-free case, the bases consist of very simple rules in
the form of a ``relativized disjunction property''; with parameters,
the rules corresponding to irreflexive tight predecessors look quite
similar, but the rules for reflexive tight predecessors get more
complex, as they need to express exactly the composition of the tight
predecessor cluster in terms of its valuation of parameters. 

We will see other interesting instances of this phenomenon in the
planned sequel to this paper, which will deal with the computational
complexity of admissible rules. We know from~\cite{ej:admcomp} that in the parameter-free case,
admissibility in the logics in question can be only
$\cxt{coNP}$-complete or $\cxt{coNEXP}$-complete according to whether
the logic is linear or not. In contrast, there
will be several more possibilities for admissibility with parameters,
and they will depend on sizes of clusters occurring in frames of the logic.

Finally, let us recall that the following question was left unresolved
in Section~\ref{sec:finite-indep-bases}:
\begin{Prob}\th\label{prob:infindep}
Do clx logics have independent bases of admissible rules with
infinitely many parameters?
\end{Prob}

\section{Acknowledgements}
I would like to thank Rosalie Iemhoff and Wojciech Dzik for
interesting discussions on the subject, and the anonymous referees for
helpful suggestions.

\bibliographystyle{mybib}
\bibliography{mybib}
\end{document}